\documentclass[doctor]{NTHUthesis}
\usepackage[utf8]{inputenc}

\usepackage{zhlipsum, lipsum}   
\usepackage{graphicx}           
\usepackage{booktabs}           
\usepackage[ruled]{algorithm2e} 
\usepackage{cite}               
\usepackage{multirow}
\usepackage{amsmath}
\usepackage{hyperref}
\usepackage{enumitem}
\usepackage{times}
\usepackage{latexsym}
\usepackage{xspace}
\usepackage{colortbl}
\usepackage{xcolor}         
\usepackage{subcaption}
\usepackage{tcolorbox}
\usepackage[normalem]{ulem}
\useunder{\uline}{\ul}{}
\usepackage{listings}

\usepackage{float}
\usepackage{comment}
\usepackage{mathrsfs}
\usepackage{color}

\titleZH{重新探討語音情緒辨識的建模與評量方法：\\考慮標註者的主觀性與情緒的模糊性}
\titleEN{Revisiting Modeling and Evaluation Approaches in Speech Emotion Recognition:\\ Considering Subjectivity of Annotators and Ambiguity of Emotions}
\instituteZH{電機工程學系博士班 \hspace{1cm} 組別: 系統A}
\studentID{104061701\hspace{0.5cm}}
\studentZH{周惶振}
\studentEN{Huang-Cheng Chou}
\advisorZH{李祈均\hspace{1cm}博士}
\advisorEN{Prof. Chi-Chun Lee}
\yearZH{113}
\monthZH{7}

\begin{document}

\makecover

\pagenumbering{Roman}

\begin{abstractZH}
語音情緒辨識在過去二十年裡獲得了越來越多的關注。建立語音情緒識別系統需要情緒數據庫，資料庫需要有人聲以及人類的情緒感受標記。研究人員們會訓練群眾標記者或內部標記者，在收聽或觀看情緒錄像後，通過選擇預先定義的情緒類別來描述和提供他們的情緒感知。然而，當研究人員們要求標記者們從預定義的情緒中選擇情緒時，觀察到標記者之間出現分歧是很常見的。為了處理這種標記者之間的分歧，大部分專家學者們將分歧視為雜訊，並使用標籤聚合方法來獲得單一的共識情緒標記，作為訓練語音情緒識別系統的學習目標。雖然這種通行做法將任務簡化為單一情緒標籤識別任務，但這個方法忽略了人類情緒感知的自然行為。 在本論文中，我們主張應重新檢視語音情感識別中的建模和評估方法。主要的研究問題是：(1) 我們是否應該移除少數的情感評分？(2) 我們是否應該只有讓語音情感識別系統學習少數人的情感感知？(3) 語音情感識別系統是否應該每次只預測一種情緒類別？

從心理學領域相關研究成果發現情緒感受是主觀的，每個個體對同一情感刺激的情緒感受可能有所不同。此外，人類感知中的情緒類別界限是重疊、混合且模糊的。這些情感的模糊性和情緒感受的主觀性的發現啟發我們重新審視在語音情緒辨識中的建模與評估方法。本博士論文探討了構建語音情緒識別系統的三個層面的新穎觀點。首先，我們接受情緒感受的主觀性，並考慮標記者的所有情緒標記。傳統的方法只允許每位標記者對每個樣本給予一票情緒標記，但我們藉由考慮所有標記者的所有標記，利用現有的軟標籤方法(soft-label)重新計算標籤表示方式。此外，我們直接利用個別標記者的情緒標記來訓練個別標記者的語音情緒識別系統，並聯合訓練個別標記者語音情緒識別系統和標準語音情緒識別系統(使用共識標籤)。在使用多數決所獲得的共識標籤當作最終情緒標籤進行測試時，個別標記者的建模方法提升了語音情緒辨識系統的性能。

其次，我們重新思考了評估語音情緒辨識系統的方法以及語音情緒辨識任務的制定和定義。我們主張在評估語音情緒辨識系統的性能時，不應該刪除任何數據和情緒標記。此外，我們認為語音情緒辨識任務的定義可以包含情緒的共現性（例如，悲傷和生氣）。因此，樣本的真實標籤不應該是單一情緒標籤，而應該是包含更多情緒感知多樣性的分佈式標籤。我們提出了一種新的標籤聚合規則，稱為「全包容規則」(all-inclusive rule)，用於選擇訓練集和測試集的數據和其情緒標記。對四個公開英文情緒數據庫的結果表明，使用「全包容規則」方法決定的訓練集所訓練的語音情緒辨識系統，在各種測試條件下，其性能優於使用傳統方法，包括絕對多數決和相對多數決訓練的語音情緒辨識系統。

最後但同樣重要的是，我們受到心理學研究關於情緒共現性的發現啟發。我們根據情緒資料庫訓練集中情緒標記來估計情緒共現性的頻率，並基於每種情緒類別的數量對矩陣進行標準化。接著，我們使用單位矩陣減去標準化矩陣作為懲罰矩陣。我的想法是在訓練過程中當模型預測到罕見的情緒共現時對語音情緒辨識系統進行懲罰。因此，懲罰矩陣被整合進現有的目標函數，例如交叉熵損失(cross-entropy)。在最大的英語情緒數據庫結果顯示，即使在單一情緒標籤測試條件下，懲罰矩陣也提升了語音情緒辨識系統的性能。

\end{abstractZH}

\begin{acknowledgementsZH}
我要向那些在我博士旅程中支持和指導我的人表達我真誠感謝。我的指導教授，李祈均博士，從大學專題開始指導我，從2014年到2024年，老師的專業知識、寶貴的見解和不懈的支持對這篇博士論文有著非常大的貢獻。祈均老師的指導和支持讓我可以順利完成這篇博士論文以及旅程。

我也向我的博士學位考試委員會的委員們，林嘉文教授、馬席彬教授、陳柏琳教授、冀泰石教授，以及王新民博士，感謝各位委員們抽空參與，並提供建設性的回饋與建議，以及投入時間審閱我的博士論文。也謝謝指導過我的Carlos Busso教授、李宏毅教授、劉奕汶教授、Albert Ali Salah教授、阮大成博士，和Alexander Visheratin博士，他們不僅是合作者，還是我的指導員；有他們的帶領和建議，讓我有不斷解決問題的勇氣和想法。

在七年的博士生涯，有參與過數次的國際研討會，而「傑出人才發展基金會」與「計算語言學與中文語言處理學會」在我就讀博士班期間提供慷慨資金與鼓勵。在他們的財務支持下，這項工作得以完成。我同樣感謝聯詠博士獎學金、國家科學技術委員會博士生赴國外計畫、Google東亞學生旅行獎勵、中華扶輪獎學金以及國立清華大學校長博士生卓越獎學金；這些獎學金讓我能夠專心研究而無財務之憂。

我也感謝ACII 2017和國際口語溝通學會獎勵（INTERSPEECH 2022），讓我能夠在國際會議上親自到現場介紹我們的研究，這些機會提供了寶貴的經驗和新的視角。同時，能夠在享有盛譽的期刊上發表我的研究成果，包括APSIPA《Transactions on Signal and Information Processing》及IEEE《Transactions on Affective Computing》期刊，也是一個非常有價值的經歷，也謝謝 ISCA Student Advisory Committee，讓我可以成為委員，為國際會議貢獻，也讓我認識到非常多來自世界各地和各領域的專家和學者們。

在此，我想向我的家人們、所有實驗室的夥伴們、朋友們（林維誠、陳思睿、Seong-Gyun Leem、Lucas Goncalves、吳姿瑩、Ali N. Salman、張凱為、林羿成、吳海濱、任文澤、Andrea Vidal、許德丞、陳志杰和林旻萱）和人生中的教練和老師們(徐碩鴻老師、林靜枝老師、曹昱博士、張俊盛教授、徐桂平老師、吳德成教練、黃錫瑜教授、陳榮順老師、劉素貌教練、黃老師、蔡文祺老師、范光榮老師、楊硯茗教練、Richard Lee老闆、洪光燦教練、張炳煌教練和王禮章老師)表示感謝，感謝他們無微不至的支持和鼓勵。他們的耐心和理解讓這段旅程變得可承受且充實。感謝你們成為這段旅程的一部分。

\end{acknowledgementsZH}

\begin{abstractEN}

Over the past twenty years, there has been a growing focus on speech emotion recognition (SER). To develop SER systems capable of identifying emotions in speech, researchers need to gather emotional databases for training purposes. This process involves training crowdsourced raters or in-house annotators to express their emotional responses after experiencing emotional recordings by selecting from a set list of emotions. Nevertheless, it is common for raters to disagree on emotion selection from these predefined categories. To address this issue, many researchers consider such disagreements noise and apply label aggregation techniques to produce a unified consensus label, which serves as the target for training SER systems. While the common practice simplifies the task as a single-label recognition task, it ignores the natural behaviors of human emotion perception. In this dissertation, we contend that we should revisit the modeling and evaluation approaches in SER. The driving research questions are (1) Should we remove the minority of emotional ratings? (2) Should we only let the SER systems learn the emotional perceptions of a few people? (3) Should SER systems only predict one emotion per speech?

Based on the findings of psychological studies, emotion perception is subjective. Each individual could have varying responses to the same emotional stimulus. Additionally, boundaries of emotions in human perception are overlapped, blended, and ambiguous. Those ambiguities of emotions and subjectivity of emotion perceptions inspire us to revisit modeling and evaluation approaches in SER. This dissertation explores novel perspectives on three main views of building SER systems. First, we embrace the subjectivity of emotional perception and consider every emotional rating from annotators. Also, the conventional approach only allows each rater to provide one vote for each sample. Still, we re-calculate label representation in the distributional format with the existing soft-label method by considering all ratings from all raters. Moreover, we directly utilize ratings of individual annotators to train SER systems and jointly train the individual SER systems and the standard SER systems. The modeling of individual annotators improves the performances of SER systems on the test sets with the consensus labels obtained by the majority vote. 

Secondly, we rethink the determination of methods to evaluate SER systems and the formulation and definition of the SER task. We argue that we should not remove any data and emotional ratings when assessing the performances of SER systems. Also, we think the definition of SER task can have a co-occurrence of emotions (e.g., sad and angry). Therefore, the ground truth of samples should not be the one-hot single label, and it can be distributional to include more diversity of emotion perception. We propose a novel label aggregation rule, named the ``all-inclusive rule,'' to use all data and include the maximum emotional rating for the test set. The results across 4 public English emotion databases show that the SER systems trained by the train set decided by the proposed method outperformed the ones trained by the conventional techniques, including majority rule and plurality rule on the various testing conditions. 

Finally, we draw inspiration from psychological research on the co-occurrence of emotions. We assess the frequency with which different categorical emotions occur together, using emotional ratings from the training data of emotion databases. This matrix is then normalized, considering the frequency of each emotion class. We derive a penalization matrix by subtracting the normalized matrix from an identity matrix. We aim to apply penalties to SER systems during training when they predict rarely occurring combinations of emotions. This penalization matrix is integrated into objective functions like cross-entropy loss. The findings from the largest English emotion database indicate that using the penalization matrix enhances the performance of SER systems, even under single-label testing conditions.

With the extensive results, we conclude that (1) we should involve the minority of emotional ratings instead of removing them to build better-performance SER systems, (2) we should consider emotional ratings from more people instead of fewer people during training SER systems to get better-performance SER systems; (3) we should allow SER systems to predict multiple emotions to handle the possibility of co-occurring emotions in the real-life scenarios. In future work, we plan to investigate training emotion recognition systems with multi-modalities (e.g., video, text, and audio) to process signals to improve the performance of SER systems. Also, we are interested in the relationship between the number of training human-labeled data and the performances of SER systems. Furthermore, we aim to understand the performance bias in the demographic groups, such as gender, race, and age. Last but not least, we plan to build a multi-lingual emotion recognition system.

\end{abstractEN}

\begin{acknowledgementsEN}
I am deeply grateful to everyone who has supported and guided me during my Ph.D. journey. I extend special thanks to my advisor, Professor Chi-Chun Lee, who has been a guiding force since my undergraduate years, from 2014 to 2024. Professor Lee's expertise, invaluable insights, and unwavering support have been instrumental in completing this doctoral dissertation. His guidance has played a crucial role in the successful conclusion of my dissertation and my entire Ph.D. journey.

I am also profoundly grateful to the members of my dissertation committee, Professor Chia-Wen Lin, Professor Hsi-Pin Ma, Professor Berlin Chen, Professor Tai-Shih Chi, and Dr. Hsin-Min Wang. Thank you all for taking the time to participate, providing constructive criticism, and reviewing my work. Big thanks go to my co-advisors, Professor Carlos Busso, Professor Hung-yi Lee, Professor Yi-Wen Liu, Professor Albert Ali Salah, Dr. Da-Cheng Juan, and Dr. Alexander Visheratin, who have not only been collaborators but also respected mentors. Their support has created a positive and motivating research environment.

I am indebted to the Foundation for the Advancement of Outstanding Scholarship and the Association for Computational Linguistics and Chinese Language Processing for the generous funding and resources provided during my research. With their financial support, this work was possible. I acknowledge the scholarships I received from the NOVATEK Fellowship and the National Science and Technology Council Ph.D. Students Study Abroad Program, Google East Asia Student Travel Grants, The Rotary Foundation Excellence Scholarship, and National Tsing Hua University Dean Ph.D. Student Excellence Scholarship: These scholarships allowed me to concentrate fully on my research without financial worry. 

I am also thankful for the opportunities to present my work at international conferences, such as the ACII 2017 and International Speech Communication Association Grants (INTERSPEECH 2022), which provided valuable feedback and new perspectives. Publishing my work in esteemed journals, including APSIPA Transactions on Signal and Information Processing and IEEE Transactions on Affective Computing, has also been an enriching experience.

I want to express my gratitude to my family, my lab-mates (BIICers), and friends (Wei-Cheng Lin, Szu-Jui Chen, Seong-Gyun Leem, Lucas Goncalves, Tz-Ying Wu, Ali N. Salman, Kai-Wei Chang, Yi-Cheng Lin, Haibin Wu, Wenze Ren, Andrea Vidal, Te-Cheng Hsu, Jhih-Jie Chen, and Min-Hsuan Lin), and the coaches and teachers in my life (Professor Shawn S. H. Hsu, Teacher Ching-Chin Lin, Dr. Yu Tsao, Professor Jason S. Chang, Teacher Kuei-Ping Hsu, Coach Te-Cheng Wu, Professor Shi-Yu Huang, Teacher Rong-Shun Chen, Coach Su-Mao (Tammy) Liu, Teacher Huang, Teacher Wen-Qi Tsai, Teacher Guang-Rong Fan, Coach Yan-Ming Yang, CEO Richard Lee, Coach Guang-Can (Michael) Hung, Coach Bing-Huang Zhang, and Teacher Li-Zhang Wang) for their support and encouragement. Their help and understanding have made this journey bearable and fulfilling. Thank you all for being part of this journey.

\end{acknowledgementsEN}

\maketoc


\pagenumbering{arabic}

\chapter{Introduction}
\begin{figure}[!b]
    \centering
    \includegraphics[width=0.8\linewidth]{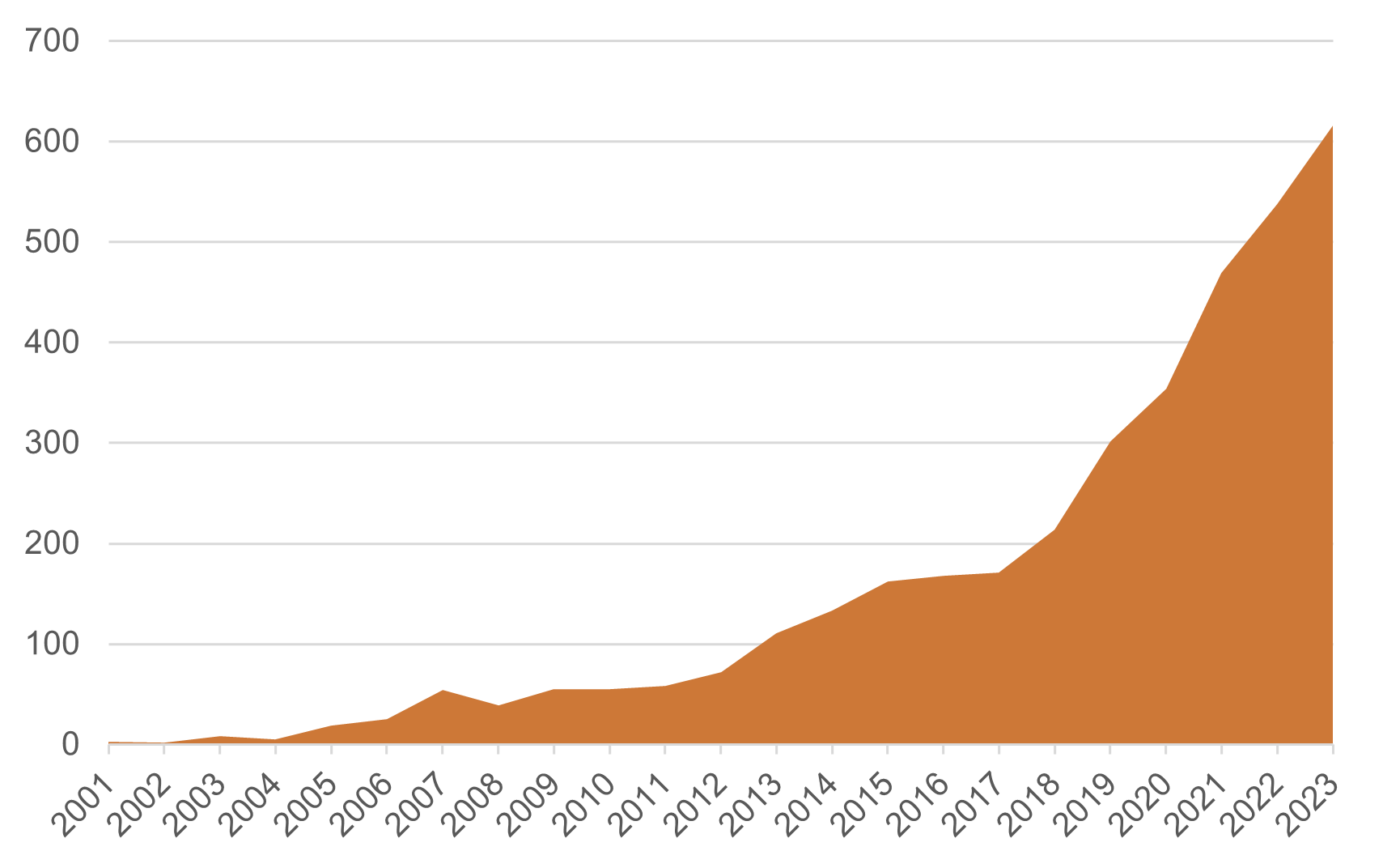}
    \caption{The figure illustrates the trends of papers whose title includes speech emotion recognition based on the Google Scholar website. The x-axis means years; the y-axis is the number of papers.}
    \label{fig:trends}
\end{figure}

\section{Motivation}

Recent research has led to the development of automated systems that perform tasks traditionally done by humans like automatic speech recognition \cite{Radford_2023} and speech translation \cite{Seamless_2023}. Moreover, Figure~\ref{fig:trends} shows increasing research studies on speech emotion recognition (SER) to understand emotions from human voice from 2001 through 2023. In recent years, SER has been a potential indicator to predict customer satisfaction \cite{Parra_Gallego_2022}. Moreover, SER is crucial for voice assistants as it enables them to generate speech with appropriate emotions by predicting the emotional tone of users' voices. I also observe some startups contribute to building SER solutions to improve users' experiences, such as HUME, BEHAVIORAL SIGNALS, UNIPHORE, and COGNOVI LABS. Those companies provide emotion-aware solutions for their clients to understand users' emotions and improve user experiences. 

Additionally, SER needs interdisciplinary collaboration, such as studies of psychology, spoken language, and natural language understanding. Most researchers follow psychological studies to design and collect emotional corpus. In common practice, researchers provide pre-defined categorical emotions for raters to choose from after listening to or watching emotional stimuli. In most public emotion databases, each sample was annotated by at least 3 annotators. However, it is expected to observe disagreement among raters on the same sample. The prior SER studies utilize label aggregation methods, such as majority vote or plurality vote, to obtain a single consensus label as the ground truth to train SER systems. This common practice removes a minority of emotional perceptions and the data without a consensus label. 

\begin{figure}[!b]
    \centering
    \includegraphics[width=\linewidth]{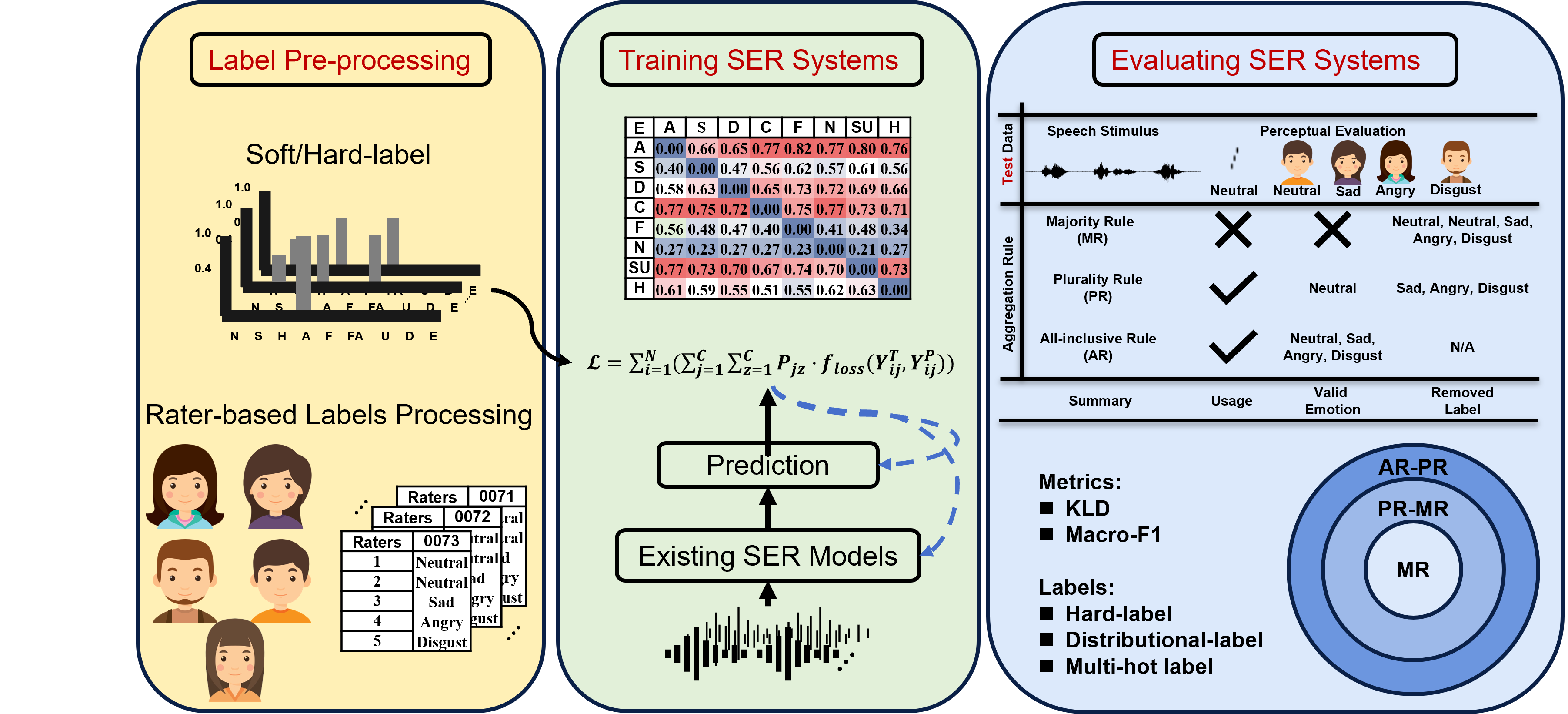}
    \caption{The figure illustrates three main contributions to modeling and evaluation approaches in SER systems.}
    \label{fig:three_contributions}
\end{figure}

The recent findings of psychological studies \cite{Cowen_2017,Cowen_2021} reveal that the emotion perception of humans is high-dimensional, and boundaries of emotions among emotion perception of humans are blended and overlap. The critical findings inspire and motivate me to revisit standard SER systems' whole modeling and evaluation methods. We have come up with the following research questions to answer: (1) Should we remove those minority emotional ratings? (2) Should we only let SER systems learn the emotional perception of a few annotators? (3) Should SER systems only predict one single emotion for each sample? Those questions could be split into two main factors, the subjectivity of emotion perception and ambiguity of emotions, contributing to the disagreement of emotion perception among raters because of human bias, including gender \cite{Hall_2004}, culture \cite{David_1992}, and age \cite{Suzuki_2007}. This dissertation dives into the whole process of modeling and evaluating SER systems, and the entire process is threefold, as shown in Figure~\ref{fig:three_contributions}: (1) label preprocessing,  (2) evaluation of SER systems, and (3) training of SER systems. We propose three novel methods to contribute to each process separately to improve the modeling and assessment of SER systems based on the existing SER frameworks.

\begin{figure}[!b]
    \centering
    \includegraphics[width=\linewidth]{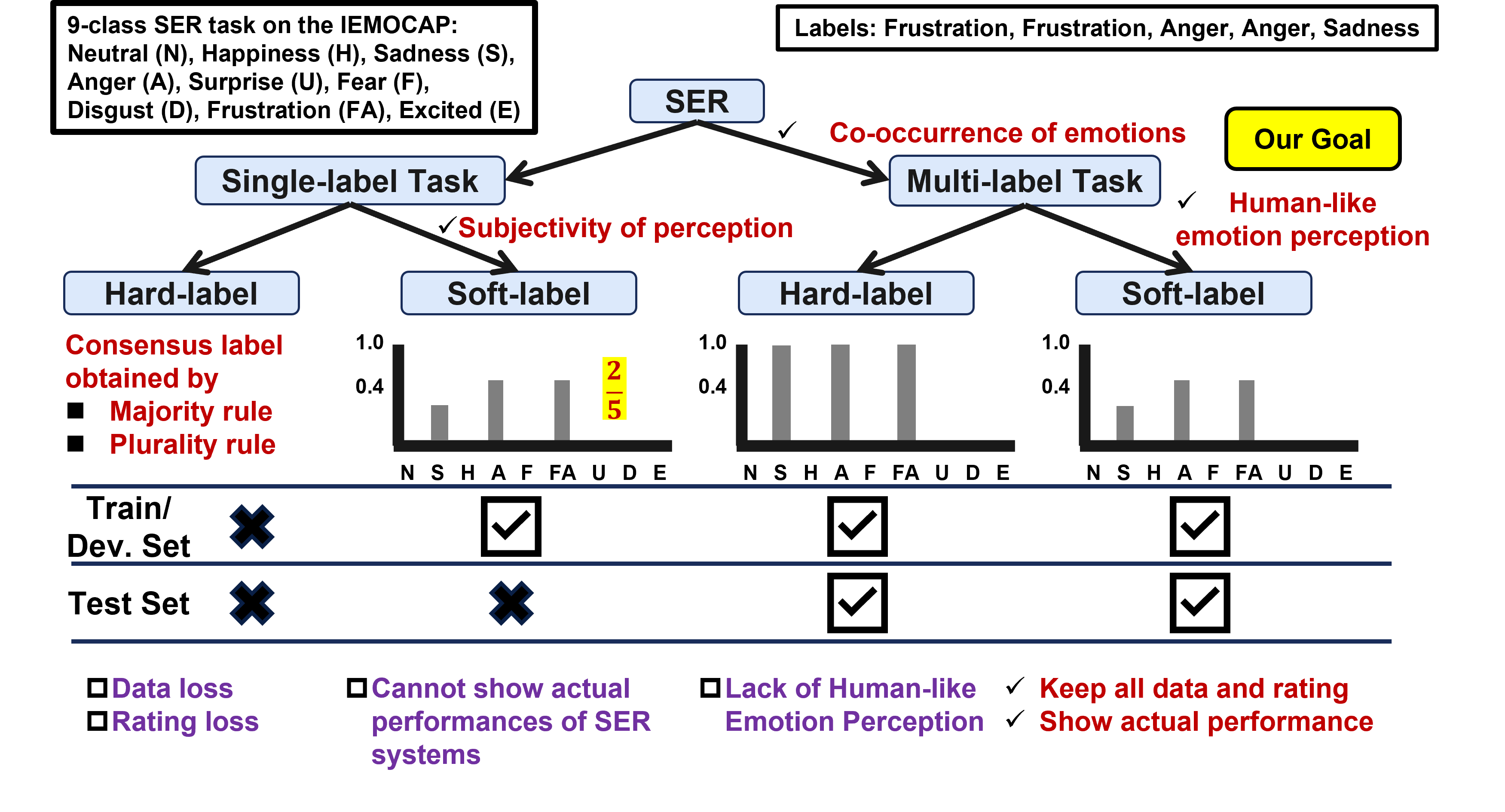}
    \caption{The figure illustrates the current trends of prior works on SER.}
    \label{fig:challenges}
\end{figure}

\section{Background, Related Works, and Challenges}
This section introduces an overview of the background and challenges of the prior SER studies.

\subsection{Emotion Representations}
There are two main ways to represent emotion perception. One is a dimensional attribute that assumes every attribute is independent of each other, like arousal \cite{Russell_2003} or valence \cite{Barrett_2006}. The other one is categorical emotions \cite{Cowen_2017,Cowen_2021}, such as anger or sadness. This dissertation only focuses on categorical emotions since the perception of categorical emotions is better perceived across cultures than dimensional attributions \cite{Cowen_2019}.

\subsection{Emotion Recognition Systems using Multi-/Uni-modality}
Emotion can be expressed through various modalities, including facial expressions, body movements, vocal speech, and text. According to Cowen and Keltner \cite{Cowen_2021}, humans can recognize at least 27 distinct emotions from video, 24 from vocal expression, and 28 from facial and bodily cues. Furthermore, humans can discern approximately 13 emotions from music. Consequently, different modalities lead to varying emotional perceptions in humans.

Previous research has employed various modalities to develop emotion recognition systems. For example, Goncalves et al. \cite{Goncalves_2022_2} utilized audio-visual data to build such systems. Almedia et al. \cite{Almeida_2021} focused on training systems to identify emotions from facial expressions. Additionally, studies \cite{Juan_2021, Juan_2021_2} have successfully detected emotions through music. This dissertation, however, is concerned solely with speech-based emotion recognition systems \cite{Chou_2019, Chou_2022_2}.

\subsection{Evaluation of SER Systems}

Many research studies on SER primarily focus on predicting a single emotion, using methods like majority voting \cite{Mirsamadi_2017} or plurality rules \cite{Parthasarathy_2020} to derive a consensus label for evaluating SER systems. This approach often discards emotional ratings with minority opinions and data lacking a consensus label, resulting in a simpler test set. Consequently, SER system performance evaluations may only partially represent their true capabilities due to excluding some data and emotional ratings. Figure~\ref{fig:challenges} showcases the trends in previous SER research, highlighting that many studies approach SER as a single-label task with hard-label (one-hot encoding) targets. For instance, in Figure~\ref{fig:challenges}, a sample from IEMOCAP was rated as frustration, frustration, anger, anger, and sadness. Since frustration and anger received equal votes, no consensus label was achieved, leading to the sample being removed from the training set (development set), resulting in data and emotional rating loss. Moreover, defining SER as a single-label task neglects the frequent co-occurrence of emotions in real-world situations \cite{Vansteelandt_2005}. Although some research considers the subjective nature of emotion perception by using all emotional ratings as soft labels for training \cite{Steidl_2005}, the "tie" samples are still excluded from the test set, so the evaluation of SER systems does not fully reflect their actual performance.

Few studies treat Speech Emotion Recognition (SER) tasks as multi-label tasks, contrary to other emotion recognition fields, which do. For example, in text emotion recognition \cite{Zhang_2020,Kang_2023}, image emotion classification \cite{Wang_2023}, and audio-visual emotion recognition \cite{Xincheng_2020,Zhang_2021}, emotions are considered valid if they receive any vote. They use multi-hot labels, as illustrated in Figure~\ref{fig:challenges}. A close study by \cite{Shan_2019} on facial expression recognition calculated soft labels based on vote frequency for each emotion, converting these into binary vectors (either multi-hot or single-hot) based on the threshold ($1/(C-1)$, where C is the number of emotion classes). However, this misses key information about primary and secondary emotions. In label distribution learning \cite{Geng_2016}, \cite{Ying_2015} used facial expression databases to collect emotion intensity scores from multiple annotators, averaging these to normalize the distributional labels for system training and evaluation. Similarly, \cite{Zhou_2016} did this for text emotion recognition. However, these methods are challenging to apply in SER due to the lack of emotional intensity scores, as SER databases typically ask raters to select pre-defined emotions. This is possible because querying raters on intensity could be overwhelming and bias-inducing \cite{Yannakakis_2017,Yannakakis_2021}. Our objective is to incorporate all data and emotional ratings in test sets to assess SER systems' performance accurately. We employ a soft-label format as ground truth, convertible to a binary vector to highlight traditional accuracy. Thus, we propose metrics to evaluate the distribution similarity between predictions and the actual data. This dissertation will report results using two such metrics.

\begin{table*}[!t]
\fontsize{7}{9}\selectfont
\centering
\caption{Table summarizes the loss of data and emotion rating across the public emotion databases.}
\begin{tabular}{@{}c|cc|cc|cc@{}}
\toprule
Label Aggregation Method & \multicolumn{2}{c}{Majority Rule} & \multicolumn{2}{c}{Plurality Rule} & \multicolumn{2}{c}{\textbf{Our Goal}} \\ \midrule
Emotion   Database       & Data            & Rating          & Data             & Rating          & Data         & Rating        \\ \midrule
IMPROV (P)               & 9.18\%          & 28.52\%         & 4.63\%           & 26.41\%         & 0\%          & 0\%           \\
CREMA-D                  & 35.80\%         & 52.96\%         & 8.55\%           & 40.57\%         & 0\%          & 0\%           \\
PODCAST (P)              & 44.81\%         & 59.87\%         & 19.85\%          & 49.24\%         & 0\%          & 0\%           \\
IEMOCAP                  & 31.37\%         & 49.44\%         & 25.32\%          & 45.70\%         & 0\%          & 0\%           \\
IMPROV (S)               & 54.18\%         & 76.91\%         & 12.32\%          & 56.70\%         & 0\%          & 0\%           \\
PODCAST (S)              & 92.01\%         & 96.99\%         & 33.72\%          & 78.13\%         & 0\%          & 0\%           \\ \midrule
Average                  & 44.56\%         & 60.78\%         & 17.40\%          & 49.46\%         & 0\%          & 0\%           \\ \bottomrule
\end{tabular}
\label{tab:loss}
\end{table*}

\subsection{Label Prepossessing for Training SER Systems}
The most common approaches to obtain consensus labels are majority rule and plurality rule, whose definitions are introduced below. Table~\ref{tab:loss} summarizes the loss of data and emotional ratings using two conventional label aggregation methods, majority rule and plurality rule. We aim to try our best to retain all data and emotional ratings to train the SER systems. 

\begin{itemize}
    \item Majority Rule (MR): it selects one of the pre-defined emotion classes only if more than half of the votes select that class. 
    \item Plurality Rule (PR): it selects a class if one emotional class obtains more votes than others.  
\end{itemize}

We argue that the two above conventional aggregation methods lose variations of emotion perception and the number of data during SER systems, leading to the poor ability to predict the samples that have co-occurrence of emotions.  Therefore, we propose two ways to model the variations of emotion perception. (1) The first is to model individual raters' SER systems since each person has a different sensitivity to various emotions \cite{Martin_1996}. For instance, some raters are good at perceiving sad emotions; some can easily sense happy emotions. (2) The second approach proposes a novel label aggregation rule, named the ``all-inclusive'' rule, incorporating all the labeled data with the emotional ratings in the emotion databases.

\subsection{Co-occurrence of Emotions}

In everyday situations, emotions often occur simultaneously \cite{Vansteelandt_2005}. However, most previous research in SER does not account for the possibility of concurrent emotions like "contempt and anger," treating the task as one requiring a single label. Although soft-label or multi-label approaches can accommodate emotion co-occurrence, they do not effectively depict the relationships between different emotions. For example, a person is more likely to feel simultaneously sad and neutral than both sad and happy. In text emotion recognition, a study by \cite{Alhuzali_2021} adapted a loss function initially proposed by \cite{Yeh_2017} to account for label correlation among raters, thereby quantifying dependency between emotions. Additionally, research by \cite{Deng_2023} introduced a multi-label focal objective function designed to enhance the differentiation between positive and negative emotions to improve emotion recognition perceptivity in text systems. This approach, however, overlooks the possibility of both negative and positive emotions occurring together. Unlike the studies mentioned earlier, we directly examine the relationships between co-existing categories of emotions based on their co-occurrence frequencies, allowing for mixed emotional states such as happiness and anger. For instance, consider a scenario where a girl was upset at her boyfriend for being over an hour late, but upon his arrival, she saw he was holding her a present, her favorite dress. By conceptualizing the co-occurrence frequency of emotions in a matrix and normalizing it to create a penalty matrix, we can integrate this into existing objective functions like cross-entropy to penalize models during training. The penalty matrix assigns greater loss values when SER systems predict rare emotional co-occurrences, as indicated by the training set annotations. Importantly, this approach allows for the recognition of both positive and negative emotions happening concurrently.

\subsection{Disagreement between Raters on the Emotion Datasets}

Unlike common classification tasks (e.g., speaker identification or laughter/cry detection) with well-established ``gold standard'' labels, subjective tasks like emotion recognition lack clear labels and often rely on perceptual evaluations. Researchers frequently use crowd-sourcing platforms such as Amazon Mechanical Turk to collect labels rapidly and extensively \cite{Sabou_2014}. While cost-effective, this method often compromises label quality. This trade-off is particularly significant in subjective tasks, where the inherent ambiguity amplifies annotation variability. Subjective tasks, such as emotion perception \cite{Lotfian_2019} or hate speech tagging \cite{Waseem_2016,Davani_2022}, pose unique challenges due to their fundamentally subjective nature. Obtaining labels for these tasks is complex because they heavily depend on individual interpreters. Annotator disagreements can result from various factors \cite{Prabhakaran_2021,Sandri_2023,Oluyemi_2024}, such as diverse backgrounds leading to different interpretations, lack of interest in providing accurate labels, emotional biases, and contextual differences \cite{Martinez_Lucas_2023}. These variances introduce substantial noise into the labeling process, particularly problematic in crowd-sourced evaluations \cite{Hube_2019}.

Annotation noise poses a major challenge, and various strategies have been developed to lessen its effects. For speech emotion classification tasks, it's important to understand that while noise can lead to discrepancies in labels, it's not the only cause of disagreement. In line with the methodologies applied in the MSP-PODCAST corpus \cite{Lotfian_2019}, efficient noise reduction approaches include excluding evaluators with persistently low agreement rates, pausing crowd-sourcing efforts when agreement falls below a certain threshold, and employing in-house staff who can undergo specialized training to enhance label consistency. Yet, it's crucial to note that perceptual variations are not merely noise; they can offer valuable insights that a speech emotion recognition system should utilize.

This dissertation aims to demonstrate that traditional methods of label aggregation, such as majority or plurality voting, often overlook the nuanced nature of subjective perception and may not be suitable for speech emotion classification. We propose an alternative aggregation method for the speech emotion recognition task, aiming for a more comprehensive and inclusive approach to label aggregation.

\section{Contributions}
The research goals of this dissertation aims to answer three questions: 

\begin{itemize}[label=]
    \item (1) Should we remove those minority emotional ratings?
    \item (2) Should we only let SER systems learn the emotional perception of a few annotators?
    \item (3) Should SER systems only predict one emotion for each sample?
\end{itemize}

This dissertation proposes three methods to improve the process for label preprocessing, evaluation of SER systems, and training of SER systems, respectively. 

\subsection{Every Rating Matters Considering the Subjectivity of Annotators}

We first explore the inherent subjectivity of how each annotator perceives emotions to improve the performances of SER systems. We aim to maximize the emotional ratings using the existing soft-label method introduced by \cite{Steidl_2005} for training SER systems. Additionally, we develop an individual SER model for each annotator based on their respective ratings. To integrate all possible emotional data, we merged the embeddings obtained from pre-trained SER models using the traditional hard-label and the existing soft-label methods across five individual annotator SER systems for late-fusion. The results indicate that the proposed framework improves performance when assessed on a test set determined by a majority vote.

\subsection{Novel Evaluation Method by an All-Inclusive Aggregation Rule}

In addition, we implemented an innovative label aggregation approach that incorporates all annotated data from the test phase, ensuring no data is overlooked. We propose maintaining all emotional ratings within test data samples to precisely evaluate the performance of SER systems, since determining consensus labels isn't practical in real-world situations. Our ground truth is structured as a distribution reflecting the frequency ratio of emotional votes for each emotion class. Like the study conducted by \cite{Steidl_2005}, we believe that using a distributional similarity metric offers a more accurate assessment of SER systems' performance compared to accuracy-based metrics like the macro-F1 score, due to the subjective nature of SER. Distributional metrics better match how humans perceive emotions \cite{Cowen_2017, Cowen_2021}. Nevertheless, since the SER field is more accustomed to accuracy-based metrics, we also provide results using those. However, we've noticed that transforming distributional labels into binary vectors for accuracy evaluation might lead to losing some emotional ratings, which is different from our goal. Still, accuracy-based metrics give the community and reviewers a more precise understanding.

\subsection{Training Loss Using Co-occurrence Frequency of Emotions}
To model the connection between co-occurrence of emotions, we counted the counts of co-occurrence of emotions based on labeled data in the train set of the emotion database as a matrix and normalized the frequency matrix by the number of individual emotion classes. Then, we use the unit matrix to subtract the normalized frequency matrix as a penalization matrix. To penalize the SER systems during training when the models predict rare co-occurrence of emotions, we integrate the designed penalization matrix into the current common objective functions, e.g., cross-entropy. Considering the link to the co-occurrence of emotions, this method improves the SER performances on the test conditions, including single-label and multi-label tasks. 

\section{Outline of the Dissertation}
The rest of the dissertation is structured as follows. It primarily focuses on three key aspects of modeling and evaluating SER systems. We begin by presenting the public emotion databases (Chapter~\ref{ch:database}), and then introduce the three proposed methods: modeling the subjectivity of annotators (Chapter~\ref{ch:subjectivity}), a novel evaluation method (Chapter~\ref{ch:ch_evaluation}), and a training loss that accounts for the co-occurrence of emotion classes (Chapter~\ref{ch:training_loss}).

\chapter{Emotion Databases}
\label{ch:database}
The chapter discusses four public emotional databases leveraged in this dissertation. We might use one or multiple databases in different chapters, but we introduce them here.

\section{IEMOCAP}
\label{s:iemocap}
The IEMOCAP dataset \cite{Busso_2008} was carefully constructed from motion capture, audio, and visual data. Ten professional actors speak English. This unique dataset includes scripted and spontaneous dialogues, primarily focusing on romantic relationship scenarios to evoke diverse emotions. Each session involved one male and one female actor. To guarantee an expressive variation, performers utilized scripts to elicit distinct feelings. The final recordings were segmented into 10,039 utterances. All utterances have human-typed transcripts. Raters watched segmented clips and selected emotions from a predefined list of ten categories: neutral, happy, sad, angry, surprised, fear, disgusted, frustrated, excited, and "other." Addressing issues related to results reproducibility, mentioned in prior research \cite{Antoniou_2023}, we have provided meticulous details on dataset splits in Section~\ref{ss:improv_cv}. Given the original dataset's deficiency of standardized split sets, our documentation aims to bridge that gap, ensuring greater consistency and reproducibility. The IEMOCAP is used in Chapter~\ref{ch:subjectivity} and Chapter~\ref{ch:ch_evaluation}.

\section{IMPROV}

The MSP-IMPROV dataset, also known as IMPROV \cite{Busso_2017}, collects audio-visual recordings, and 12 actors are English speakers. All dyadic interactions represent four distinct emotional states: angry, happy, sad, and neutral. The sessions are thoroughly segmented into 8,438 individual clips by humans, with each clip being assessed by a minimum of five annotators using crowdsourcing methods. To maintain the high quality of annotations, the dataset integrates a quality control technique outlined by \cite{Burmania_2016}, aimed at identifying and excluding unreliable annotators.

The labeling process for the dataset involves two distinct procedures: primary (P) and secondary (S) emotions. Raters are instructed to select one emotion from five pre-defined emotions in the primary procedure: angry, happy, sad, neutral, or "other." The secondary emotions span a wider spectrum, including frustrated, depressed, disgusted, excited, fear, and surprised. Since the dataset lacks predefined sets for training, development, and testing, we elucidate the suggested division of these sets in Section~\ref{ss:improv_cv} to ensure clarity and maintain consistency in follow-up analyses. The IMPROV is only used in Chapter~\ref{ch:ch_evaluation}.

\section{CREMA-D}
The CREMA-D dataset \cite{Cao_2014} collects high-fidelity audio-visual recordings from 91 professional actors, comprising forty-three women and forty-eight men. They were directed to deliver one of six unique emotions with the given scripts: angry, disgusted, fearful, happy, sad, or neutral. A key highlight is the comprehensive labeling procedure; spanning over 7,442 segments, each segment was evaluated by more than two thousand distinct crowdsourcing raters. All data were subjected to assessments from no fewer than six annotators, who identified one of the six defined emotions for each performance.

The perceptual annotation process operates within three contexts: voice- and face-only and audio-visual. Raters focus exclusively on listening to the segments' audio in the voice-only context. They watch the actors' faces without audio input in the face-only context. Lastly, in the audio-visual context, annotators evaluate both the facial expressions and the audio concurrently.

For this dissertation on SER, we specifically concentrate on the emotional labels gathered from the voice-only scenario. Unlike numerous previous SER studies that leveraged labels from the audio-visual context or omitted annotation specifics entirely, we decided to depend exclusively on voice-only labels for the evaluation. Furthermore, our paper includes detailed specifications regarding the dataset splits we utilized, which can be found in Section \ref{ss:cremad_cv}. The CREMA-D is only used in Chapter~\ref{ch:ch_evaluation}.

\section{MSP-PODCAST}
\label{s:msp_podcast}

The MSP-PODCAST dataset \cite{Lotfian_2019} is a comprehensive, emotionally varied voice compiled from licensed podcast resources. These recordings are initially split into utterances and subsequently labeled through a crowdsourcing website. The labeling protocol encompasses primary (P) and secondary (S) methods. Raters choose from nine categorized emotions in the primary emotion: angry, sad, happy, surprised, fear, disgusted, contempt, neutral, and "other."  The secondary emotion broadens this scope to the primary emotions, plus eight additional ones: amused, frustrated, depressed, concerned, disappointed, excited, confused, and annoyed, making a total of 17 emotional categories. At least five contributors meticulously annotate each utterance to ensure robust and reliable annotations. Different dataset versions certify various utterance quantities within the training, development, and complementary test groups (test1 and test2). The MSP-PODCAST is used in Chapter~\ref{ch:ch_evaluation} and Chapter~\ref{ch:training_loss}.

\section{Standard Partition}
\label{s:partition}
The preceding study \cite{Antoniou_2023} indicates that 80.77\% of SER research papers produce irreproducible results with the widely recognized IEMOCAP dataset. The primary obstacle to reproducibility is the absence of standardized data splits (e.g., training, development, and test sets) within the database. Previous studies each defined their partitions; however, they often withheld detailed partitioning methodologies or source code, complicating repeatability. Consequently, this dissertation aims to make SER more transparent and reproducible for everyone. We establish and define standard partitions for four prominent and publicly available emotion databases, facilitating future SER advancements. 

\begin{table}[!b]
\fontsize{7}{9}\selectfont
\centering
\caption{Table overviews the defined standard partitions for the IEMOCAP and the \textbf{Ses.} means the session in the database.}
\begin{tabular}{@{}cccc@{}}
\toprule
Partition & Training Set  & Development Set & Test Set \\ \midrule
1     & Ses. 1,2,3 & Ses. 4          & Ses. 5   \\
2     & Ses. 2,3,4 & Ses. 5          & Ses. 1   \\
3     & Ses. 3,4,5 & Ses. 1          & Ses. 2   \\
4     & Ses. 1,4,5 & Ses. 2          & Ses. 3   \\
5     & Ses. 1,2,4 & Ses. 3          & Ses. 4   \\ \bottomrule
\end{tabular}
\label{tab:cviemocap}
\end{table}

\subsection{Standard Partition of the IEMOCAP}
\label{ss:iemocap_cv}
In a speaker-independent scenario, models are trained using data from specific individuals and evaluated using data from entirely different individuals who were not part of the training set. This approach ensures an unbiased and robust assessment of the model's performance. For instance, in the IEMOCAP study, we summarize the data partitioning approach in Table \ref{tab:cviemocap}. Each session involves two speakers in interactive dialogues and allows us to define five speaker-independent splits, referred to as Ses. 1 through Ses. 5. We perform a five-fold cross-validation, as detailed in Table \ref{tab:cviemocap}, whereby every fold comprises different sets for training, development, and testing to guarantee a thorough assessment of the model's effectiveness across sessions.

\subsection{Standard Partition of the MSP-IMRPOV}
\label{ss:improv_cv}
The IMPROV dataset is segmented into 6 distinct folds for cross-validation purposes. All folds feature a specific blend of development, training, and test data as detailed in Table \ref{tab:cvimprov}. The splitting method provides the SER system with information on interactions featuring various pairs of speakers and testing on completely new speaker pairings. Consequently, this strategy systematically evaluates the model’s capacity to generalize to various dyadic exchanges within the IMPROV dataset.

\begin{table}[!t]
\fontsize{7}{9}\selectfont
\centering
\caption{Table overviews the defined standard partitions for the IMPROV and the \textbf{Ses.} means the session in the database.}
\begin{tabular}{@{}cccc@{}}
\toprule
Partition & Training Set    & Development Set & Test Set \\ \midrule
1     & Ses. 1,2,3,4 & Ses. 5          & Ses. 6   \\
2     & Ses. 1,2,3,6 & Ses. 4          & Ses. 5   \\
3     & Ses. 1,2,5,6 & Ses. 3          & Ses. 4   \\
4     & Ses. 1,4,5,6 & Ses. 2          & Ses. 3   \\
5     & Ses. 3,4,5,6 & Ses. 1          & Ses. 2   \\
6     & Ses. 2,3,4,5 & Ses. 6          & Ses. 1   \\ \bottomrule
\end{tabular}
\label{tab:cvimprov}
\end{table}

\subsection{Standard Partition of the CREMA-D}
\label{ss:cremad_cv}
The CREMA-D database is split into 5 subsets according to speaker IDs for the speaker-independent context. Each subset comprises a unique blend of males and females and specific speaker IDs, elaborated in Table \ref{tab:cvcremad}. The partitioning strategy aligns with the methodology employed for the IEMOCAP dataset discussed in section~\ref{ss:iemocap_cv}.

\begin{table}[!t]
\fontsize{7}{9}\selectfont
\caption{Tables summarize the session in the CREMA-D emotion database. Notice that the \textbf{M} and \textbf{F} mean the male and F, respectively.}
\centering
\begin{tabular}{@{}ccr@{}}
\toprule
Session & Gender& \multicolumn{1}{c}{Speaker ID}                                                                                   \\ \midrule
1    &7M;11F& 1037-1054       \\\midrule
2    &12M;6F& 1001-1018       \\\midrule
3    &13M;6F& 1073-1091 \\\midrule
4    &9M;9F& 1055-1072       \\\midrule
5    &15M;3F& 1019-1036       \\ \bottomrule
\end{tabular}
\label{tab:cvcremad}
\end{table}

\chapter{Every Rating Matters Considering Subjectivity of Annotators}
\label{ch:subjectivity}

We present a methodology in which joint learning addresses emotional rating uncertainty and annotator idiosyncrasies by leveraging both hard and soft emotion label annotations and individual and crowd annotator modeling. Further analyses reveal that emotion perception heavily depends on raters. Combining hard labels with soft emotion distributions provides a well-rounded approach to affect modeling. Additionally, the integrated learning of both general emotional insights and specific rater profiles yields the highest accuracy in emotion recognition.

\section{Motivation}
Traditional SER systems \cite{Mirsamadi_2017, Ando_2018} typically use the plurality rule or majority rule from a group of raters as the learning targets, termed the hard label, to train emotion recognizers. However, factors like gender \cite{Hall_2004}, culture \cite{David_1992}, and age \cite{Suzuki_2007} significantly influence emotion perception, leading to natural disagreement and ambiguity in annotations \cite{Mower_2009,Vidhyasaharan_2019}. Consequently, the hard label approach overlooks emotion perception's diverse annotations and subjective nuances. To address this limitation, researchers \cite{Steidl_2005,Ando_2018} have proposed using soft labels—a distributional representation instead of a single definitive label—to capture blended emotion perceptions better. While the soft labeling method enhances flexibility in representing the variability of emotion perception, it still disregards the unique input of individual annotators because it creates the label distribution by aggregating inputs from all annotators. Therefore, we first build individual-based SER systems to model diverse and accurate subjectivity of emotion perception to improve aggregated emotion performance.

\section{Background and Related Works}

\subsection{Subjectivity of Emotion Perception}
Variability in how raters perceive emotions is not a new observation. Annotator modeling, which addresses this issue, has gained attention for years. For instance, the work \cite{Kim_2015} employed agreement between raters to assign weights to training instances, which facilitated the development of a speaker-dependent audio-visual emotion recognition system. Similarly, Han et al. \cite{Han_2017} introduced a model that leverages inter-rater disagreement to estimate perception uncertainty, thereby enhancing performance in continuous dimensional emotion tracking from audio and video sources. However, they still consider the ratings of all raters at that same time, but our method directly models individual raters' emotion perception, which can preserve more subjectivity of emotion perception.

\subsection{Mixture of Annotators}
Disagreement is present not just in emotion perception but also in various other areas like medical image tagging. Yan et al. \cite{Yan_2010} suggested that discrepancies arise because each annotator possesses unique medical domain knowledge. To address this, they proposed a method involving multiple annotators, which takes into account all available information by repeatedly using training data points until the models fully grasp each annotator's input. Their approach was found to be more effective than the traditional method that uses majority voting to determine ground truth.

\subsection{Soft-label Training Method for SER Systems}

Steidl et al. \cite{Steidl_2005} initially argued that using only a single emotion label as ground truth in emotion recognition tasks might not be suitable due to the subjective nature of emotion perception. They proposed using soft labels as ground truth based on count data and utilized entropy loss as an evaluation criterion. However, they re-assigned categorical emotions to enhance inter-rater agreement. In contrast, we retained the original labels to reflect the original emotion perception. Furthermore, Fayek et al. \cite{Fayek_2016} showed that training SER systems with soft labels can yield better performance than those trained with hard labels on test sets where the majority rule determines the consensus label. Additionally, Ando et al. \cite{Ando_2018} adjusted the soft labels with a small $\alpha$ value, creating an effect similar to label smoothing for training their SER systems. Their findings indicated that SER systems trained with these modified soft labels outperformed those trained with common soft labels and hard labels. Finally, Zhang \cite{Zhang_2019} was the first to demonstrate the advantages of using soft labels in cross-corpus SER, through their proposed objective function.

\section{Resource and Task Formulation}
We employ the IEMOCAP database as referenced in section~\ref{s:iemocap}, encompassing emotional ratings by 12 distinct raters across 10 categorical emotion classes. For consistency with previous research, we utilize the identical evaluation data wherein the entry is tagged with a singular emotion state, determined by a majority vote from more than three annotators. This study focuses on recognizing four primary emotion classes: sad, neutral, happy, and angry. Following the practices in \cite{Mirsamadi_2017,Ando_2018}, we consolidate the happiness and excitement categories into one: happiness. This aggregation includes 5,531 data samples used in the emotion recognition evaluation; this approach aligns with the traditional use of the IEMOCAP dataset as a benchmarking standard. The test set excludes the data without consensus labels. The emotion class distributions of data samples are sad: 19.60\%, happy: 29.58\%, neutral: 30.88\%, and angry: 19.94\%.  Half of the raters are also actors, while the other half consists of in-house raters. We selected only 5 out of the 6 in-house raters (E1, E2, E4, E5, and E6) since these five provided annotations for samples across all 5 sessions. Therefore, the proposed method only builds those 5 individual annotators.

\section{Speech Emotion Classifier}
\subsection{Input Features}
We use the openSMILE toolkit \cite{Eyben_2010} to derive frame-level acoustic features from utterances. Specifically, the "Emobase.config" configuration helps us obtain a 45-dimensional feature set. This set includes 12-dimensional Mel-Frequency Cepstral Coefficients (MFCCs), voice probability measures, zero-crossing rates, fundamental frequency (F0) values, loudness metrics, and their respective first-order derivatives. Additionally, the second-order derivatives of both loudness and MFCCs are featured. The feature extraction process is carried out with a frame length of 60ms and a step size of 10ms. These features are normalized for each speaker with z-score normalization and then downsampled by averaging over sets of five consecutive frames.

\subsection{Model Structure}
Utilizing the framework \cite{Mirsamadi_2017}, all models in this study are designed. This framework comprises an input layer, a bidirectional long short-term memory (LSTM) layer, a fully connected layer, and an output layer. Mirsamadi et al. \cite{Mirsamadi_2017} investigated the various attention mechanisms, and their proposed weighted pooling layer considering the attention weights applied over the frame-level input features achieved the best performance when the input features extracted by the ``Emobase.config'' file in the OpenSMILE toolkit. Therefore, all models used the same structure, and we denoted the model as the \textbf{BLSTM-FC}. 

\subsection{Training Labels}
\label{ss:train_labels}
All models utilize the BLSTM-FC architecture, consistent with the design proposed in prior research \cite{Mirsamadi_2017}. Our objective in this study is to address the subjectivity of emotional perception. To tackle the variability in emotion perception, we train BLSTM-FC models using two distinct label learning: hard-label (denoted as CROWD$_{H}$) and soft-label (denoted as CROWD$_{S}$). The hard label is the conventional way to obtain a single-label emotion as ground truth. To consider that emotion perception could be overlapped and blended, Steidl et al. \cite{Steidl_2005} first propose the soft labels to calculate the distributional label based on the votes for each emotion class. Fayek et al. \cite{Fayek_2016} integrated the inter-rater variability with soft-label to build SER systems. Also, Ando et al. \cite{Ando_2018} modified the formulation of the calculation of soft labels by introducing $\alpha$ to slightly change the distribution of conventional soft labels as below.

\begin{equation}
    t(c_{i}) = \frac{\alpha +\sum_{n}^{R} v_{i}^{n} }{\alpha\times C + \sum_{i}^{C} \sum_{n}^{R} v_{i}^{n}},
\end{equation}

where $c_{i}$ means the $i-$th emotion class, n represents the $n-$rater, $v_{i}^{n}$ is the binary value to check whether $n-$rater chooses $c_{i}$ emotions, $C$ is how many categorical emotions and the $R$ represents the number of annotators for $t$ samples. In this dissertation, we follow the study \cite{Ando_2018} to assign the $\alpha$ value as $0.75$, and $C$ is 4 since the number of emotion classes is 4.

\section{Proposed Method}

\subsection{Rater-Modeling}
Due to the inherent subjectivity and unique individual differences, different people may interpret the exact spoken phrase in varied ways. To enhance the SER systems, we incorporate rater modeling. We categorize annotators into two groups: \textbf{Crowd} and \textbf{E}. In this context, \textbf{Crowd} refers to their collective annotations in the used dataset, while E pertains to the annotations by each specific individual rater. For the \textbf{Crowd} category, we develop two distinct models based on whether the targets are hard or soft labels (as detailed in Section~\ref{ss:train_labels}). Conversely, for the E category, models are trained using soft labels available to each annotator, implying that each annotator's quantity of annotated data varies. Table~\ref{tab:data_distribution} and Table~\ref{tab:label_distribution} show a summation of how many data points are used for training each model, and Table~\ref{tab:label_distribution} overviews the number of data samples for models. 
With soft labels, \textbf{Crowd$_S$} has over 3,185 more samples compared to \textbf{Crowd$_H$}.

\begin{table}[!b]
\fontsize{7}{9}\selectfont
\centering
\caption{Table overviews the data samples used to train models.}
\begin{tabular}{@{}cccc@{}}
\toprule
\textbf{Model}  & \textbf{Total} & \textbf{Multiple} & \textbf{Single} \\ \midrule
\textbf{Crowd$_H$} & 5531           & 0                 & 5531            \\
\textbf{Crowd$_S$} & 7774           & 3185              & 4589            \\
\textbf{E1}     & 5954           & 44                & 5910            \\
\textbf{E2}     & 7845           & 38                & 7807            \\
\textbf{E4}     & 6429           & 212               & 6217            \\
\textbf{E5}     & 422            & 3                 & 419             \\
\textbf{E6}     & 773            & 20                & 753             \\ \bottomrule
\end{tabular}
\label{tab:data_distribution}
\end{table}

\begin{table}[!t]
\fontsize{7}{9}\selectfont
\centering
\caption{Table summarizes the label distribution for each emotion class of the models. Notice that each sample could have more than one emotional rating.}
\begin{tabular}{@{}ccccc@{}}
\toprule
\textbf{Model}  & \textbf{Neutral} & \textbf{Anger} & \textbf{Sadness} & \textbf{Happiness} \\ \midrule
\textbf{CrowdH} & 80.88\%          & 19.94\%        & 19.60\%          & 29.58\%            \\
\textbf{CrowdS} & 29.33\%          & 17.77\%        & 17.10\%          & 35.79\%            \\
\textbf{E1}     & 8.49\%           & 21.21\%        & 20.64\%          & 49.67\%            \\
\textbf{E2}     & 22.45\%          & 26.58\%        & 19.62\%          & 31.35\%            \\
\textbf{E4}     & 52.88\%          & 12.41\%        & 10.95\%          & 23.76\%            \\
\textbf{E5}     & 69.88\%          & 15.29\%        & 5.88\%           & 8.94\%             \\
\textbf{E6}     & 26.73\%          & 15.76\%        & 14.22\%          & 43.38\%            \\ \bottomrule
\end{tabular}
\label{tab:label_distribution}
\end{table}

\begin{figure*}[!b]
    \centering
    \includegraphics[width=0.8\linewidth]{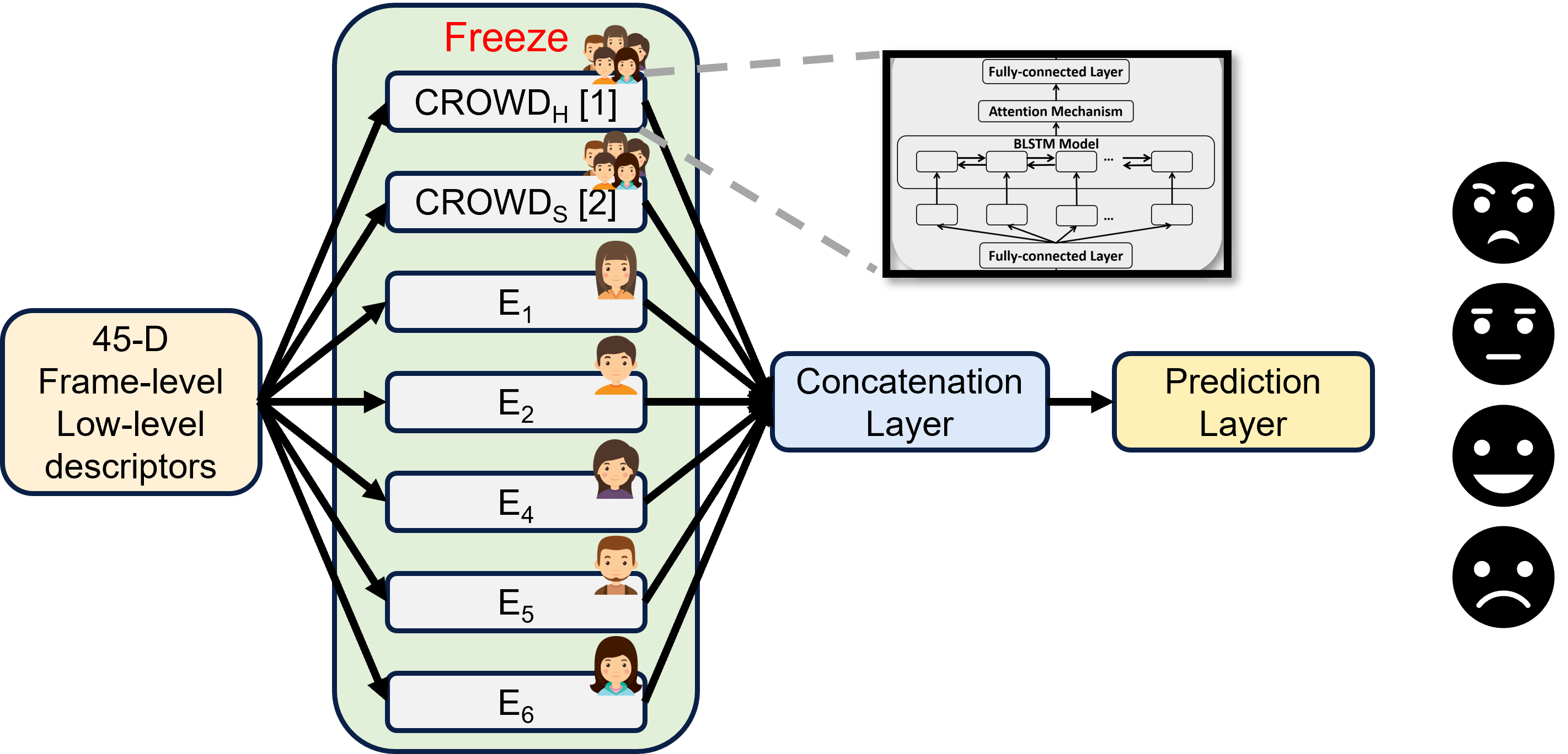}
    \caption{The figure illustrates the overall proposed model.}
    \label{fig:rater_model}
\end{figure*}

\subsection{Final Concatenation Layer}
After successfully computing all model components, including the two Crowd models and 5 E$_N$ models, we froze their respective model weights. Then, we concatenate the representation from the final layer before the softmax activation in each BLSTM-FC model (depicted within the square box in Fig.~\ref{fig:rater_model}). This concatenated layer is followed by an additional softmax layer to output the last prediction. The entire architecture is presented in Fig.~\ref{fig:rater_model}.

\section{Experimental Setup}
\subsection{Foundational Component}
The foundational component is the BLSTM-FC model equipped with an attention mechanism. This model architecture includes two fully connected (FC) layers with Rectified Linear Unit (ReLU) activation functions, a BLSTM layer enhanced using attention weights, and concludes with a fully-connected layer employing a softmax function. Specifically, the model comprises 256 hidden units in the initial dense layer, 128 hidden units in the BLSTM layer with attention, 256 hidden units in the second dense layer, and the final softmax-enabled layer differentiates into four units. Each layer also integrates a dropout mechanism with a 50\% 

\begin{figure*}[!t]
    \centering
    \includegraphics[width=0.8\linewidth]{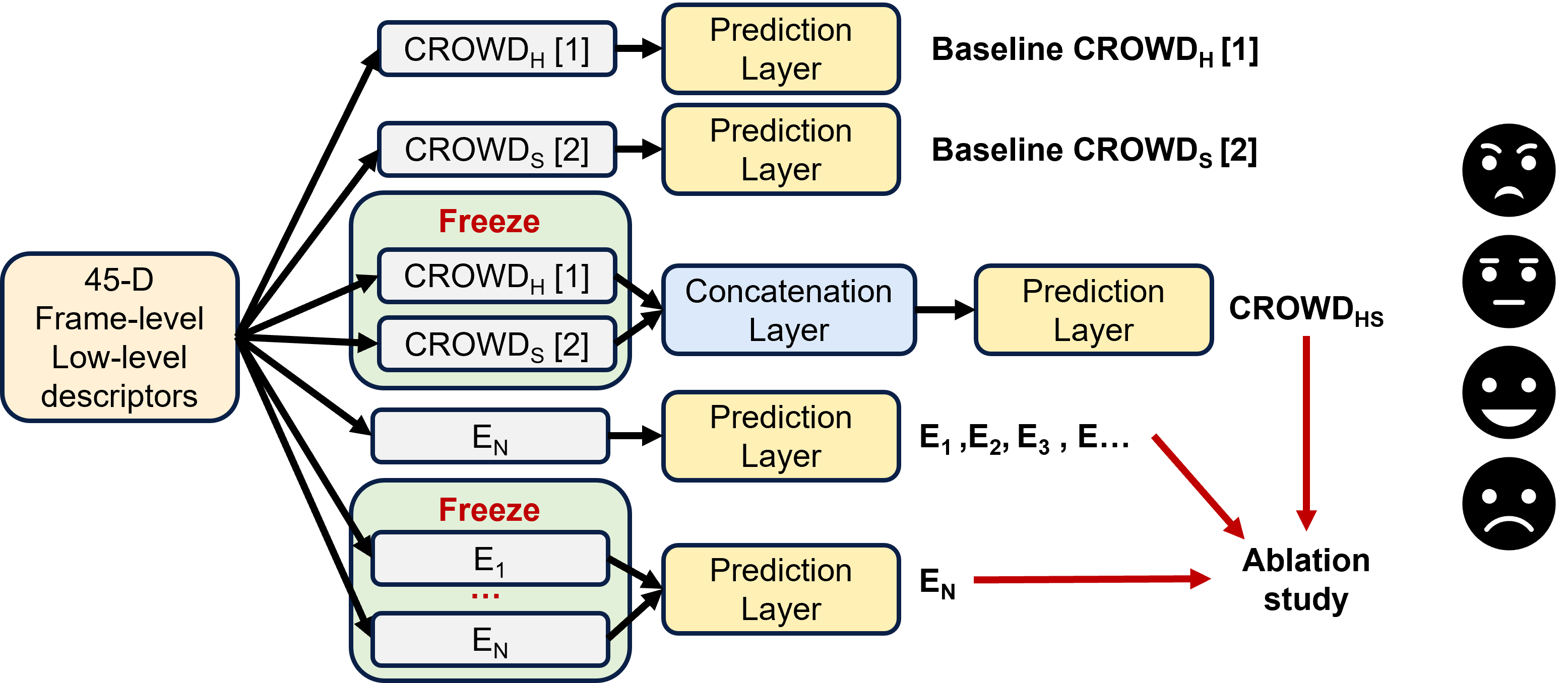}
    \caption{The figure illustrates the baselines and models for the ablation study.}
    \label{fig:rater_model_ablation}
\end{figure*}

\subsection{All Model Comparison}
Figure~\ref{fig:rater_model_ablation} illustrates the various baselines and models considered in the ablation study. Subsequently, we evaluate and compare the performance outcomes for each component of the overall architecture, as detailed below.

\begin{itemize}
    \item \textbf{Baseline CROWD$_H$}: This model closely parallels the previous proposed, but it utilizes a BLSTM-FC framework trained on hard labels in the study \cite{Mirsamadi_2017}.
    \item \textbf{Baseline CROWD$_S$}: This model employs soft label training, a method proposed by the work \cite{Ando_2018}, designed to utilize all labeled samples. 
    \item \textbf{Baseline CROWD$_{HS}$}: The model represents a fusion of Crowd$_H$ and Crowd$_S$. It combines all Crowd-relevant information by concatenating the representations from Crowd$_H$ and Crowd$_S$ before feeding them into the final softmax layer.
    \item \textbf{Proposed Rater Model, E$_N$}: Each of the E$_N$ models is trained using soft label learning based on the annotations made by individual raters.
    \item \textbf{Proposed Fusion of E$_N$}: The model integrates all individual E$_N$ components (five separate annotators). It consolidates all rater-specific information by concatenating the representations from each E$_N$ before passing them through the final softmax layer.
    \item \textbf{Proposed Model}: As depicted in Figure~~\ref{fig:rater_model}, this model represents our ultimate proposal and leverages all available Crowd and E$_N$ information. It achieves this by concatenating the representations from both Crowd$_H$ and Crowd$_S$ before feeding them into the final softmax layer. 
\end{itemize}

\begin{table*}[!b]
\fontsize{7}{9}\selectfont
\centering
\caption{Table summarizes cross-validation details of the IEMOCAP, following \cite{Mirsamadi_2017,Ando_2018}.}
\begin{tabular}{@{}c|c|c|c@{}}
\toprule
\textbf{Fold} & \textbf{Training Set} & \textbf{Development Set}                        & \textbf{Test Set} \\ \midrule
\textbf{1}    & Ses. 1,2,3,4       & \multirow{5}{*}{Randomly 10\% of training data} & Ses. 5         \\
\textbf{2}    & Ses. 2,3,4,5       &                                                 & Ses. 1         \\
\textbf{3}    & Ses. 3,4,5,1       &                                                 & Ses. 2         \\
\textbf{4}    & Ses. 1,4,5,2       &                                                 & Ses. 3         \\
\textbf{5}    & Ses. 1,2,4,3       &                                                 & Ses. 4         \\ \bottomrule
\end{tabular}
\label{tab:cv_iemocap_rater_modeling}
\end{table*}

\section{Other Training Details}
We run the experiments on the cross-validation in the leave-one-session-out condition (as shown in Table~\ref{tab:cv_iemocap_rater_modeling}), evaluating performance in an unweighted average recall (UAR). The batch size is 32; the learning rate is $0.0001$; the number of epochs is 200. Early stopping criteria to minimize the loss value on the validation set are applied during training across all configurations to prevent overfitting and ensure model effectiveness. The optimizer utilized in this study is ADAMMAX \cite{Kingma_2015}.

\begin{table*}[!t]
\fontsize{7}{9}\selectfont
\centering
\caption{Table summarizes the results in unweighted average recall (UAR) of all models evaluated on the IEMOCAP.}
\begin{tabular}{@{}cccccc@{}}
\toprule
\textbf{Model}    & \textbf{Overall} & \textbf{Neutral} & \textbf{Angry}  & \textbf{Happy} & \textbf{Sad} \\ \midrule
\textbf{Crowd$_H$\cite{Mirsamadi_2017}}   & 0.5745           & 0.5571           & 0.6329          & 0.4502             & 0.6577           \\
\textbf{Crowd$_S$\cite{Ando_2018}}   & 0.5712           & 0.4970            & 0.6298          & 0.6285             & 0.5314           \\
\textbf{Crowd$_{HS}$}  & 0.5858           & 0.5966           & 0.5931          & 0.5363             & 0.6171           \\
\textbf{E1}       & 0.5098           & 0.0804           & 0.6131          & \textbf{0.7724}    & 0.5734           \\
\textbf{E2}       & 0.5968           & 0.3878           & 0.6435 & 0.6425             & 0.6261  \\
\textbf{E4}       & 0.4859           & 0.8129           & 0.4542          & 0.3820              & 0.2944           \\
\textbf{E5}       & 0.3762           & \textbf{0.8689}  & 0.4762          & 0.1121             & 0.0475           \\
\textbf{E6}       & 0.4582           & 0.3685           & 0.4010           & 0.6039             & 0.4595           \\
\textbf{EN}       & 0.6024           & 0.4964           & 0.6364          & 0.6148             & 0.6619           \\ \midrule
\textbf{Proposed} & \textbf{0.6148}  & 0.5455           & \textbf{0.6451}          & 0.6032             & \textbf{0.6656}           \\ \bottomrule
\end{tabular}
\label{tab:rater_model_results}
\end{table*}

\section{Experimental Results and Analyses}

Table~\ref{tab:rater_model_results} provides an overview of the results across various comparative models. Our presented framework achieves the highest overall emotion classification performance, with an unweighted average recall (UAR) of 61.48\%. This method exceeds the performance of the previously leading approaches, surpassing Crowd$_H$ proposed by \cite{Mirsamadi_2017} by 3.18\% and Crowd$_S$ proposed by \cite{Ando_2018} by 4.36\% in absolute terms. These results highlight the benefits of incorporating subjective annotator emotional information to enhance emotion recognition over previous methods.

One significant finding is the differential effectiveness of soft-label vs. hard-label learning: Crowd$_S$ demonstrates better performance on happy emotions than Crowd$_H$, which performs better with neutral and sad emotions. The complementary strengths of Crowd$_H$ and Crowd$_S$ underscore the importance of their integration for advanced emotion recognition results. Historically, happiness has been a challenging class to identify \cite{Mirsamadi_2017,Ando_2018}, and it benefits from soft-label learning, suggesting that happiness has a more diffused presence within the acoustic spectrum than other emotions, such as anger and sadness.

Furthermore, individual models tend to have low recognition rates, probably because of the subjective perspectives of unique raters and the uneven distribution of emotion classes within each annotator's labeling. For instance, as Table~\ref{tab:rater_model_results} illustrates, the E$_1$ model shows low recognition accuracy for the neutral state but reasonable performance for recognizing happy emotion, while it is reverse for the E$_5$ model. Examining the emotion distribution in Table~\ref{tab:label_distribution} reveals that this discrepancy is linked to the variety and quantity of emotional data annotated by each rater. By integrating raters' models at the late-fusion level, E$_N$ effectively taps into multiple complementary viewpoints, enabling it to learn a well-rounded and nuanced understanding of emotion perception from distinct individual perspectives.

Last but not least, by employing individual rater models, our proposed method can expand the dataset used for building SER systems, compared to the traditional hard-label approach that limits the number of data in the train set to instances where consensus among raters is achieved. This allows for a more robust and comprehensive understanding of emotional nuances.

\section{Summary}
Human perception of emotions is relatively subjective and differs greatly from person to person. In this study, we introduce a model that merges dominant sentiment annotations with individual subjectivity models to advance emotion classification accuracy. The framework achieves an impressive score of 61.48\% on a task involving 4-class emotion categorization. Although numerous studies have tackled the issue of annotator subjectivity, this pioneering approach explicitly combines consensus and individual differences in emotion perception, leading to improved classification performance on a benchmark dataset.

\chapter{Novel Evaluation Method by an All-Inclusive Aggregation Rule}
\label{ch:ch_evaluation}

When choosing test data for subjective tasks, many studies form ground truth labels through aggregation techniques like the majority or plurality rules. Unfortunately, these techniques ignore data points lacking consensus, simplifying the test set compared to real-world tasks where predictions are necessary for every sample. These ignored data points often contain ambiguous signals with overlapping traits witnessed by annotators. We emphasize the need to account for all annotations and samples in the dataset, as focusing only on performance metrics derived from a test set filtered by majority or plurality rules may skew the model's performance evaluations. We specifically investigate SER tasks and note that traditional aggregation rules result in data loss ratios between 4.63\% and 92.01\%, as shown in Table~\ref{tab:dataloss}. Based on this insight, we introduce a versatile, all-inclusive rule, a label aggregation approach to appraise SER systems using comprehensive test data. We differentiate the conventional single-label approach with a multi-label methodology catering to the coexistence of various emotions. Training an SER model with data chosen by the all-inclusive rule consistently achieves better macro-F1 scores when evaluated on the whole test set, including ambiguous, non-consensus samples.

\section{Motivation and Background}
\label{sec:motivation}
Given the inherently subjective nature of these tasks, models are often evaluated using labels derived from human perceptual assessments, where multiple raters annotate each data point. The common practice for processing these annotations and creating training and testing sets relies on majority or plurality aggregation methods. These methods ignore annotations that do not correspond with the consensus label. However, co-existing emotions are frequently observed in everyday interactions \cite{Vansteelandt_2005}, making it challenging for a single label to represent a sample's emotional perception fully. Additionally, data points lacking agreement are excluded. According to the \emph{majority rule} (MR), data is dismissed if no class achieves more than 50\% of the votes. Similarly, the \emph{plurality rule} (PR) disregards data if no single class receives more votes than others. This exclusion of ambiguous samples undermines the validity of systems meant for real-world applications since these samples are not represented in the test set. Previous studies have investigated the use of all existing annotations during training, adopting soft-label learning strategies to incorporate all samples \cite{Fayek_2016, Lotfian_2017, Kim_2018, Chou_2019, Ando_2019, Sridhar_2021,Li_2023}. Nonetheless, test sets remain \emph{simplified}, considering only sentences that meet MR or PR criteria, thereby overlooking complex and ambiguous samples.

\begin{figure*}[!t]
    \centering
    \includegraphics[width=0.5\linewidth]{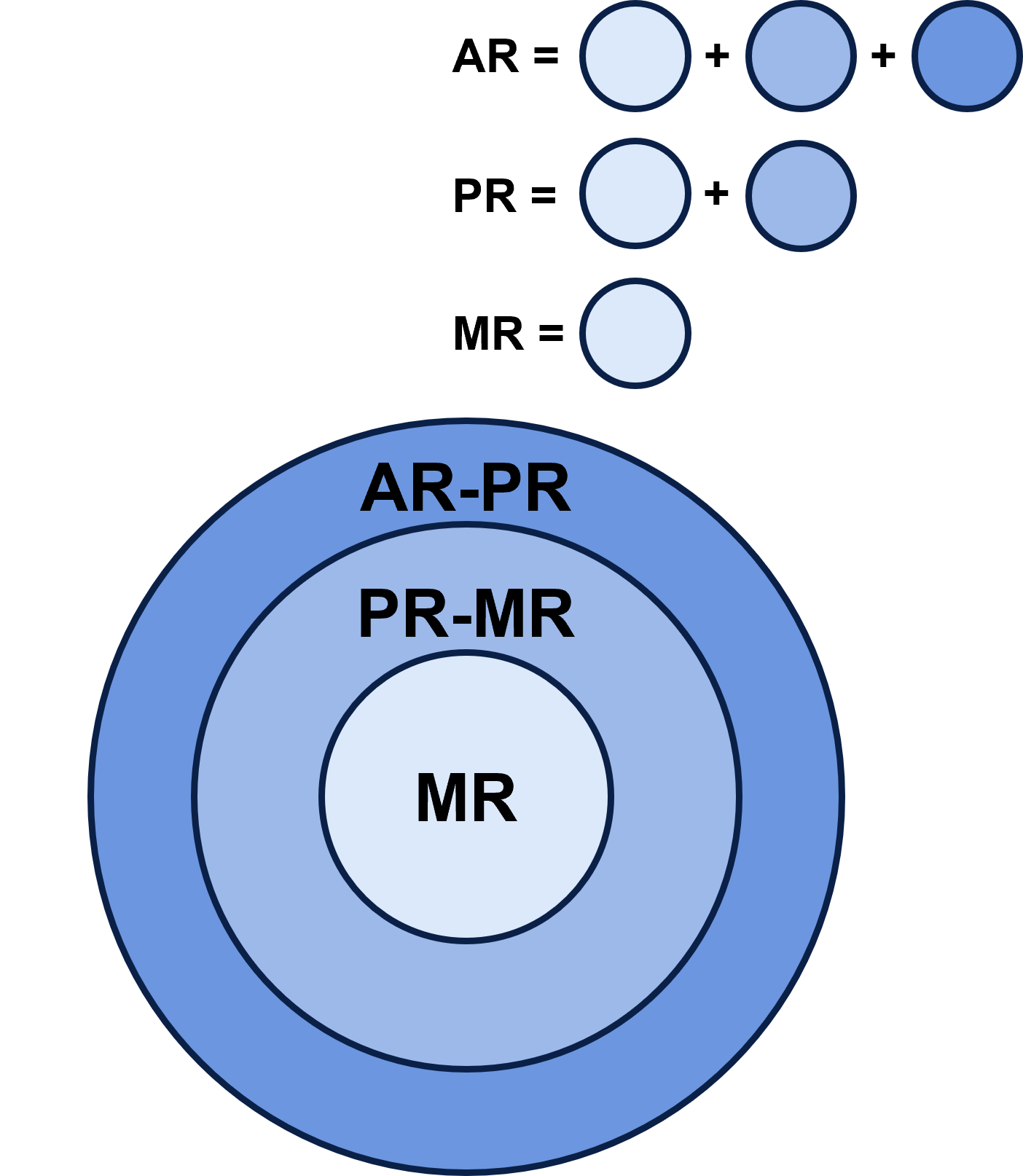}
    \caption{A diagram illustrating how much data and ratings are used in the final test set according to each aggregation method. MR contains the lowest amount of data, and AR always includes the entire test set available in the dataset.}
    \label{fig:data_rules}
\end{figure*}

We propose a practical methodology that amalgamates all annotations gathered from subjective evaluations for training and test sets, thereby enhancing the practical applicability of these systems. Although this approach is compatible with any domain requiring labels derived from perceptual assessments, our focus is on the SER task, where co-occurring emotions are common in everyday interactions. Unlike traditional methods that neglect non-consensus labels, we retain all data points in the training and test sets. This strategy enables SER models to utilize comprehensive data during training and ensures they are evaluated on all samples in the test sets, including those with divergent labels. We call this framework the \emph{all-inclusive rule} (AR) method. The ability to accommodate co-occurring emotions is a cornerstone of the approach. Figure~\ref{fig:data_rules} compares the all-inclusive aggregation strategy against the majority and plurality rules, typically applied in SER tasks to select data points for the test set. By implementing the AR method, every data point is included in the test set, thus underlining the exhaustive performance evaluation of SER systems. The main research questions driving the study are as follows.

\begin{itemize}
    \item How is the performance of SER systems influenced by using different aggregation methods for the training set annotations?
    
    \item Does utilizing data from the all-inclusive rule in training an SER system enhance performance on ambiguous emotions compared to data processed with majority or plurality rules?
    
    \item Which label learning strategy should be employed for training SER systems to achieve optimal performance when tested on the entire data set?
\end{itemize}

\section{Previous Literature}

\subsection{Evaluation of SER Systems}
Evaluating SER systems on a complete test set is crucial. However, the common approach involves discarding samples that lack consensus emotion labels. When gathering emotional annotations from multiple workers, significant disagreement often exists among annotators \cite{Devillers_2005, Mower_2009_2, Sethu_2019}, leading many studies to eliminate numerous data points from the test set. For instance, the IEMOCAP and CREMA-D corpora use the majority rule (MR) to construct ground-truth labels, discarding approximately 31.37\% and 35.8\% of the data, respectively, as shown in Table~\ref{tab:dataloss} \cite{Busso_2008_3, Busso_2008, Cao_2014}. Researchers using these corpora often adhere to this rule for testing their models \cite{Fayek_2017, Zhao_2021_2, Mekruksavanich_2020, Mocanu_2021}. Similarly, the IMPROV and PODCAST corpora use the plurality rule (PR) to annotate primary and secondary emotional labels for each speaking turn \cite{Busso_2017, Lotfian_2019}, and this default aggregation method has been widely adopted in subsequent studies \cite{Neumann_2019, Goncalves_2022, Pappagari_2021}.

The aforementioned studies assume that each speaking turn has only one emotional category, ignoring secondary emotions in the recordings. In reality, emotional states often co-exist (e.g., a person can be sad and angry) \cite{Vansteelandt_2005}. Therefore, consolidating multiple annotations into a single class and discarding non-consensus data points does not accurately capture whether SER system predictions reflect the complex emotional behaviors observed in daily interactions. Although some studies have explored using a "multiple-hot" vector to frame SER as a multi-label problem \cite{Ju_2020, Zhang_2020, Zhang_2021}, this approach does not discern dominant emotions. It treats all annotations equally valid, even if a single annotator selected a class.

To our knowledge, Riera et al. \cite{Riera_2019} is the only study advocating for including all test samples in evaluating SER systems rather than discarding non-consensus data. However, their study did not explore training SER systems with various label learning methods. It involved relabeling some emotions (e.g., labeling excited as happy and surprised as "other"), which is a significant limitation. Unlike that study, this dissertation retains the original emotional classes across all datasets. It involves empirical experiments with different label learning methods and test sets created using different aggregation methods.

\subsection{Curriculum Learning for Emotion Recognition}
Lotfian and Busso \cite{Lotfian_2019_2} rigorously employed curriculum learning, utilizing annotator disagreement on samples to enhance the performance of SER systems in predicting valence and arousal. Furthermore, Yang et al. \cite{Yang_2022} utilized emotion shift as difficulty scores to categorize samples as "easy" or "hard." They trained text-based conversational emotion recognition systems progressively, starting with easy samples and moving to more challenging ones. Their findings show that curriculum learning boosts performance in emotion recognition within conversations, as incorporating harder samples during training increases the training loss, thereby refining the system's accuracy. Similar findings have been reported in recent research \cite{Li_2024}.

\begin{table}[!t]
\fontsize{7}{9}\selectfont
\centering
\caption{Table is an overview of the number of utterances, emotion classes, and data loss ratios in several prominent emotional datasets. Here, P indicates primary emotions, and S indicates secondary emotions.}
\begin{tabular}{@{}cc|cc|cc|cc@{}}
\toprule
\multicolumn{2}{c}{Label Aggregation} & \multicolumn{2}{c}{MR} & \multicolumn{2}{c}{PR} & \multicolumn{2}{c}{AR} \\ \midrule
Database              & \# Utterances          & Data               & Rating            & Data               & Rating             & Data                  & Rating                \\ \midrule
IMPROV (P)            & 8438                   & 9.18\%             & 28.52\%           & 4.63\%             & 26.41\%            & 0.00\%                & 5.12\%                \\
CREMA-D               & 7442                   & 35.80\%            & 52.96\%           & 8.55\%             & 40.57\%            & 0.00\%                & 0.00\%                \\
PODCAST (P)           & 90978                  & 44.81\%            & 59.87\%           & 19.85\%            & 49.24\%            & 0.00\%                & 6.15\%                \\
IEMOCAP               & 10039                  & 31.37\%            & 49.44\%           & 25.32\%            & 45.70\%            & 0.00\%                & 3.10\%                \\
IMPROV (S)            & 8438                   & 54.18\%            & 76.91\%           & 12.32\%            & 56.70\%            & 0.00\%                & 4.23\%                \\
PODCAST (S)           & 90978                  & 92.01\%            & 96.99\%           & 33.72\%            & 78.13\%            & 0.00\%                & 1.64\%                \\ \midrule
\multicolumn{2}{c}{Average}                    & 44.56\%            & 60.78\%           & 17.40\%            & 49.46\%            & 0.00\%                & 3.37\%                \\ \bottomrule
\end{tabular}
\label{tab:dataloss}
\end{table}

\section{Methodology}
We present a new aggregation methodology called the \emph{all-inclusive rule} (AR), designed to facilitate the training and evaluation of SER systems using an exhaustive test set. This includes data points lacking Majority Rule (MR) or Plurality Rule (PR) consensus. The definition, significance, and application of this rule are thoroughly explained.

\subsection{Proposed All-inclusive Rule}
The All-Inclusive Rule (AR) is an aggregation methodology that retains all annotated samples within a dataset, regardless of vote frequencies. This method ensures data points are never disregarded. Initially, AR collects all classifications assigned to each data point, creating the ground truth. When forming the training set, the ground truth is represented either by a one-hot encoding or the vote distribution based on the selected label learning strategy. For the hard-label approach, AR identifies the emotional class with the most votes as the ground truth—akin to the plurality rule. In cases where a clear majority is absent, one of the top-voted classes is randomly selected as the ground truth. An example is shown in Table \ref{tab:labelexamples} Case (C1), where the hard label could be (1,0,0,0) or (0,0,1,0). AR produces the ground truth for the soft-label or distribution-label approach by reflecting the vote distribution among emotional classes.

AR consistently utilizes the distributional ground truth when producing the test set, irrespective of the chosen label-learning strategy, which is highlighted in the rightmost column of Table \ref{tab:labelexamples}. This comprehensive approach ensures every annotated data point and all its annotations are integral to the test set. The all-inclusive rule enhances the label descriptor, more accurately capturing the emotional nuances of data points by integrating sentences that reflect ambiguous emotions into the test set.

\begin{table}[!t]
\fontsize{7}{9}\selectfont
\centering
\caption{Here is an overview of how label vectors are constructed for the \emph{all-inclusive rule} (AR) using three examples, each with five annotations, in a four-class emotion classification task. The four emotions include neutral (N), happy (H), angry (A), and sad (S). Label vectors are created in the format: (N, H, A, S). We show three examples. For instance, (C1) N,N,A,A,S demonstrates that the five emotional annotations for Case (C1) selected two for neutral, two for angry, and one for sad.}
\begin{tabular}{@{}l@{\hspace{0.1cm}}|c@{\hspace{0.1cm}}c@{\hspace{0.1cm}}c@{\hspace{0.1cm}}|c@{}}
\toprule
\multirow{2}{*}{\textbf{Case}}  & \multicolumn{3}{c|}{\textbf{Training Set}}                                                                                        & \textbf{Test Set}                                                                   \\ 
                       & Hard-label                                                                     & Soft-label        & Distribution-label & \begin{tabular}[c]{@{}c@{}}Label\end{tabular} \\ \cmidrule(r){1-5}
(C1) N,N,A,A,S       & \begin{tabular}[c]{@{}c@{}}(1,0,0,0)\\    \\ OR\\    \\ (0,0,1,0)\end{tabular} & (0.4,0.0,0.4,0.2) & (0.4,0.0,0.4,0.2)  & (0.4,0.0,0.4,0.2)                                                                          \\ \midrule
(C2) N,N,H,A,S  & (1,0,0,0)                                                                      & (0.4,0.2,0.2,0.2) & (0.4,0.2,0.2,0.2)  & (0.4,0.2,0.2,0.2)                                                                             \\ \midrule
(C3) N,N,N,A,S      & (1,0,0,0)                                                                      & (0.6,0.0,0.2,0.2) & (0.6,0.0,0.2,0.2)  & (0.6,0.0,0.2,0.2)                                                                             \\ \bottomrule
\end{tabular}
\label{tab:labelexamples}
\end{table}

\subsection{Employing the All-Inclusive Rule for Test Set Construction}
We include all data samples and prioritize every available emotion as a learning target to capture the full range of opinions gathered during perceptual evaluation. For instance, previous research using the IEMOCAP corpus has aggregated annotated emotions into a 4-class emotion classification task (e.g., combining excitement and happiness while disregarding less frequent classes such as fear, surprise, and disgust). Unlike this methodology, we do not overlook any emotional states nor confine the SER models to be trained or tested solely on a few selected emotions.

In addition, our all-inclusive rule allows SER models to be tested on the entire test set, including secondary emotions. Previous studies have often disregarded secondary emotional annotations due to considerable data loss from standard aggregation methods (up to 92.01\% as noted in Table \ref{tab:dataloss}). As the AR method utilizes the fully annotated test set, we can assess the SER model with secondary emotions, which have not been examined before. Table \ref{tab:dataloss} highlights the proportion of data loss introduced by different aggregation methods across the four datasets used in this study for primary and secondary emotions—utilization of MR and PR results in discarding up to 92.01\% and 33.72\% of the data, respectively. The most significant data loss occurs when classifying secondary emotions in the MSP-Podcast corpus.

\section{Experimental Setup}

\subsection{Resource}

Four publicly available emotion databases, as detailed in Chapter \ref{ch:database}, are utilized in our research. The corpus version 1.10, containing 104,267 annotated utterances, is employed. However, the ``Test2'' set is excluded, narrowing the dataset to 90,978 utterances, as shown in Table \ref{tab:dataloss}.

\subsection{Speech Emotion Classifier}

To assess the performance of various aggregation methods, we utilize the Wav2vec2.0 architecture \cite{Baevski_2020}, which has demonstrated strong performance in SER tasks in multiple studies \cite{Wang_2021, Wagner_2023}. Specifically, we implement the "wav2vec2-large-robust" variant, as proposed in Hsu et al. \cite{Hsu_2021_3}, which has proven to be the best-performing model in the study by Wagner et al. \cite{Wagner_2023}. The model processes raw audio signals directly, preceding spectrograms or Mel-frequency cepstral coefficients (MFCC). The datasets are sampled at a rate of 16k Hz to match the sampling rate of the pre-trained model data.

For efficiency and reproducibility, we've optimized the model by removing the top 12 of the 24 transformer layers from the architecture, which preserves recognition performance while reducing the number of parameters \cite{Wagner_2023}. Two hidden layers and a softmax output layer are attached to this trimmed Wav2vec2.0 model. Each hidden layer with the rectified linear unit (ReLU) activation function comprises 1,024 nodes. Implemented with average pooling per utterance, the outputs from the Wav2vec2.0 layers feed into the classification layers. We apply a dropout function with ( p = 0.5 ) to the first and second classification layers for regularization.

Our implementation is based on the HuggingFace library \cite{Wolf_2019} and uses a pre-trained "wav2vec2-large-robust" model. During fine-tuning, convolutional and transformer layers of the Wav2vec2.0 model are frozen—a strategy that has shown better performance than fully fine-tuning all parameters \cite{Wang_2021}. We employ the Adam optimizer \cite{Kingma_2014_2} and set up a learning rate as $0.0001$, structuring mini-batches from 32 utterances and training the model over 100 epochs. The best recognition performance model is chosen from the development set after the training epochs. The entire implementation is carried out in Pytorch \cite{Paszke_2019} and executed on an NVIDIA Tesla V100 GPU.

For the POSCAST corpus, we utilize pre-defined training (63,076 samples), development (10,999 samples), and test (16,903 samples) sets. Due to the smaller nature of other corpora, we implement a cross-validation strategy. Speaker-independent sessions are used for IEMOCAP (five sessions) and IMPROV (six sessions). As the CREMA-D corpus lacks pre-defined sessions, we manually divide it into five speaker-independent sessions. For IEMOCAP, IMPROV, and CREMA-D, a $K$-fold cross-validation is applied, where K means how many sessions. Each fold includes one session as the test set, one as the development set, and the remaining is the training set. More details about partitions are in Section~\ref{s:partition}

Table \ref{tab:dataloss} reflects data loss ratios derived from each label aggregation method across various databases. We evaluate data loss ratios in every partition (train, development, test set). Observations indicate that data loss trends are relatively consistent across all four datasets. Thus, Table \ref{tab:dataloss_partition} exemplifies these distributions specifically for the PODCAST database, and the trends are similar in Table~\ref{tab:dataloss}.

\begin{table}[t]
\fontsize{7}{9}\selectfont
\centering
\caption{Here is an overview of the data loss ratios introduced by the label aggregation method on the PODCAST development, test, and train sets. P represents primary emotions, and S represents secondary emotions.}
\begin{tabular}{@{}crrr@{}}
\toprule
Rule                & Set         & PODCAST (P) & PODCAST (S) \\ \midrule
\multirow{3}{*}{MR} & Training       & 47.95\%     & 88.63\%     \\
                    & Development & 47.90\%     & 89.64\%     \\
                    & Test        & 47.08\%     & 90.90\%     \\  \midrule
\multirow{3}{*}{PR} & Training       & 18.76\%     & 29.46\%     \\
                    & Development & 19.62\%     & 28.90\%     \\
                    & Test        & 17.14\%     & 27.76\%     \\  \midrule
\multirow{3}{*}{AR} & Training       & 0.00\%      & 0.00\%      \\
                    & Development & 0.00\%      & 0.00\%      \\
                    & Test        & 0.00\%      & 0.00\%      \\ \bottomrule
\end{tabular}
\label{tab:dataloss_partition}
\end{table}

\subsection{Train/Test Set Defined by Aggregation Rules}

In this evaluation, the models are trained and tested using both matching and mismatching aggregation rules. For both the training and development sets, ground truth is established using MR, PR, and AR, denoted as $MR_{Train}$, $PR_{Train}$, and $AR_{Train}$, respectively. The models are assessed on sets derived from MR, PR, and AR rules during testing. Additionally, two extra test conditions are defined, illustrated by the \emph{donuts} in Figure \ref{fig:data_rules}: PR-MR includes test samples accepted by PR but not by MR, whereas AR-PR encompasses those accepted by AR but not by PR. The AR-PR condition is the most ambiguous set, as it includes samples with non-consensus annotations. To our knowledge, this study is the first to evaluate SER models using such non-consensus annotations.

\subsection{Label Learning for SER}
Alongside evaluating different aggregation methods, we assess the performance using various label-learning approaches. The experiments employ three label learning strategies: hard-, soft-, and distribution-label learning.

For hard-label learning, the ground truth is constructed with a one-hot encoding, where the class receiving the highest number of votes from annotators is represented as "1.0". When utilizing the training set aggregated with AR, if there is no clear consensus, one of the top-voted emotions is randomly chosen as the ground-truth emotion. We smooth the one-hot encoding ground truth vector using the smoothing strategy proposed by Szegedy et al. \cite{Szegedy_2016} with a parameter set of 0.05. This method slightly adjusts probabilities for classes assigned initially a zero value. The SER systems are then trained using the cross-entropy (CE) objective function.

For both soft-label learning and distribution-label learning, the ground-truth vector reflects the distribution of annotator votes. This is achieved by dividing the vote count for each class by the total number of votes for each data point. We also apply the label-smoothing strategy used in hard-label learning. Soft-label learning targets are optimized using the CE loss function, while distribution-label learning uses the Kullback–Leibler divergence (KLD) as the cost function.

\subsection{Evaluation Metrics and Statistical Significance}

\subsubsection{Hard-decision-based assessment}

This study employs macro-F1 scores to evaluate SER performance, which involves calculating precision and recall rates. The MR, PR, and PR-MR test sets are structured by selecting a single class, making them compatible with macro-F1 scoring. The class receiving the most votes is chosen as the target for these test sets. An output is valid if the emotion category with the highest predicted probability matches the target class.

For test sets gathered under AR and AR-PR conditions, which contain non-consensus labels, we permit the coexistence of multiple emotions to compute the macro-F1 score. Targets are selected based on using the threshold to the binarized vectors. A prediction is valid if the proportion for a specific category exceeds $1/C$, with $C$ representing how many emotion classes. This method is in line with those employed in previous studies \cite{Riera_2019,Chou_2022}.
  
Consider an emotion recognition task that distinguishes between four emotions: neutral, anger, sadness, and happiness. In one instance, five different reviewers each gave their rating based on the sample, resulting in the following annotations: angry (A), sad (S), sad (S), neutral (N), and angry (A). To determine the label distributions, we categorize the data into (N, A, S, H) and get the proportions (0.2, 0.4, 0.4, 0.0).

The threshold is set to (1/4 $=$ 0.25), so we convert the ground truth to (0,1,1,0). During the inference, suppose we have predictions from three different models: (0.1, 0.45, 0.45, 0.0), (0.2, 0.35, 0.35, 0.1), and (0.45, 0.1, 0, 0.45). Applying the (0.25) threshold, these outputs are converted into (0,1,1,0), (0,1,1,0), and (1,0,0,1), respectively. In this example, the (0,1,1,0) and (0,1,1,0) fully match the ground truth.

\subsubsection{Distribution-based Assessment}
Following the approach proposed by Steidl et al. \cite{Steidl_2005}, where results are assessed using an entropy-based metric, we employ the \emph{Kullback-Leibler divergence} (KLD) to determine the similarity between the model's predicted distribution and the subjective annotations. This method checks whether an SER model aligns with human emotional perception. Unlike the macro-F1 evaluations, which binarize the model's output for single-label or multi-label tasks, we utilize the model's probability distributions across all test sets and measure them using KLD. As illustrated in Table \ref{tab:kld}, lower KLD values indicate better performance in this context.

\subsubsection{Distribution-based Assessment Versus Hard-decision-based Assessment}

Distribution-based assessments, such as KLD, are suitable metrics for evaluating distributions. In contrast, hard-decision-based assessments like the F1-score necessitate categorical decisions, making them less appropriate for comparing distribution similarities. Distribution-based assessments keep all data and maximum usage of emotional ratings while evaluating the SER performances. Every assessment has different advantages and disadvantages, so both are presented in the paper. Table \ref{tab:metrics_comparison} outlines the benefits and limitations of using KLD versus traditional accuracy metrics (e.g., macro-F1, micro-F1, and weighted-F1). Reporting both metrics offers complementary and more detailed insights.

\begin{table*}[!t]
 \fontsize{7}{9}\selectfont
    \centering
    \caption{Comparison of distribution-based and hard-decision-based assessment metrics.}
    \begin{tabular}{| m{1.3cm} | m{6cm} | m{6cm} |}
        \hline
        \textbf{Metric} & \textbf{Distribution-based assessment} & \textbf{Hard-decision-based assessment} \\
        \hline
        Advantages & 

        \begin{itemize}[leftmargin=3.5mm]
            \item We can directly compare the models’ predictions to human perception without needing extra thresholding methods. 
            \item It is tuned to the overall shape and structure of the distributions, reflecting both the accuracy and confidence of the predictions.
        \end{itemize}    
         
        & 
        \begin{itemize}[leftmargin=3.5mm]
            \item The macro-F1 score has a defined range between 0 and 1, facilitating straightforward interpretation and comparison across various models and datasets.
            \item By applying thresholds to the labels, we can mitigate the influence of noisy annotations from raters who were not attentive.
        \end{itemize} 
        \\
        \hline
        Limitations & 

        \begin{itemize}[leftmargin=3.5mm]
            \item The performance differences between the baseline and proposed models are difficult to interpret because the scale differences are minimal.
            \item It lacks a fixed range, which makes it challenging to interpret and compare absolute values across different datasets or models.
            \item It is susceptible to slight distribution variations, especially when dealing with sparse or high-dimensional data.
        \end{itemize}
        & 
        \begin{itemize}[leftmargin=3.5mm]
            \item All predictions and ground truth need to be binary (0 or 1)—this requirement can lead to information loss and reduced granularity, as it oversimplifies the distributional nature of the ground truth.
            \item It is challenging to assess model performance on less common emotions, such as fear, due to limited available data. 
        \end{itemize}
        \\
        \hline
    \end{tabular}
    \label{tab:metrics_comparison}
\end{table*}

\subsubsection{Statistical Significance}
We assess the statistical significance of the results per each aggregation method utilized. For cross-validation experiments (IEMOCAP, CREMA-D, and IMPROV), the predictions for each condition across all folds are concatenated, ensuring that every piece of data is considered (i.e., each sample appears in one fold of the test set). Predictions from all pre-defined test sets using a single model are employed directly for the PODCAST experiments. After collecting these predictions, they are segmented into 40 folds to average the macro-F1 score. A two-tailed t-test is then conducted to determine statistical significance and to confirm whether the $p$-value is below $0.05$. The notations $\ast$, $\dag$, and $\star$ are used to indicate when a model's performance is significantly superior to models trained using $MR_{Train}$, $PR_{Train}$, and $AR_{Train}$ sets, respectively.

\section{Experimental Results and Analyses}
The experimental analysis begins by benchmarking our presented method against state-of-the-art (SOTA) baselines, showcasing the advantages of the SER strategy utilized in this research (Section \ref{ssec:SOTA}). Following that, we address the three research questions outlined in Section \ref{sec:motivation} (Sections \ref{ssec:IncompleteData}, \ref{ssec:AmbiguousSet}, and \ref{ssec:labellearning}).

\begin{table*}[!b]
\centering
\fontsize{7}{9}\selectfont
\caption{Table shows the comparative evaluation between our proposed model and existing SOTA baselines for the IMPROV(P), CREMA-D, IEMOCAP, and PODCAST(P) databases. The performance measurements are in the macro-F1 score, capturing the effectiveness of models by grouping labels in the test sets following the “majority rule” (MR) or “plurality rule” (PR).}
\begin{tabular}{l|l|lll|l}
\toprule
&& \multicolumn{3}{c|}{MR}  &  \multicolumn{1}{c}{PR} \\
Aggregation & Method & IMPROV(P) & CREMA-D & IEMOCAP & PODCAST(P)  \\ 
\midrule
\multirow{4}{*}{MR$_{Train}$/PR$_{Train}$} & Li et al. \cite{Li_2019}        & 0.398     & 0.311   & 0.256      &   0.150          \\
& Pepino et al.  \cite{Pepino_2021}   & 0.331     & 0.223   & 0.191      & 0.142            \\
& Goncalves et al. \cite{Goncalves_2022} & \textbf{0.539}     & 0.574   & 0.261           & 0.161            \\
& The proposed        & 0.512     & \textbf{0.591}   & \textbf{0.269}      & \textbf{0.184}       \\
\hline
$AR_{Train}$ & The proposed        & \textbf{0.562}     & \textbf{0.585}   & \textbf{0.279}      & \textbf{0.166}      \\
 \bottomrule
\end{tabular}
\label{tab:sota_comparison}
\end{table*}

\subsection{Comparison of Results with Prior SOTA Methods}
\label{ssec:SOTA}
We assess the SER model’s performance against three existing SOTA benchmarks using the IMPROV(P), CREMA-D, PODCAST(P), and IEMOCAP corpora. As discussed by Li et al. \cite{Li_2019}, the first reference model establishes an end-to-end system that converts speech into spectrograms and utilizes a self-attention mechanism to highlight emotional elements within sentences. At its time of publication, this model set the SOTA performance for identifying four primary emotions from the IEMOCAP database. The second baseline is presented in the work by Pepino et al. \cite{Pepino_2021}, which leverages wav2vec 2.0 for extracting speech characteristics and combines them with hand-crafted features from eGeMAPS \cite{Eyben_2016}, achieving top-tier classification outcomes on the IEMOCAP collection. The third model framework introduced by Goncalves and Busso \cite{Goncalves_2022} highlights a transformer-based architecture with multimodal losses, setting new SOTA records on the CREMA-D and IMPROV(P) databases. For a fair comparison, we have focused only on their "audio-only" mode version, using 65 discrete acoustic low-level descriptors for input.

All these baseline models were reconstructed as described in their respective research studies and evaluated under the same experimental conditions as the model. Following previous research, the investigations treat SER as a single-label problem, providing performance results under MR or PR test paradigms while training and evaluating models across all primary emotion categories. Table \ref{tab:sota_comparison} details the comparative results. We outlined comparisons with the other SOTA models trained on MR$_{Train}$ datasets for the CREMA-D and IEMOCAP repositories and PR$_{Train}$ datasets for the IMPROV(P) and PODCAST(P). The model, trained on the AR$_{Train}$ set, surpasses all examined SOTA results on the IMPROV(P), CREMA-D, IEMOCAP, and PODCAST(P) databases. These findings substantiate that the SER approach is robust against the SOTA references.

Specifically, on the IMPROV(P) corpus, our method achieves a macro-F1 score of 0.562 with AR${Train}$, outstripping the result from Goncalves and Busso \cite{Goncalves_2022} which had a score of 0.539. Nevertheless, training with PR${Train}$ resulted in a lower macro-F1 score of 0.512 compared to 0.539. For the CREMA-D set, we report a result of 0.591 using the MR$_{Train}$ set and 0.585 with the AR$_{Train}$ scenario in a macro-F1 score, both excelling beyond the SOTA score of 0.574. Moreover, our approach similarly outperforms other SOTA methodologies \cite{Li_2019, Pepino_2021, Goncalves_2022} in both the IEMOCAP and PODCAST(P) datasets.

\begin{table*}[!t]
\fontsize{5.5}{9}\selectfont
\centering
\caption{Table illustrates the Kullback-Leibler divergence (KLD) when training and testing with each aggregation method under each label-learning strategy for each database. We highlight in bold the best performance for each condition. We denote $\ast$, $\dag$, and $\star$ when a model has significantly better performance than a model training with $MR_{Train}$, $PR_{Train}$, and $AR_{Train}$, respectively.}
\begin{tabular}{@{}l@{\hspace{0.1cm}}c@{\hspace{0.1cm}}|l@{\hspace{0.1cm}}l@{\hspace{0.1cm}}l@{\hspace{0.1cm}}l@{\hspace{0.1cm}}l@{\hspace{0.1cm}}|l@{\hspace{0.1cm}}l@{\hspace{0.1cm}}l@{\hspace{0.1cm}}l@{\hspace{0.1cm}}l@{\hspace{0.1cm}}|l@{\hspace{0.1cm}}l@{\hspace{0.1cm}}l@{\hspace{0.1cm}}l@{\hspace{0.1cm}}l@{\hspace{0.1cm}}}
\toprule
                            & Aggregation         & \multicolumn{5}{c|}{Hard-label learning} & \multicolumn{5}{c|}{Soft-label learning} & \multicolumn{5}{c}{Distributional-label learning} \\
Database                    & (training)          & MR & PR & AR & PR - MR    &   AR - PR & MR & PR & AR & PR - MR    &   AR - PR    & MR & PR & AR & PR - MR    &   AR - PR \\ \midrule
\multirow{3}{*}{IMPROV(P)}  &$MR_{Train}$&0.235$\dag$&0.231$\dag$&0.229$\dag$&0.143&0.190&0.175&0.172&0.171&\textbf{0.108}&\textbf{0.150}&\textbf{0.232}&\textbf{0.233}&\textbf{0.236}&\textbf{0.271}&0.285\\
&$PR_{Train}$&0.256&0.252&0.249&\textbf{0.138}&0.193&\textbf{0.172}&\textbf{0.169}&\textbf{0.169}&0.117&0.162&0.240&0.242&0.244&0.289&0.286\\
&$AR_{Train}$&\textbf{0.211$\ast\dag$}&\textbf{0.208$\ast\dag$}&\textbf{0.206$\ast\dag$}&0.139&\textbf{0.180}&0.182&0.180&0.178&0.115&0.157&0.237&0.238&0.241&0.277&\textbf{0.283}\\
\hline
\multirow{3}{*}{CREMA-D}  &$MR_{Train}$&0.078&0.091&0.094&0.122&0.129&\textbf{0.055}&0.058&0.059&0.066&0.066&0.112&0.136&0.142&0.192&0.203\\
&$PR_{Train}$&\textbf{0.064$\ast$}&\textbf{0.073$\ast$}&\textbf{0.075$\ast$}&0.094$\ast$&0.099$\ast$&0.058&0.057&0.057&0.056$\ast$&0.059$\ast$&0.109&0.131&0.136&0.185&0.188$\ast$\\
&$AR_{Train}$&0.068$\ast$&0.075$\ast$&0.076$\ast$&\textbf{0.092$\ast$}&\textbf{0.095$\ast$}&0.058&\textbf{0.056}&\textbf{0.056$\ast$}&\textbf{0.053$\ast$}&\textbf{0.055$\ast$}&\textbf{0.103$\ast$}&\textbf{0.122$\ast\dag$}&\textbf{0.126$\ast\dag$}&\textbf{0.166$\ast\dag$}&\textbf{0.178$\ast$}\\
\hline
\multirow{3}{*}{PODCAST (P)}  &$MR_{Train}$&\textbf{0.150}&0.142&0.141&0.128&0.136&\textbf{0.157}&\textbf{0.142}&\textbf{0.139}&0.114&0.125&0.203&0.219&0.223&0.248&\textbf{0.242}\\
&$PR_{Train}$&0.150&\textbf{0.137}&\textbf{0.135}&\textbf{0.112$\ast\star$}&\textbf{0.125$\ast\star$}&0.164&0.144&0.140&\textbf{0.110}&\textbf{0.122}&0.203&0.218&0.222&0.244&0.243\\
&$AR_{Train}$&0.170&0.153&0.150&0.123&0.134&0.172&0.151&0.146&0.113&0.125&\textbf{0.200}&\textbf{0.214}&\textbf{0.220}&\textbf{0.240}&0.245\\
\hline
\multirow{3}{*}{IEMOCAP}  &$MR_{Train}$&0.203&0.201&0.202&0.180&0.205&0.162&0.160&0.156&0.128&0.147&0.211&0.212&0.221&0.223&0.245\\
&$PR_{Train}$&0.202&0.201&0.201&0.186&0.201&0.150$\ast$&0.148$\ast$&0.145$\ast$&0.125&0.135$\ast$&\textbf{0.204}&\textbf{0.206}&0.214&0.225&0.237\\
&$AR_{Train}$&\textbf{0.185}&\textbf{0.183$\ast\dag$}&\textbf{0.182$\ast\dag$}&\textbf{0.159$\ast\dag$}&\textbf{0.178$\ast\dag$}&\textbf{0.143$\ast$}&\textbf{0.141$\ast$}&\textbf{0.138$\ast$}&\textbf{0.120}&\textbf{0.130$\ast$}&0.208&0.209&\textbf{0.214}&\textbf{0.217}&\textbf{0.227$\ast$}\\
\hline
\multirow{3}{*}{IMPROV (S)}  &$MR_{Train}$&0.118&0.128&0.130&0.139&0.147&\textbf{0.090$\dag\star$}&0.088&0.089&0.087&0.094&\textbf{0.116}&0.158&0.163&0.205&0.196\\
&$PR_{Train}$&\textbf{0.103$\ast\star$}&\textbf{0.103$\ast$}&\textbf{0.104$\ast$}&0.102$\ast$&\textbf{0.109$\ast$}&0.103&0.089&0.088&0.074$\ast$&0.082$\ast$&0.119&0.150&0.155&0.184$\ast$&0.190\\
&$AR_{Train}$&0.116&0.108$\ast$&0.108$\ast$&\textbf{0.099$\ast$}&0.110$\ast$&0.100&\textbf{0.087}&\textbf{0.086}&\textbf{0.072$\ast$}&\textbf{0.080$\ast$}&0.120&\textbf{0.146$\ast$}&\textbf{0.150$\ast$}&\textbf{0.174$\ast$}&\textbf{0.180$\ast$}\\
\hline
\multirow{3}{*}{PODCAST (S)}  &$MR_{Train}$&0.085&0.098&0.102&0.100&0.113&\textbf{0.074$\dag\star$}&0.071&0.073&0.071&0.078&0.079&0.144&0.151&0.153&0.171\\
&$PR_{Train}$&\textbf{0.075}&\textbf{0.062$\ast$}&\textbf{0.063$\ast$}&\textbf{0.060$\ast$}&0.067$\ast$&0.091&\textbf{0.060$\ast$}&\textbf{0.060$\ast$}&\textbf{0.056$\ast$}&\textbf{0.058$\ast$}&\textbf{0.067}&0.123$\ast$&0.132$\ast$&0.131$\ast$&0.156$\ast$\\
&$AR_{Train}$&0.084&0.064$\ast$&0.064$\ast$&0.061$\ast$&\textbf{0.065$\ast$}&0.097&0.062$\ast$&0.061$\ast$&0.057$\ast$&0.058$\ast$&0.073&\textbf{0.117$\ast$}&\textbf{0.126$\ast$}&\textbf{0.124$\ast$}&\textbf{0.149$\ast$}\\
 \bottomrule
\end{tabular}
\label{tab:kld}
\end{table*}

\begin{table*}[!b]
\fontsize{5.5}{9}\selectfont
\centering
\caption{The table presents the macro-F1 scores achieved when models are trained and tested using each aggregation method across various label-learning strategies for each database. The highest performance for each condition is highlighted in bold. Symbols such as $\ast$, $\dag$, and $\star$ are used to indicate when a model significantly surpasses the performance of those trained with MR$_{Train}$, PR$_{Train}$, and AR$_{Train}$, respectively.}
\begin{tabular}{@{}l@{\hspace{0.1cm}}c@{\hspace{0.1cm}}|l@{\hspace{0.1cm}}l@{\hspace{0.1cm}}l@{\hspace{0.1cm}}l@{\hspace{0.1cm}}l@{\hspace{0.1cm}}|l@{\hspace{0.1cm}}l@{\hspace{0.1cm}}l@{\hspace{0.1cm}}l@{\hspace{0.1cm}}l@{\hspace{0.1cm}}|l@{\hspace{0.1cm}}l@{\hspace{0.1cm}}l@{\hspace{0.1cm}}l@{\hspace{0.1cm}}l@{\hspace{0.1cm}}}
\toprule
                            & Aggregation         & \multicolumn{5}{c|}{Hard-label learning} & \multicolumn{5}{c|}{Soft-label learning} & \multicolumn{5}{c}{Distributional-label learning} \\
Database                    & (train/test set)          & MR & PR & AR & PR - MR    &   AR - PR & MR & PR & AR & PR - MR    &   AR - PR    & MR & PR & AR & PR - MR    &   AR - PR \\ \midrule
\multirow{3}{*}{IMPROV(P)}  &MR$_{Train}$&0.512$\dag$&0.507$\dag$&0.555$\dag$&0.300&\textbf{0.516$\dag$}&0.595&0.587&0.613&\textbf{0.346}&0.530&\textbf{0.612}&\textbf{0.604}&0.599&\textbf{0.401}&0.440\\
&PR$_{Train}$&0.450&0.448&0.513&0.305&0.465&\textbf{0.600}&\textbf{0.593}&\textbf{0.623}&0.341&\textbf{0.531}&0.601&0.596&0.590&0.359&0.436\\
&AR$_{Train}$&\textbf{0.562$\ast\dag$}&\textbf{0.555$\ast\dag$}&\textbf{0.593$\ast\dag$}&\textbf{0.335}&0.498&0.576&0.569&0.602&0.339&0.518&0.602&0.594&\textbf{0.600}&0.340&\textbf{0.441}\\
\hline
\multirow{3}{*}{CREMA-D}  &MR$_{Train}$&0.591&0.532&0.551&0.381&0.500&0.640&0.575&0.671&0.409&0.651&0.518$\star$&\textbf{0.474$\star$}&0.411&\textbf{0.357}&0.368\\
&PR$_{Train}$&\textbf{0.600}&\textbf{0.545}&0.595$\ast$&\textbf{0.390}&0.572$\ast$&0.667&0.594&0.699$\ast$&0.416&0.688&\textbf{0.518$\star$}&0.473$\star$&\textbf{0.419}&0.357&\textbf{0.374}\\
&AR$_{Train}$&0.585&0.528&\textbf{0.607$\ast$}&0.386&\textbf{0.593$\ast$}&\textbf{0.673}&\textbf{0.615$\ast$}&\textbf{0.710$\ast$}&\textbf{0.444}&\textbf{0.706$\ast$}&0.486&0.442&0.414&0.340&0.370\\
\hline
\multirow{3}{*}{PODCAST (P)}  &MR$_{Train}$&0.214$\star$&0.184$\star$&0.303&0.143&0.300&0.215&0.185&0.326&0.145&0.328&0.161&0.137&0.162&0.102&0.159\\
&PR$_{Train}$&\textbf{0.259$\ast\star$}&\textbf{0.232$\ast\star$}&\textbf{0.403$\ast\star$}&\textbf{0.187$\ast\star$}&\textbf{0.420$\ast\star$}&\textbf{0.241$\star$}&\textbf{0.207$\ast\star$}&\textbf{0.397$\ast\star$}&\textbf{0.160$\star$}&\textbf{0.408$\ast\star$}&0.195$\ast$&0.166$\ast$&0.192$\ast$&0.126$\ast$&0.184$\ast$\\
&AR$_{Train}$&0.192&0.166&0.330$\ast$&0.129&0.351$\ast$&0.199&0.174&0.355$\ast$&0.138&0.367$\ast$&\textbf{0.204$\ast$}&\textbf{0.175$\ast$}&\textbf{0.200$\ast$}&\textbf{0.139$\ast$}&\textbf{0.192$\ast$}\\
\hline
\multirow{3}{*}{IEMOCAP}  &MR$_{Train}$&0.269&0.260&0.339&0.203&0.351&0.346&0.343&0.412&0.257&0.426&0.354&0.341&0.299&0.253&0.287\\
&PR$_{Train}$&0.259&0.254&0.345&0.186&0.355&0.369&0.359&0.433&\textbf{0.279}&0.453&\textbf{0.377}&0.361&0.320&0.253&0.306\\
&AR$_{Train}$&\textbf{0.279}&\textbf{0.268}&\textbf{0.365}&\textbf{0.238$\dag$}&\textbf{0.378}&\textbf{0.390$\ast$}&\textbf{0.383$\ast$}&\textbf{0.464$\ast\dag$}&0.266&\textbf{0.479$\ast$}&0.369&\textbf{0.361}&\textbf{0.325$\ast$}&\textbf{0.265}&\textbf{0.317}\\
\hline
\multirow{3}{*}{IMPROV (S)}  &MR$_{Train}$&0.424&0.254&0.229&0.234&0.245&\textbf{0.451}&0.299&0.379&0.278&0.386&0.361&0.185&0.137&0.149&0.150\\
&PR$_{Train}$&\textbf{0.455$\star$}&\textbf{0.340$\ast$}&0.328$\ast$&\textbf{0.318$\ast$}&0.360$\ast$&0.433&0.353$\ast$&0.483$\ast$&0.342$\ast$&0.505$\ast$&0.397&0.248$\ast$&0.181$\ast$&0.219$\ast$&0.189$\ast$\\
&AR$_{Train}$&0.391&0.315$\ast$&\textbf{0.337$\ast$}&0.311$\ast$&\textbf{0.365$\ast$}&0.410&\textbf{0.360$\ast$}&\textbf{0.491$\ast$}&\textbf{0.343$\ast$}&\textbf{0.522$\ast$}&\textbf{0.431$\ast$}&\textbf{0.306$\ast\dag$}&\textbf{0.216$\ast\dag$}&\textbf{0.282$\ast\dag$}&\textbf{0.227$\ast\dag$}\\
\hline
\multirow{3}{*}{PODCAST (S)}  &MR$_{Train}$&0.344&0.078&0.138&0.076&0.141&\textbf{0.389$\star$}&0.080&0.199&0.076&0.198&0.352&0.051&0.060&0.047&0.059\\
&PR$_{Train}$&\textbf{0.392$\star$}&0.113$\ast$&0.327$\ast$&0.111$\ast$&0.328$\ast$&0.321&0.122$\ast$&0.450$\ast$&0.122$\ast$&0.457$\ast$&0.412&0.076$\ast$&0.078$\ast$&0.072$\ast$&0.074$\ast$\\
&AR$_{Train}$&0.283&\textbf{0.125$\ast$}&\textbf{0.352$\ast\dag$}&\textbf{0.124$\ast\dag$}&\textbf{0.357$\ast\dag$}&0.237&\textbf{0.139$\ast$}&\textbf{0.457$\ast$}&\textbf{0.142$\ast$}&\textbf{0.466$\ast$}&\textbf{0.425}&\textbf{0.078$\ast$}&\textbf{0.091$\ast\dag$}&\textbf{0.075$\ast$}&\textbf{0.088$\ast\dag$}\\
 \bottomrule
\end{tabular}
\label{tab:cv}
\end{table*}

\begin{table}[!t]
\fontsize{7}{9}\selectfont
\centering
\caption{The table shows averaged macro-F1 scores across 18 experiments listed in Table \ref{tab:cv} and Table~\ref{tab:kld} with different databases and label-learning strategies on the different evaluation sets generated by three rules, \emph{majority rule} (MR), \emph{plurality rule} (PR), and \emph{all-inclusive rule} (AR). The $\downarrow$ means the lower values mean the higher performance; the $\uparrow$ is the opposite.}
\begin{tabular}{@{}c|ccccc|ccccc@{}}
\toprule
\textbf{Metric}                    & \multicolumn{5}{c|}{\textbf{KLD $\downarrow$}}                                          & \multicolumn{5}{c}{\textbf{Macro-F1 Score $\uparrow$}}                               \\ \midrule
\textbf{Train\textbackslash{}Test} & \textbf{MR} & \textbf{PR} & \textbf{AR} & \textbf{PR-MR} & \textbf{AR-PR} & \textbf{MR} & \textbf{PR} & \textbf{AR} & \textbf{PR-MR} & \textbf{AR-PR} \\ \midrule
\textbf{MR$_{Train}$}                   & 0.1408      & 0.1491      & 0.1512      & 0.1488         & 0.1623         & 0.4082      & 0.3153      & 0.3546      & 0.2309         & 0.3353         \\
\textbf{PR$_{Train}$}                   & 0.1406      & 0.1425      & 0.1433      & 0.1382         & 0.1507         & \textbf{0.4192}      & 0.3378      & 0.4098      & 0.2524         & 0.3947         \\
\textbf{AR$_{Train}$}                   & \textbf{0.1404}      & \textbf{0.1397}      & \textbf{0.1402}      & \textbf{0.1334}         & \textbf{0.1461}         & 0.4052      & \textbf{0.3418}      & \textbf{0.4172}      & \textbf{0.2576}         & \textbf{0.4019}         \\ \bottomrule
\end{tabular}
\label{tab:kld_maf1}
\end{table}

\subsection{Assessment with Full and Partial Test Data}
\label{ssec:IncompleteData}
Table~\ref{tab:cv} and Table~\ref{tab:kld} present the macro-F1 scores and KLD values for the different combinations of aggregation methods and label-learning strategies, respectively. These scores are derived from 18 experiments using various databases, where models were trained with the $MR_{Train}$, $PR_{Train}$, or $AR_{Train}$ sets (6 databases × 3 learning strategies). Figures \ref{fig:avg_barplot} and \ref{fig:avg_barplot_ambiguous} illustrate the average performance for each evaluation set (MR, PR, AR, PR-MR, or AR-PR). A small-sample test of the hypothesis (matched pairs) was conducted on these results. Table~\ref{tab:kld_maf1} summarizes the overall averaged results from 18 experiments, and the bold numbers mean the better performance on each test set. Based on the KLD metric, the SER system trained with the training set selected by the proposed AR can perform overall better than those trained with MR$_{Train}$ and PR$_{Train}$. 

{Our first research question investigates: \textbf{How is the performance of SER systems influenced by different aggregation methods used in the training set?} We address this question by evaluating the models trained under varying conditions on complete and incomplete test data in a cross-corpus setting.

\subsubsection{Assessment on the Complete Test Set (AR)}
When evaluating using the AR approach with all annotated data in the test set, Figure \ref{fig:ar_set} demonstrates that the macro-F1 score obtained with the $AR_{Train}$ set is significantly higher than those with the $MR_{Train}$ and $PR_{Train}$ sets across the 18 conditions. Remarkably, models trained with the $AR_{Train}$ set achieved the highest macro-F1 scores in 14 of 18 experiments, as detailed in Table \ref{tab:cv}. These results indicate that training with the AR approach can substantially enhance performance compared to the MR or PR criteria. This underscores the value of incorporating more annotated samples during the training process of SER tasks.

\begin{figure}[!h]
     \centering
     \begin{subfigure}[b]{0.4\textwidth}
         \centering
         \includegraphics[width=\textwidth]{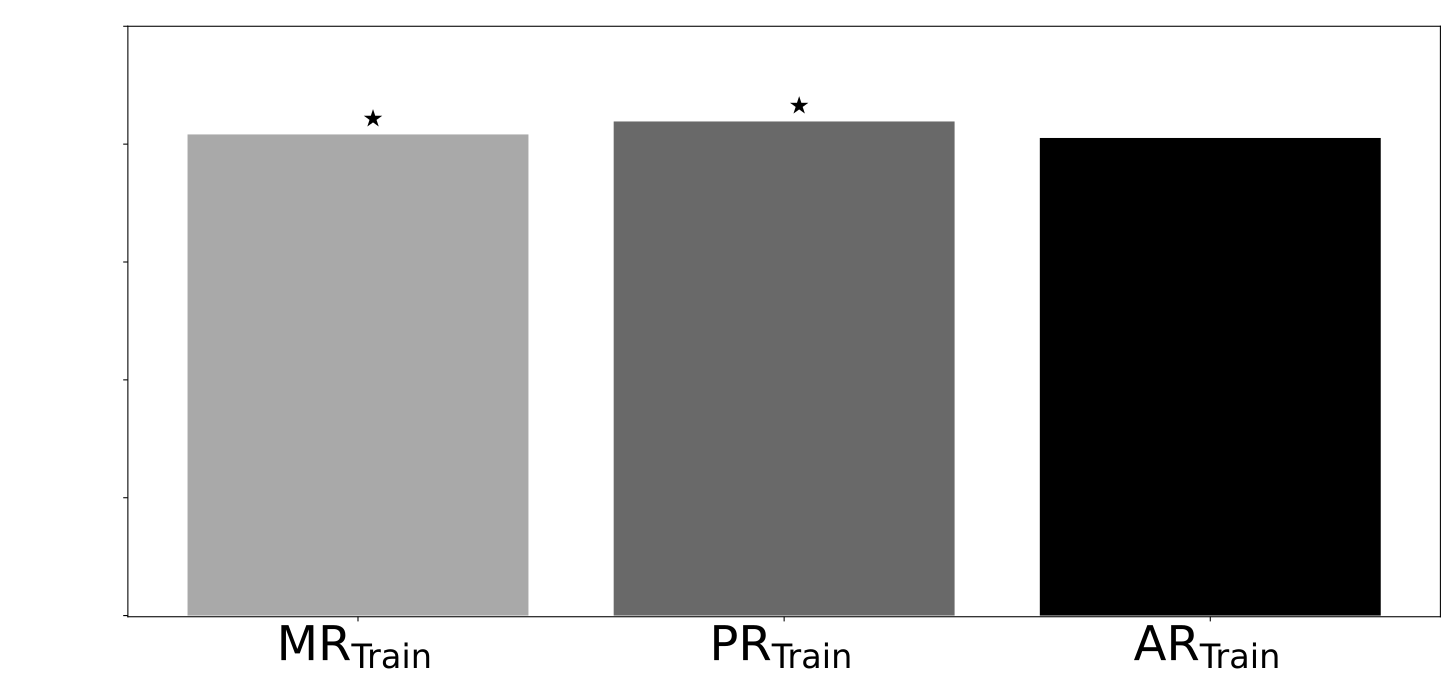}
         \caption{Macro-F1 scores on MR set.}
         \label{fig:mr_set}
     \end{subfigure}
     \vfill
     \begin{subfigure}[b]{0.4\textwidth}
         \centering
         \includegraphics[width=\textwidth]{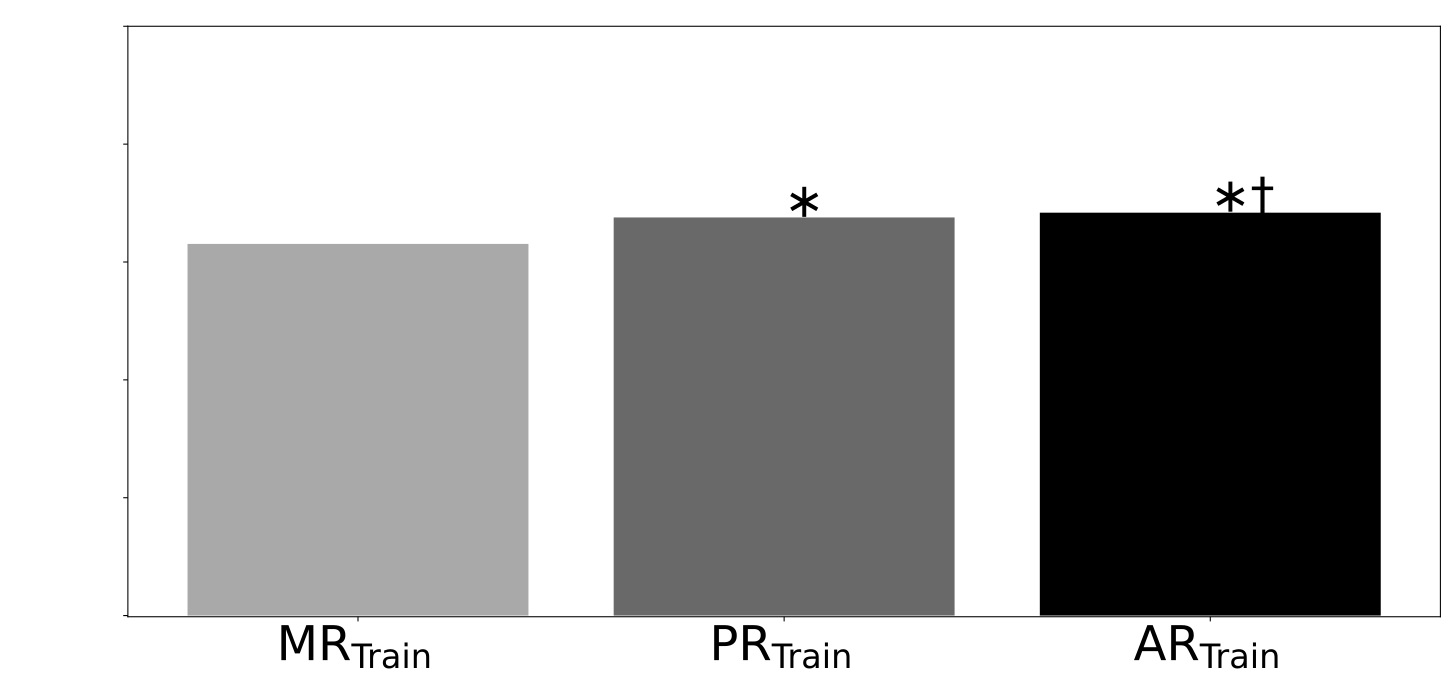}
         \caption{Macro-F1 scores on PR set.}
         \label{fig:pr_set}
     \end{subfigure}
     \vfill
     \begin{subfigure}[b]{0.4\textwidth}
         \centering
         \includegraphics[width=\textwidth]{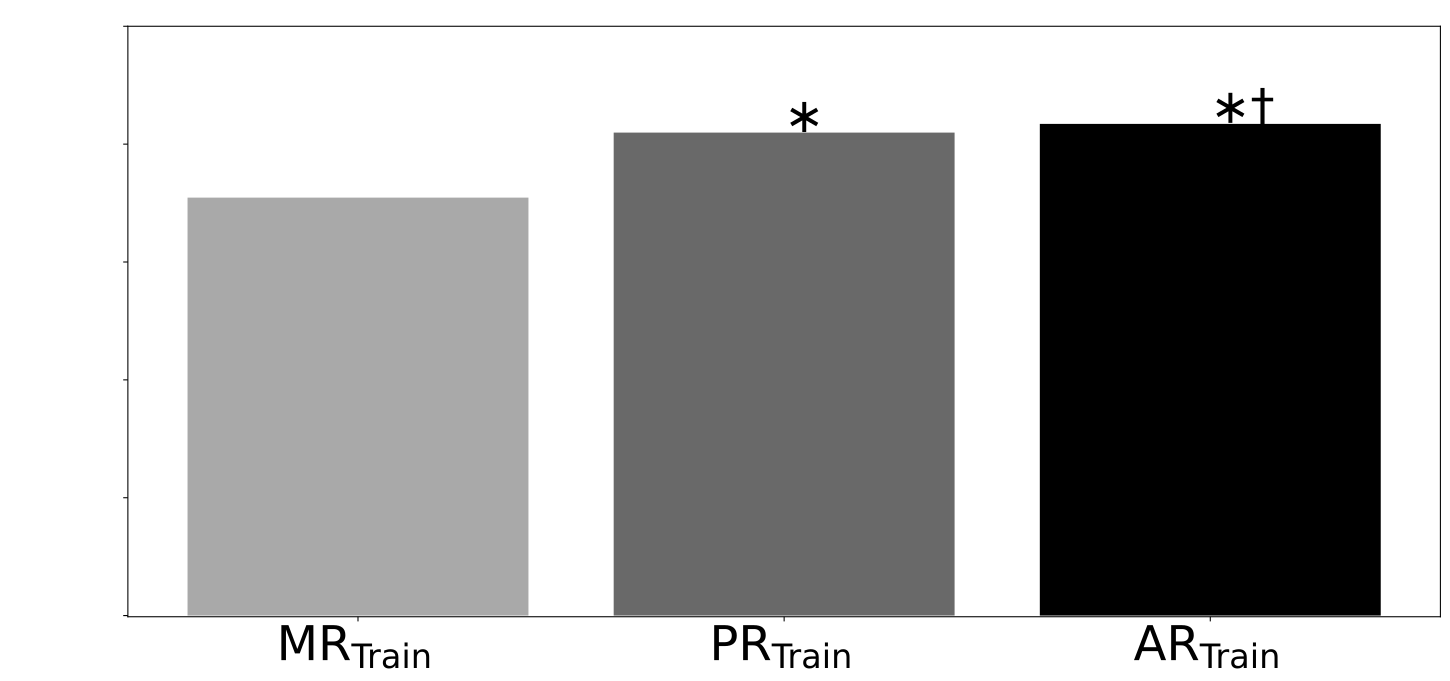}
         \caption{Macro-F1 scores on AR set.}
         \label{fig:ar_set}
     \end{subfigure}
        \caption{Averaged macro-F1 scores across 18 experiments (as shown in Table \ref{tab:cv}) involving different databases and label-learning strategies on various evaluation sets generated by the three rules: \emph{majority rule} (MR), \emph{plurality rule} (PR), and \emph{all-inclusive rule} (AR). The notations $\ast$, $\dag$, and $\star$ are used to indicate when a model achieves significantly better performance than models trained with $MR_{Train}$, $PR_{Train}$, and $AR_{Train}$, respectively.}
        \label{fig:avg_barplot}
\end{figure}

\subsubsection{Assessment of the Incomplete Test Sets (MR \& PR)}

The single-label SER performance was assessed under the MR and PR test conditions. As detailed in Table \ref{tab:cv}, testing with the PR set consistently yielded lower performance than testing with the MR set, as the MR set omits more ambiguous samples. Similarly, performance in the PR-MR condition was generally worse than in the PR conditions. These results highlight that including more ambiguous samples (lacking majority consensus) in the test set—common in practical scenarios—can degrade SER model performance. Therefore, using only PR or MR to define the test set might not accurately reflect the outcomes likely to be encountered in real-world deployments where recognition of every sentence is necessary.

Out of the 18 conditions analyzed, Table \ref{tab:cv} indicates that when tested with the PR set, training with the $AR_{Train}$ configuration resulted in the best performance in 11 out of 18 cases (approximately 61\%), and 14 out of 18 cases (approximately 78\%) when tested with the AR set. Figure \ref{fig:pr_set} presents statistically significant evidence that the average macro-F1 scores for models trained with $AR_{Train}$ were superior to those trained with the $PR_{Train}$ set. These findings suggest that using the AR approach for training aggregation may enhance SER performance on samples with lower annotation agreement.

For tests evaluated on the MR condition, models trained with the $AR_{Train}$ set outperformed others in only 7 out of 18 experiments (approximately 39\%). Figure \ref{fig:mr_set} shows a decline in performance for the $AR_{Train}$ trained model compared to those trained with either the $MR_{Train}$ or $PR_{Train}$ sets.

Including more complex samples in the training set ($AR_{Train}$) appears to reduce accuracy for the most straightforward samples; however, this trade-off bolsters the model's resilience in real-world contexts where both ambiguous and unambiguous samples are commonly encountered. This suggests that training SER systems with the $AR_{Train}$ set could be more efficient for real-life applications, reflecting the genuine mix of sample types in practical environments. Additionally, increased training samples with the $PR_{Train}$ set displayed better performance than those trained with $MR_{Train}$, aligning with findings reported by Chou et al. \cite{Chou_2019}.

\subsubsection{Assessment in Cross-Corpus Scenarios}

The past experimental outcomes were obtained using within-corpus settings. Our interest is to investigate the impact of training models with various aggregation strategies in cross-corpus settings. We’ve opted to train the model with the PODCAST (P) corpus and test its performance using the IMPROV (P) corpus to showcase the advantages of the AR method in cross-corpus experiments. The IMPROV (P) corpus comprises anger, sadness, happiness, and neutral emotions. On the other hand, the PODCAST (P) corpus includes the additional emotions of surprise, fear, disgust, and contempt, along with the emotions present in IMPROV (P). Given the emotional overlap, we can perform this cross-corpus evaluation where a SER model trained with the PODCAST (P) corpus aims to predict emotions within the IMPROV (P) corpus. Our approach involves directly utilizing the models trained on the PODCAST (P) corpus and assessing their efficacy on the IMPROV (P) set. We specifically focus on predictions for anger, sadness, happiness, and the neutral state. We transform the distribution predictions into binary labels by applying a threshold and then compute the results using the macro F1 score. For instance, let’s consider a sample prediction for the IMPROV (P) dataset as: (angry, sad, happy, surprised, fear, disgusted, contempt, neutral) = (0.2, 0.2, 0.1, 0.1, 0.2, 0.1, 0.0, 0.1). We extract the four key predictions without renormalizing them: (angry, sad, happy, neutral) = (0.2, 0.2, 0.1, 0.1). The next stage involves using the threshold $1/C=$1/8 to convert predictions into a binary form: (1, 1, 0, 0). Assuming the ground truth of the sample is: (anger, sadness, happiness, neutral) = (0.4, 0.4, 0.1, 0.1). Given a threshold of 1/4 for the IMPROV (P) corpus, the binary conversion is (1, 1, 0, 0). In this case, the prediction accuracy equates to 100\%.

Table \ref{tab:cross-corpus} presents the macro-F1 scores for cross-corpus testing, generated using the MR, PR, AR, PR-MR, and AR-PR labels. The results indicate that using the AR$_{Train}$ set for training yields superior performance on the MR, PR, AR, and AR-PR test sets. This assessment also highlights the presented approach's efficacy for cross-corpus evaluations.

\begin{table}[!t]
\fontsize{7}{9}\selectfont
\centering
\caption{The table presents the cross-corpus macro-F1 scores for models trained using the 8-class MSP-PODCAST (P) dataset, applied to predict emotions in the 4-class IMPROV (P) dataset.}
\begin{tabular}{@{}l|lllll@{}}
\toprule
Test Set   & MR             & PR             & AR             & PR-MR          & AR-PR          \\ \midrule
MR$_{Train}$ & 0.445          & 0.441          & 0.520          & 0.271          & 0.506          \\
PR$_{Train}$ & 0.445          & 0.448          & 0.521          & \textbf{0.295} & 0.495          \\
AR$_{Train}$ & \textbf{0.458} & \textbf{0.459} & \textbf{0.523} & 0.276          & \textbf{0.520} \\ \bottomrule

\end{tabular}
\label{tab:cross-corpus}
\end{table}

\begin{figure}
     \centering
     \begin{subfigure}[b]{0.4\textwidth}
         \centering
         \includegraphics[width=\textwidth]{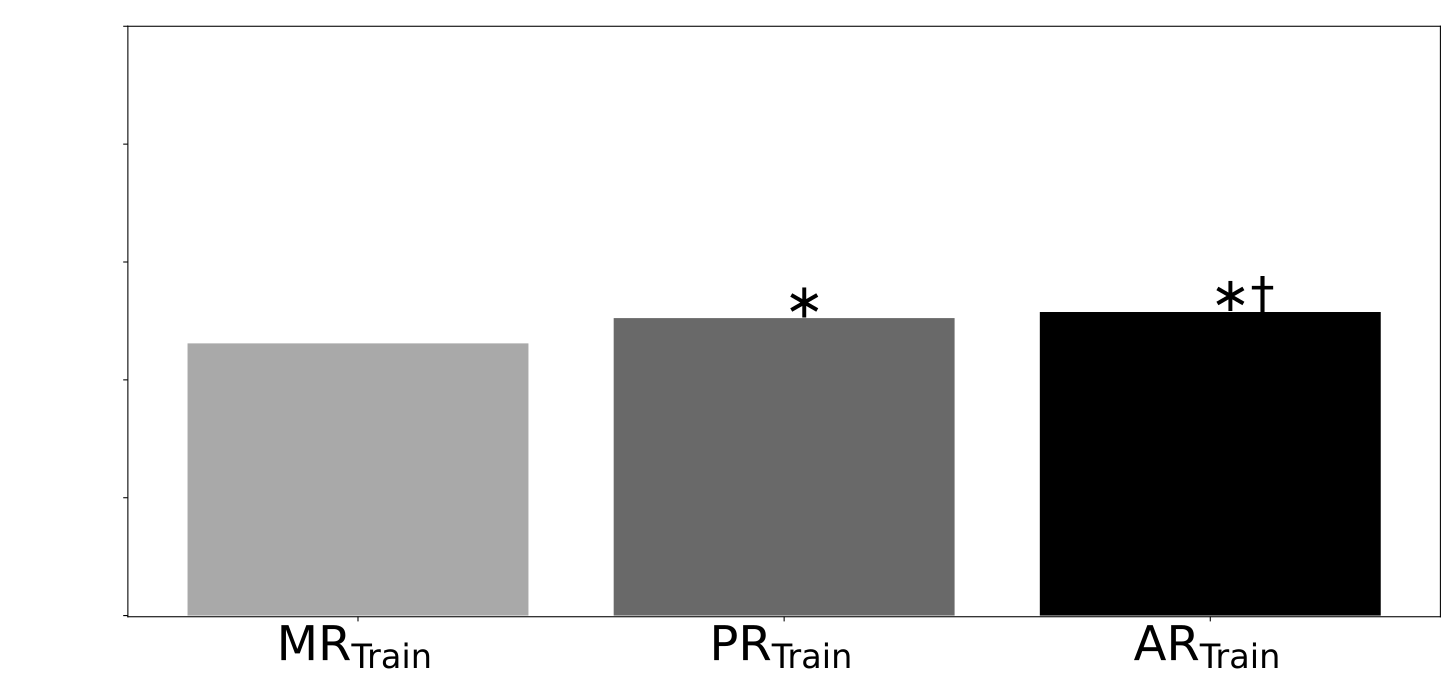}
         \caption{Macro-F1 scores on PR-MR set.}
         \label{fig:pr_mr_set}
     \end{subfigure}
     \vfill
     \begin{subfigure}[b]{0.4\textwidth}
         \centering
         \includegraphics[width=\textwidth]{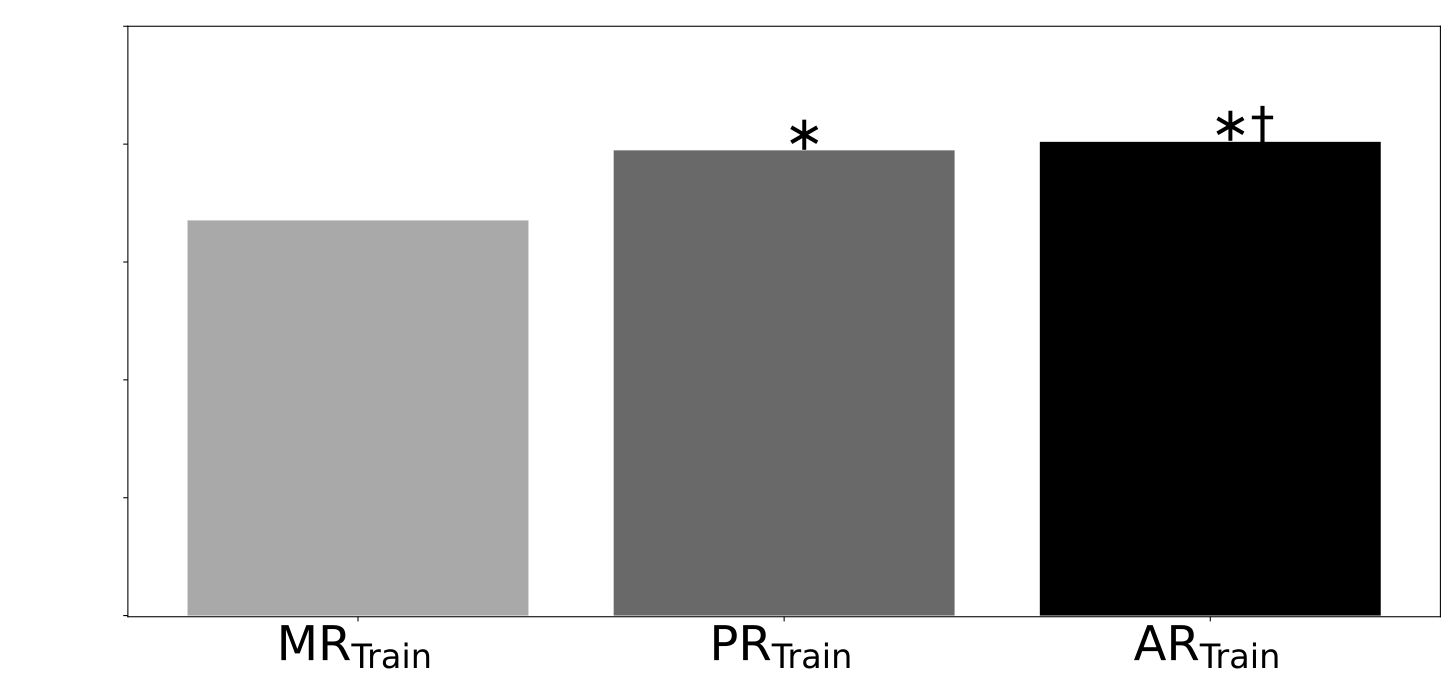}
         \caption{Macro-F1 scores on AR-PR set.}
         \label{fig:ar_pr_set}
     \end{subfigure}
        \caption{Averaged macro-F1 scores across 18 experiments (detailed in Table \ref{tab:cv}) with different databases and label-learning strategies on the varied evaluation sets generated by the PR-MR and AR-PR rules. The notations $\ast$, $\dag$, and $\star$ indicate significantly better performance compared to models trained using the $MR_{Train}$, $PR_{Train}$, and $AR_{Train}$ sets, respectively.}
        \label{fig:avg_barplot_ambiguous}
\end{figure}

\subsection{Evaluation on the Ambiguous Set}
\label{ssec:AmbiguousSet}

We address the second research question: \textbf{is the performance of an SER system on ambiguous emotions enhanced when trained with data derived from the all-inclusive rule as opposed to data obtained using the majority or plurality rules?}

\subsubsection{Performance on the $AR-PR$ Condition}

We examine the results under the $AR-PR$ test condition, which includes only the samples from the AR dataset that are not part of the PR dataset. As detailed in Table \ref{tab:cv}, models trained with the $MR_{Train}$ set did not yield the best performances in 17 out of 18 cases (approximately 94\%) when evaluating this condition. In these evaluations, models trained with the $PR_{Train}$ set demonstrated significantly better performance in 10 out of 18 experiments (approximately 56\%). Further, Figure \ref{fig:ar_pr_set} indicates that the model trained with the $MR_{Train}$ set produced the lowest classification performance in the averaged macro-F1 score. These findings suggest that including more ambiguous data in the training set (i.e., models trained with either the $PR_{Train}$ or $AR_{Train}$ sets) enhances performance on sentences embodying more unclear emotions. Consequently, we conclude that training SER models exclusively with the MR dataset is ineffective in predicting ambiguous emotions.

Additionally, Figure \ref{fig:avg_barplot_ambiguous} reveals that models trained with the $AR_{Train}$ set achieve significantly higher averaged macro-F1 scores compared to those taught with either the $MR_{Train}$ set or the $PR_{Train}$ set on both $AR-PR$ and $PR-MR$ test conditions. Therefore, we recommend employing the AR approach to select training data for SER tasks.

\begin{figure*}[!b]
    \centering
    \includegraphics[width=0.8\linewidth]{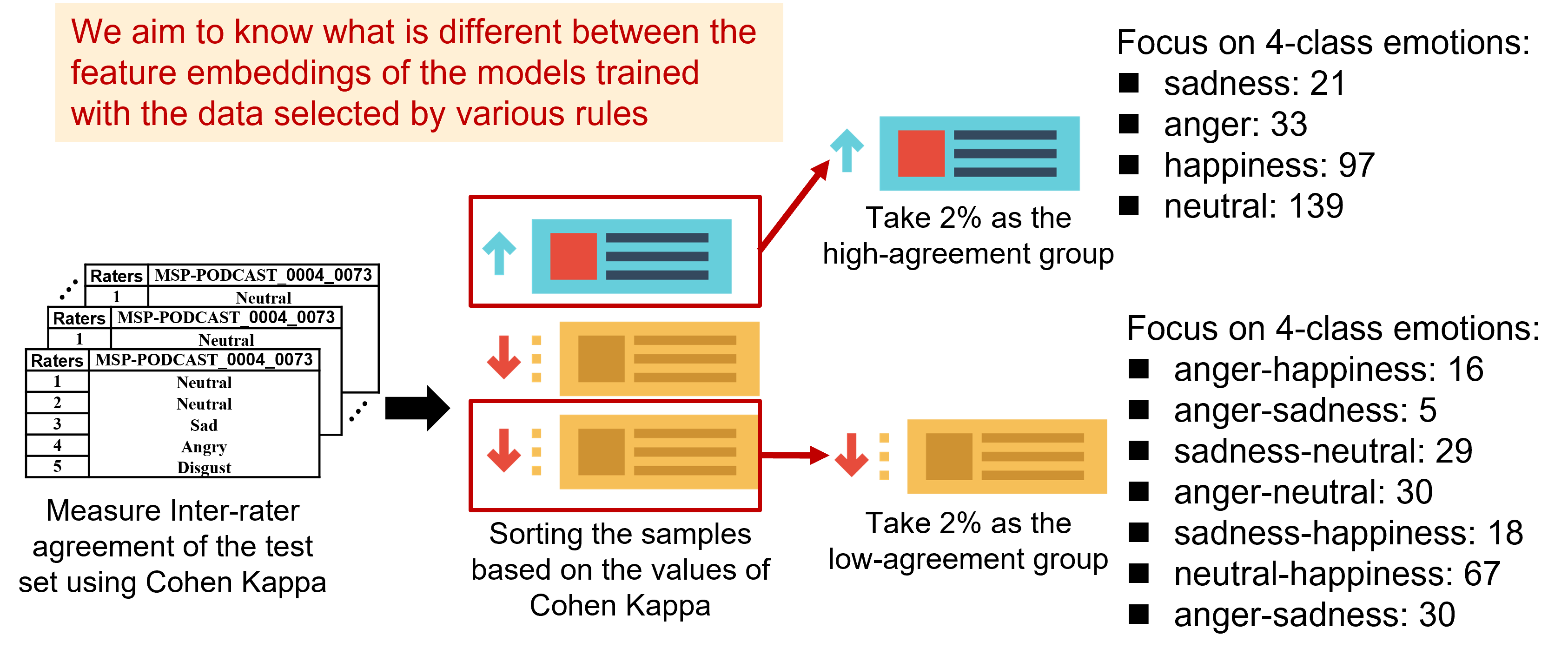}
    \caption{The figure depicts the procedure of visualizing feature embeddings.}
    \label{fig:embeddings_diagram}
\end{figure*}

\begin{figure}[!h]
  \centering
  \includegraphics[width=6.0cm]{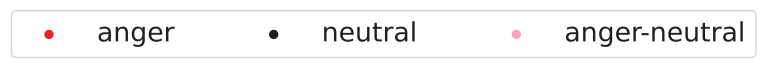}
  \begin{subfigure}{.5\columnwidth}
  \centering
    \includegraphics[width=4.0cm]{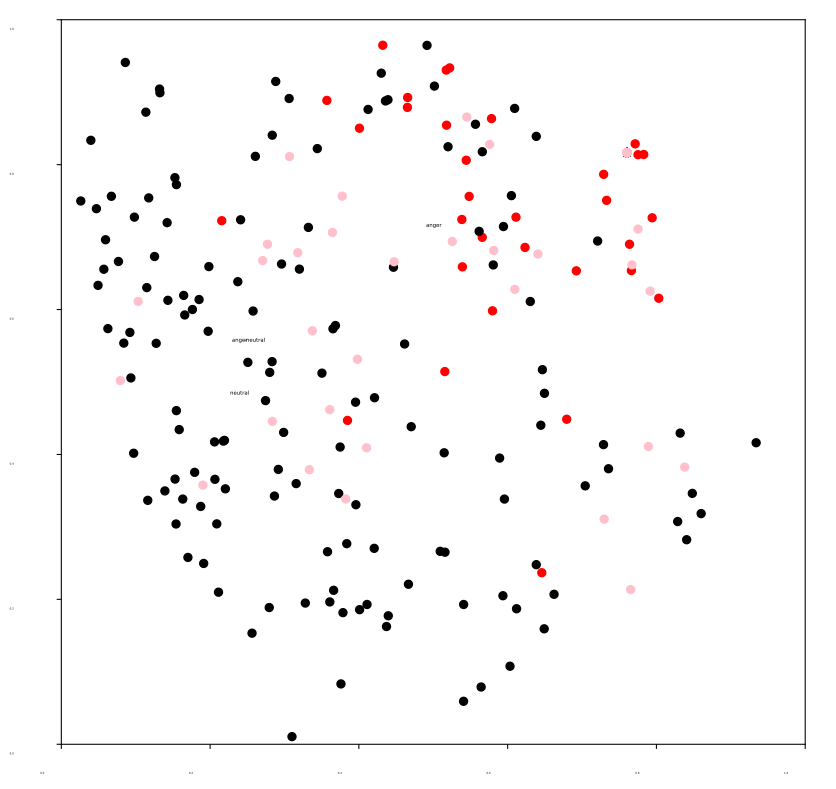}
    \caption{anger-neutral,AR$_{Train}$}
  \end{subfigure}%
  \begin{subfigure}{.5\columnwidth}
  \centering
    \includegraphics[width=4.0cm]{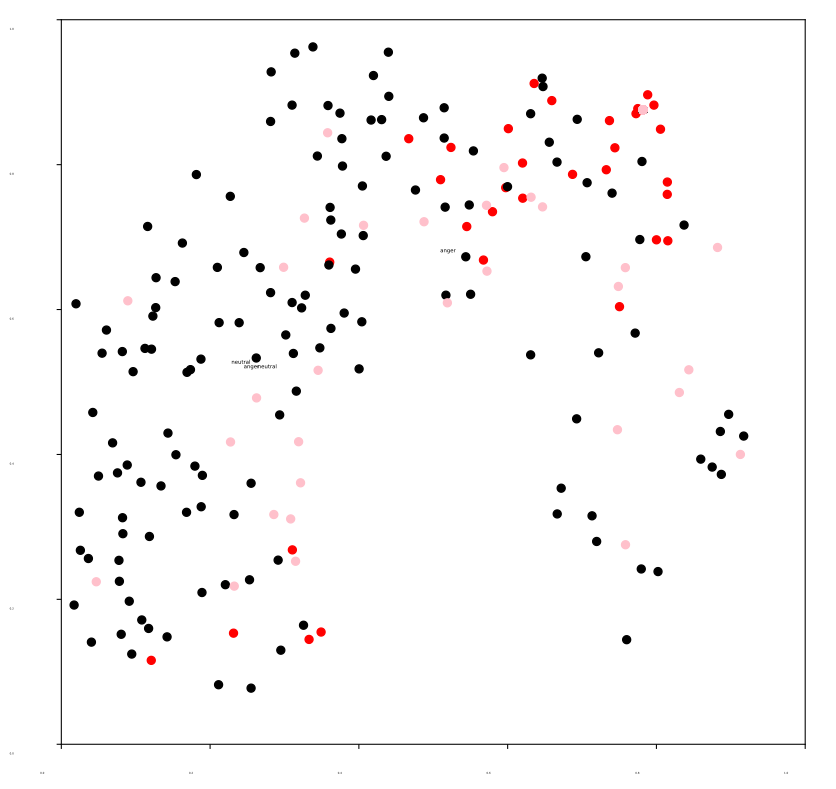}
    \caption{anger-neutral,MR$_{Train}$}
  \end{subfigure}

  \includegraphics[width=6.0cm]{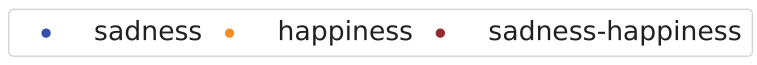}
  \begin{subfigure}{.5\columnwidth}
  \centering
    \includegraphics[width=4.0cm]{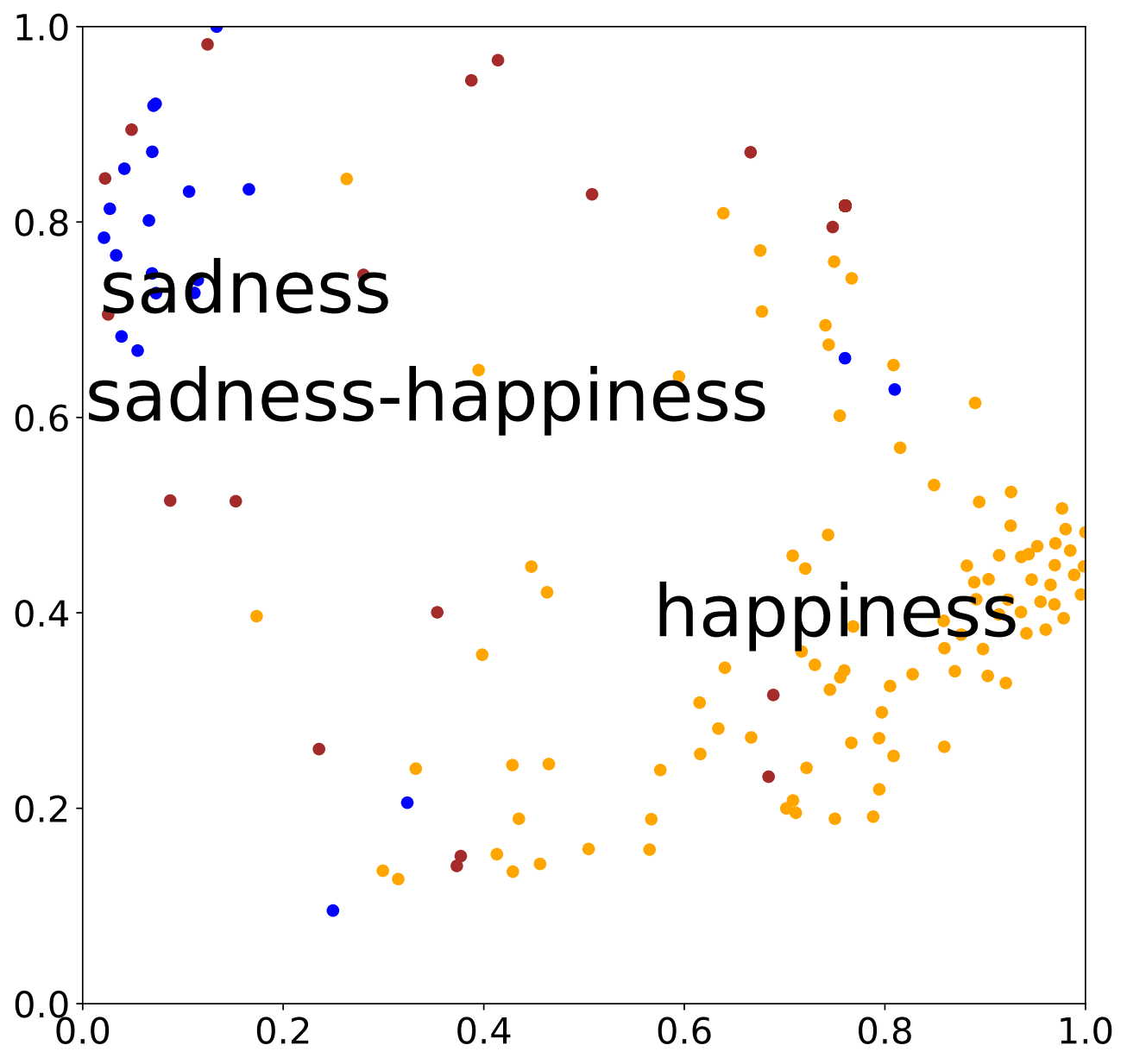}
    \caption{sadness-happiness,AR$_{Train}$}
  \end{subfigure}%
  \begin{subfigure}{.5\columnwidth}
  \centering
    \includegraphics[width=4.0cm]{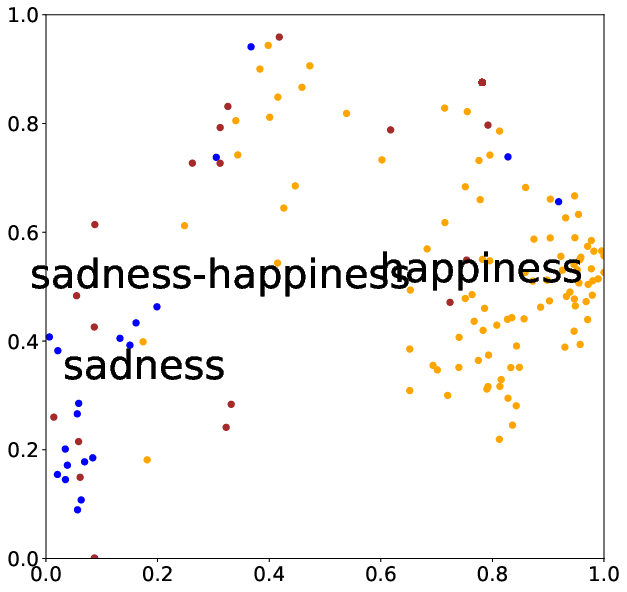}
    \caption{sadness-happiness,MR$_{Train}$}
  \end{subfigure}

    \includegraphics[width=6.0cm]{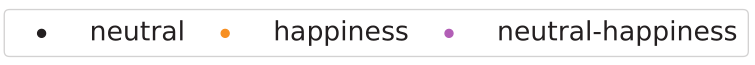}
  \begin{subfigure}{.5\columnwidth}
  \centering
    \includegraphics[width=4.0cm]{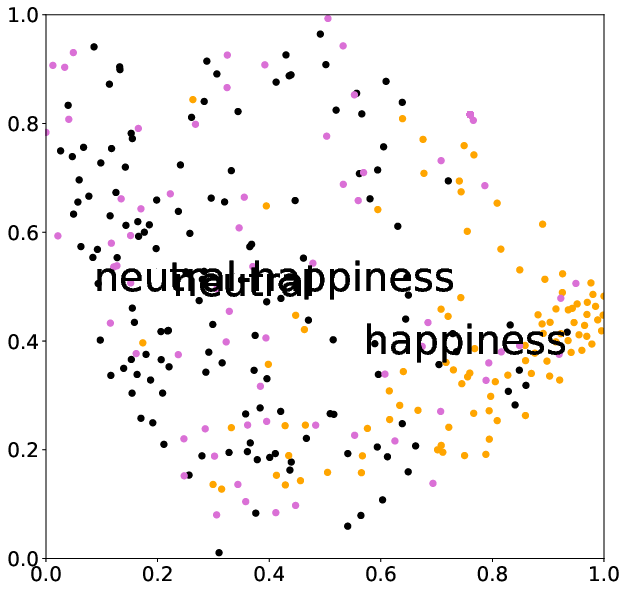}
    \caption{neutral-happiness,AR$_{Train}$}
  \end{subfigure}%
  \begin{subfigure}{.5\columnwidth}
  \centering
    \includegraphics[width=4.0cm]{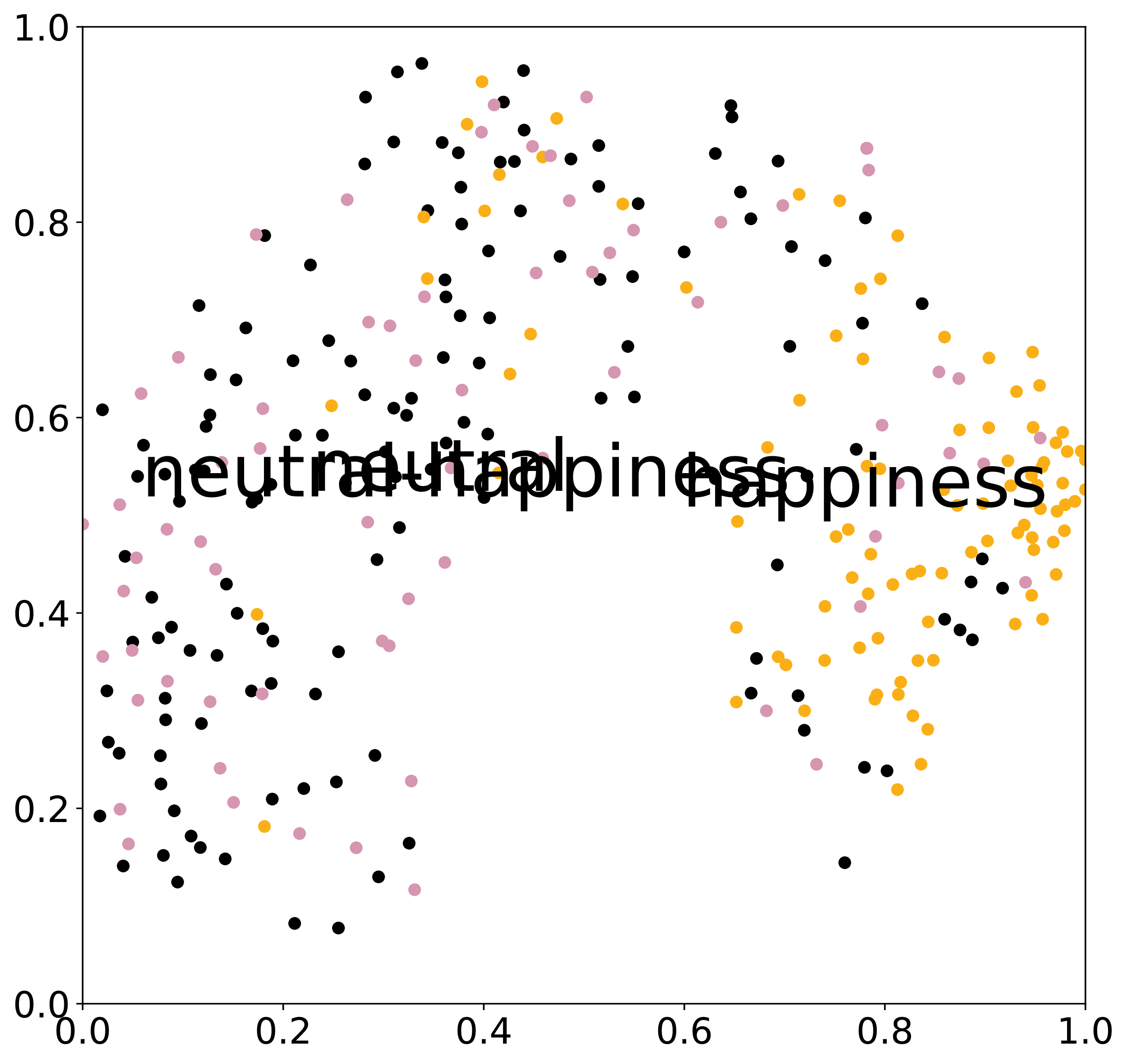}
    \caption{neutral-happiness,MR$_{Train}$}
  \end{subfigure}

  \includegraphics[width=6.0cm]{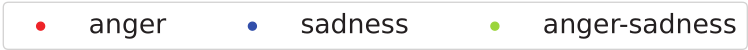}
  \begin{subfigure}{.5\columnwidth}
  \centering
    \includegraphics[width=4.0cm]{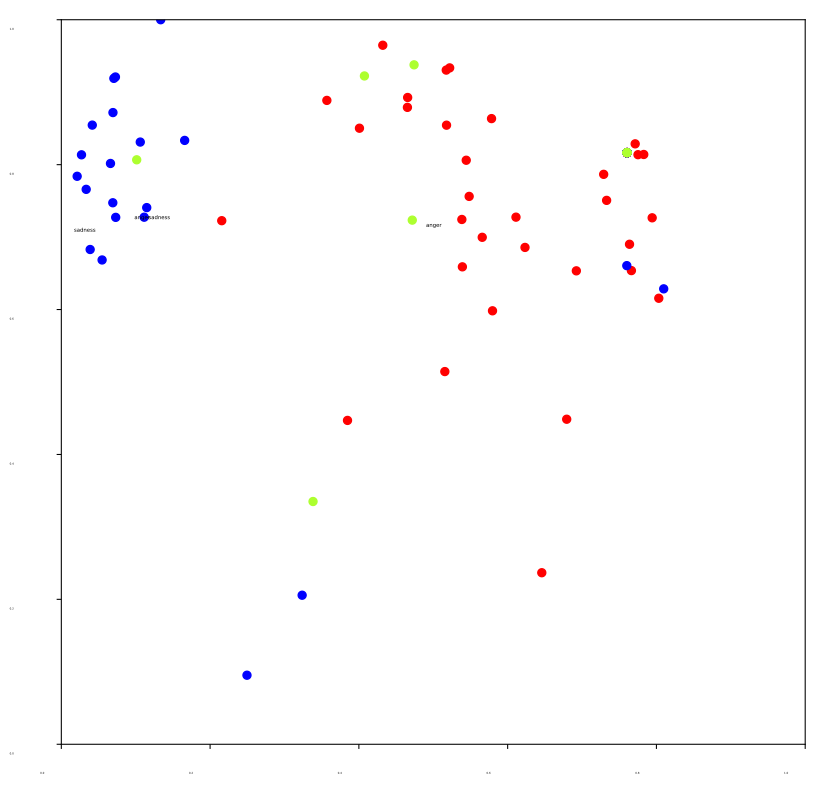}
    \caption{anger-sadness,AR$_{Train}$}
  \end{subfigure}%
  \begin{subfigure}{.5\columnwidth}
  \centering
    \includegraphics[width=4.0cm]{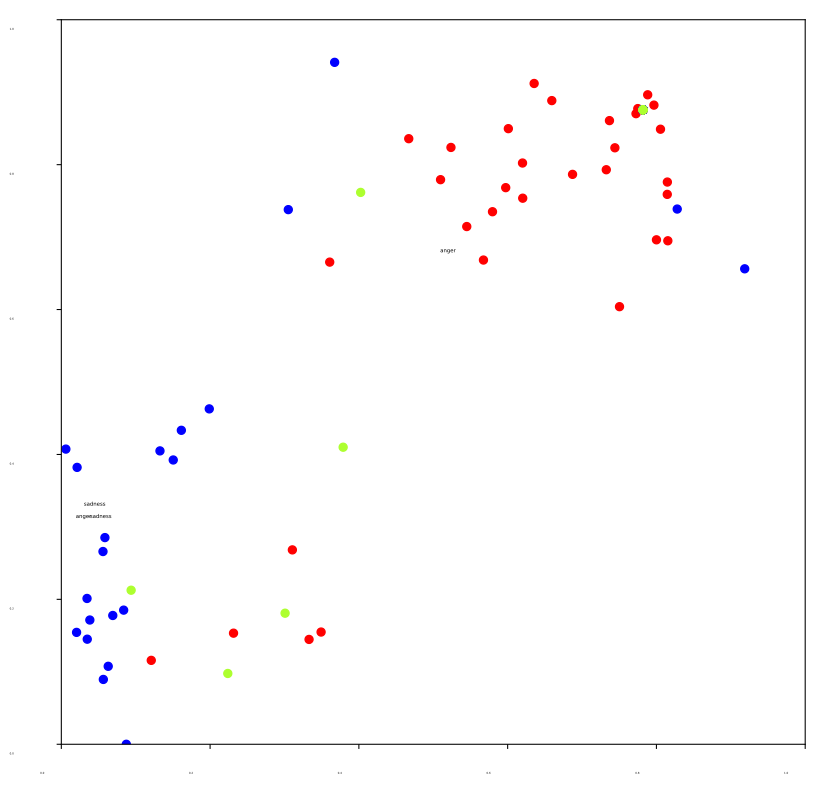}
    \caption{anger-sadness,MR$_{Train}$}
  \end{subfigure}
  
  \caption{T-SNE visualizations using embeddings generated from models trained with the MR${Train}$ and AR${Train}$ sets show the distribution of feature embeddings. This analysis includes the following emotion pairs: anger-neutral, sadness-happiness, neutral-happiness, and anger-sadness.}
  \label{fig:tsne}
\end{figure}

\subsubsection{Analysis of the Feature Embeddings}

The goal is to display model embeddings trained with different sentence aggregation methods, as illustrated in Figure~\ref{fig:embeddings_diagram}. The investigation utilizes the PODCAST (P) dataset, concentrating on anger, sadness, happiness, and neutral emotions for clearer visual representation. Our focus is on test set segments with either high or low levels of agreement—Cohen's Kappa statistic \cite{Cohen_1960} is used to define high and low agreement groups. From the test samples, the top 2\% showing high agreement is chosen: 21 samples for "sadness," 33 for "anger," 97 for "happiness," and 139 for "neutral," as these are assumed to represent speech with emotional solid consensus.

Additionally, we explore ambiguous cases by selecting the top 2\% of test samples with low agreement, meaning these samples largely lack clear consensus. We analyze sentences containing a mix of two emotions along with their respective quantities in brackets: anger-neutral (30), sadness-happiness (18), neutral-happiness (67), and anger-sadness (30). These cases are indicative of low agreement among annotators.

We utilized T-SNE (\emph{T-distributed stochastic neighbor embedding}) to project the 1,024-dimensional feature vectors onto two-dimensional plots, aiming to visualize the data distribution. The visualizations include two primary emotions along with one composite emotion; for instance, the plot may feature "anger," "sadness," and the composite "anger-sadness" emotions. Each plot centers the emotional label around the mean values of the sentences from each class. Figure \ref{fig:tsne} illustrates the embeddings produced by models trained on the $AR_{Train}$ and $MR_{Train}$ sets. For conciseness, analogs for the $PR_{Train}$, $AR-PR_{Train}$, and $PR-MR_{Train}$ sets are omitted. The T-SNE plots reflect distinct separations between high-agreement emotions, with the two-emotion complex samples often appearing between pure emotions with high agreement.

\begin{table}[!t]
\fontsize{7}{9}\selectfont
\centering
\caption{Silhouette scores for emotional clusters observed in the embeddings are analyzed. The feature embedding analysis includes the following emotion pairs: anger-neutral (ang.-neu.), sadness-happiness (sad.-hap.), neutral-happiness (neu.-hap.), and anger-sadness (ang.-sad.). The highest silhouette score for each pair is emphasized in bold.}
\begin{tabular}{l|cccc}
\toprule
Case         & ang.-neu.   & sad.-hap.& neu.-hap.& ang.-sad.       \\ \midrule
MR$_{Train}$    & -0.0366         & 0.4085          & 0.0819          & 0.0819          \\
PR$_{Train}$    & 0.0502          & 0.3994          & 0.1291          & 0.0492          \\
AR$_{Train}$    & \textbf{0.0618} & \textbf{0.4571} & \textbf{0.1369} & 0.1627          \\
AR-PR$_{Train}$  & -0.1371         & 0.1597          & 0.0166          & 0.2695          \\
PR-MR$_{Train}$ & -0.1379         & 0.17            & 0.0108          & \textbf{0.4395} \\ \bottomrule
\end{tabular}
\label{tab:silhouette}
\end{table}

Comparing embeddings from models trained on the $AR_{Train}$ versus $MR_{Train}$ sets, it is apparent that the $AR_{Train}$ set offers superior separation between classes, as evidenced by the centralized labels of emotions in the plots. Further validation comes from the silhouette score \cite{Rousseeuw_1987}, a metric assessing how healthy clusters are formed within the embedding space (ranging from -1 for poor clustering to +1 for ideal clustering). We derived 1,024-dimensional feature representations using models trained under differing consensus-agreement conditions. Table \ref{tab:silhouette} provides the silhouette scores across three clusters: the first emotion, the second emotion, and their respective composite cases. This analysis incorporates embeddings generated from models trained on $PR_{Train}$, $AR-PR_{Train}$, and $PR-MR_{Train}$ sets.

According to the table, the model trained with the $AR_{Train}$ set ranks highest in silhouette score for "anger-neutral," "sadness-happiness," and "neutral-happiness" clusters. Intriguingly, the model trained with $PR-MR_{Train}$ had the top silhouette score for the "anger-sadness" cluster. Generally, models trained on datasets containing more ambiguous samples are better able to cluster responses to complex emotion scenarios than models trained solely on the $MR_{Train}$ engagement.

\begin{table*}[!b]
\fontsize{7}{9}\selectfont
\centering
\caption{Analysis of the impact of additional data introduced through the AR approach. The table outlines two strategies: the oversampling approach, which leverages data augmentation, and the undersampling approach, which involves randomly removing samples to achieve consistency in the dataset.}
\begin{tabular}{@{}cccccccccc@{}}
\toprule
Experiments                    & Train     & Real Data & Synthetic Data & Reduce Data & MR             & PR             & AR             & PR-MR          & AR-PR          \\ \midrule
\multirow{3}{*}{Oversampling}  & MR$_{Train}$ & 32,831    & 30,245         & 0           & 0.217          & 0.188          & 0.343          & 0.145          & 0.345          \\
                               & PR$_{Train}$ & 51,243    & 11,833         & 0           & 0.206          & 0.178          & 0.368          & 0.136          & 0.368          \\
                               & AR$_{Train}$ & 63,076    & 0              & 0           & \textbf{0.234} & \textbf{0.211} & \textbf{0.398} & \textbf{0.172} & \textbf{0.407} \\ \midrule
\multirow{3}{*}{Undersampling} & MR$_{Train}$ & 32,831    & 0              & 0           & \textbf{0.237} & \textbf{0.207} & 0.366          & 0.164          & 0.372          \\
                               & PR$_{Train}$ & 32,831    & 0              & 18,412      & 0.214          & 0.186          & 0.380           & 0.142          & 0.388          \\
                               & AR$_{Train}$ & 32,831    & 0              & 30,245      & 0.227          & 0.201          & \textbf{0.387} & \textbf{0.164} & \textbf{0.399} \\ \bottomrule
\end{tabular}
\label{tab:extradata}
\end{table*}

\subsubsection{Impact of Extra Data Incorporated by the AR Approach}

One advantage of using the $AR_{Train}$ set is the larger amount of data incorporated in training, as it utilizes all available samples, unlike the $MR_{Train}$ or $PR_{Train}$ sets. However, this extra data isn't the only contributor to the effectiveness of this strategy. We performed experiments comparing models trained on datasets of similar size using both oversampling and undersampling strategies.

For the oversampling approach, we generated synthetic data following the method proposed by Pappagari et al. \cite{Pappagari_2021}. The data generation continued until it matched the quantity used in the $AR_{Train}$ set. Table \ref{tab:extradata} presents the results, with the “Real Data” column indicating the number of data samples in the training set and the “Synthetic Data” column showing how many utterances were generated as synthetic samples. Table~\ref{tab:extradata} illustrates that the models trained with the $AR_{Train}$ set consistently outperform both the "$MR_{Train}$ + synthetic data" and "$PR_{Train}$ + synthetic data" models, underscoring the advantage of including additional samples from the $AR-PR_{Train}$ set in predicting ambiguous samples.

For the undersampling strategy, we conducted experiments to reduce training set sizes for $PR_{Train}$ and $AR_{Train}$ to match the sample size of $MR_{Train}$. The samples to be removed were selected randomly. Table \ref{tab:extradata} summarizes the macro-F1 scores for this approach, showing that models trained with the $AR_{Train}$ set deliver superior performance on the $AR$, $PR-MR$, and $AR-PR$ test sets.

\begin{table}[!b]
\fontsize{7}{9}\selectfont
\centering
\caption{The results for the undersampling strategy compare training sets consisting of either 20,000 samples, following the designated aggregation rules, or 32,831 samples formed by randomly adding 12,831 samples regardless of consensus. This table shows results in a macro-F1 score, underscoring the advantages of using the $AR_{Train}$ set.}
\begin{tabular}{@{}ccccccc@{}}
\toprule
Train Set                 & \# & MR             & PR             & AR             & PR-MR          & AR-PR          \\ \midrule
\multirow{2}{*}{MR$_{Train}$} & 20,000                  & 0.303          & 0.320          & 0.317          & 0.334          & 0.309          \\
                          & 32,831                  & \textbf{0.333} & \textbf{0.350} & \textbf{0.349} & \textbf{0.362} & \textbf{0.345} \\ \midrule
\multirow{2}{*}{PR$_{Train}$} & 20,000                  & 0.336          & 0.375          & 0.377          & 0.404          & 0.382          \\ 
                          & 32,831                  & \textbf{0.350} & \textbf{0.391} & \textbf{0.394} & \textbf{0.424} & \textbf{0.404} \\ \midrule
\multirow{2}{*}{AR$_{Train}$} & 20,000                  & 0.353          & 0.400          & 0.402          & 0.438          & 0.408          \\
                          & 32,831                  & \textbf{0.367} & \textbf{0.415} & \textbf{0.418} & \textbf{0.454} & \textbf{0.430} \\ \bottomrule
\end{tabular}
\label{tab:extradata_v2}
\end{table}

Additionally, a secondary undersampling strategy was implemented, training models under two conditions of uniform size across $MR_{Train}$, $PR_{Train}$, and $AR_{Train}$. The first condition included 20,000 randomly chosen training points under their respective aggregation rules (MR, PR, or AR). The second condition added 12,831 random points to result in 32,831 total samples. These extra points, picked randomly from the rest of the training set, do not meet consensus criteria. Since not all 32,831 samples have consensus for MR and PR rules, a soft-label learning strategy was applied. Table \ref{tab:extradata_v2} lists the macro-F1 scores for both conditions, revealing that adding more samples is consistently beneficial. Notably, the $AR_{Train}$ models frequently achieved the highest performance in both conditions (20,000 and 32,831 samples). These findings suggest that the AR method boosts the performance of SER models by including ambiguous data, with its benefits extending beyond simply enlarging the training set.

\begin{figure}[!h]
     \centering
     \begin{subfigure}[b]{0.5\textwidth}
         \centering
         \includegraphics[width=0.95\columnwidth]{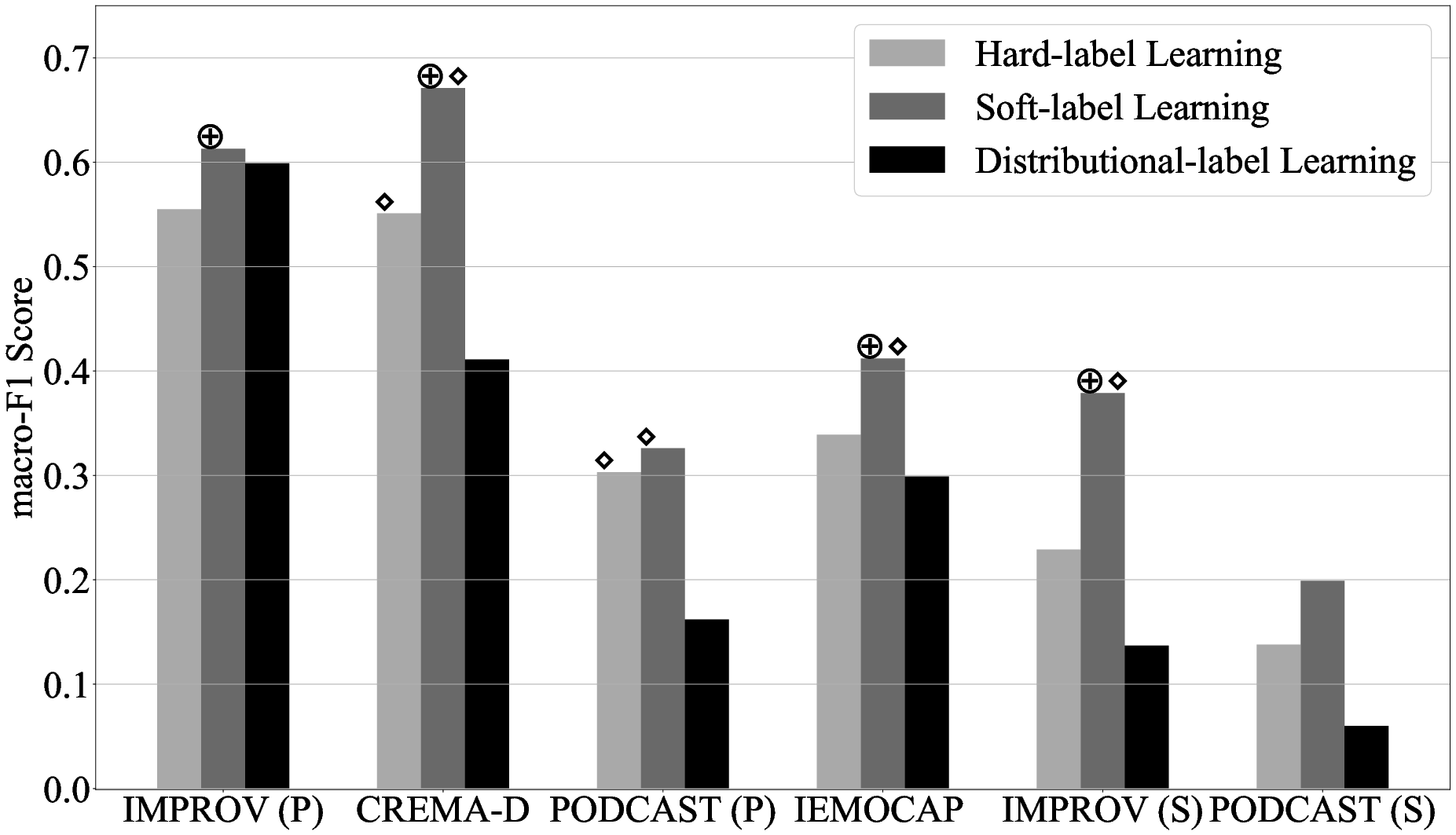}
         \caption{Training with MR.}
         \label{fig:mr_train}
     \end{subfigure}
     \vfill
     \begin{subfigure}[b]{0.5\textwidth}
         \centering
         \includegraphics[width=0.95\columnwidth]{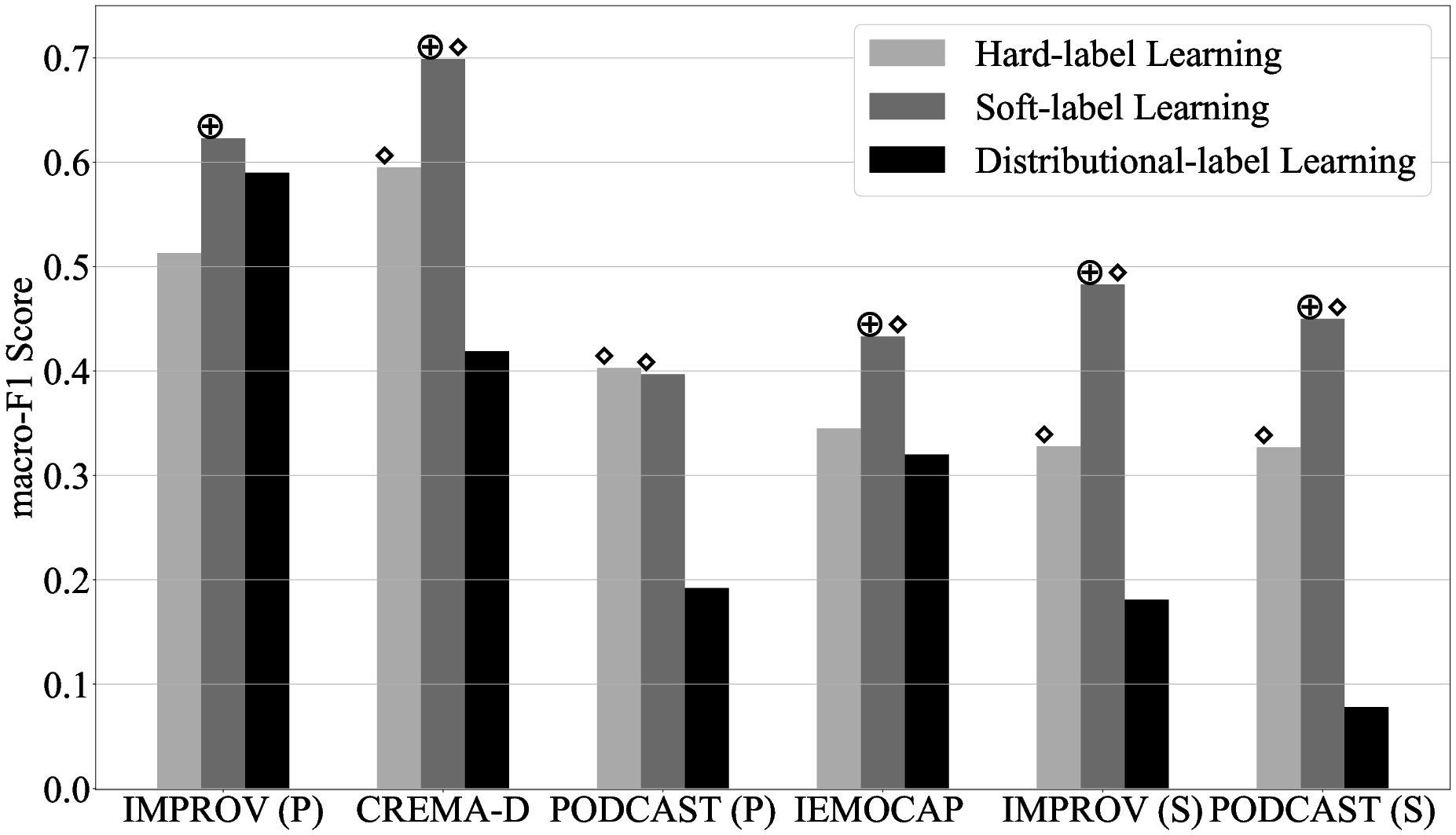}
         \caption{Training with PR.}
         \label{fig:pr_train}
     \end{subfigure}
     \vfill
     \begin{subfigure}[b]{0.5\textwidth}
         \centering
         \includegraphics[width=0.95\columnwidth]{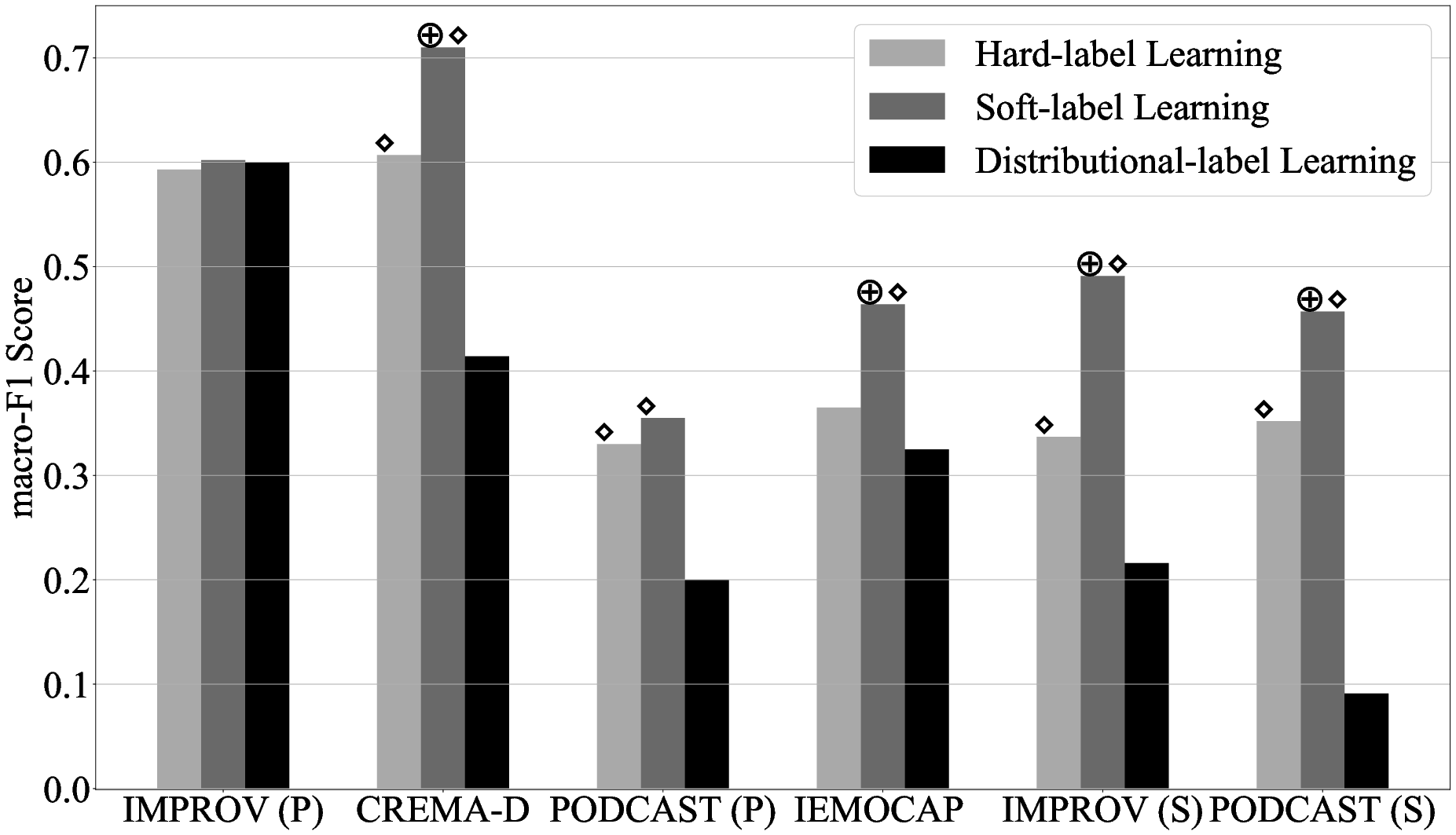}
         \caption{Training with AR.}
         \label{fig:ar_train}
     \end{subfigure}
        \caption{The bar plots depict the macro-F1 scores achieved with distributional-label learning, hard-label learning, and soft-label learning strategies. All models are assessed on the entire test set and aggregated by the AR strategy for each database. We use the symbols $\oplus$, $\ddagger$, and $\diamond$ to indicate when a model significantly outperforms those trained with distributional, soft, and hard-label learning strategies, respectively.}
        \label{fig:label_learning_bar_plots}
\end{figure}

\subsection{What is the most effective label learning method for SER?}
\label{ssec:labellearning}

We focus on the third research question: \textbf{Which label learning strategy is the most effective for training SER systems when evaluated on the entirety of the test set?} 

Figure \ref{fig:label_learning_bar_plots} illustrates the performance of SER systems trained with various aggregation techniques and label learning methods. These results are evaluated on the entire test set utilizing the AR method. Among the label-learning techniques, the strategy that employs KLD as the cost function performs the worst. Utilizing the CE loss function yields superior SER performance compared to KLD. Within the methods that use CE, soft-label learning outperforms hard-label learning. Table \ref{tab:cv} shows that SER systems using soft-label learning surpassed those using hard-label learning in 17 out of 18 cases (around 94\%). This finding is consistent with prior studies, which have demonstrated that representing emotions with soft-encoding and employing the CE loss function is a more effective label learning strategy for training SER models \cite{Fayek_2016,Han_2017,Lotfian_2017,Ando_2018,Li_2023}.

\begin{figure}[h]
     \centering
     \begin{subfigure}[b]{0.5\textwidth}
         \centering
         \includegraphics[width=0.95\columnwidth]{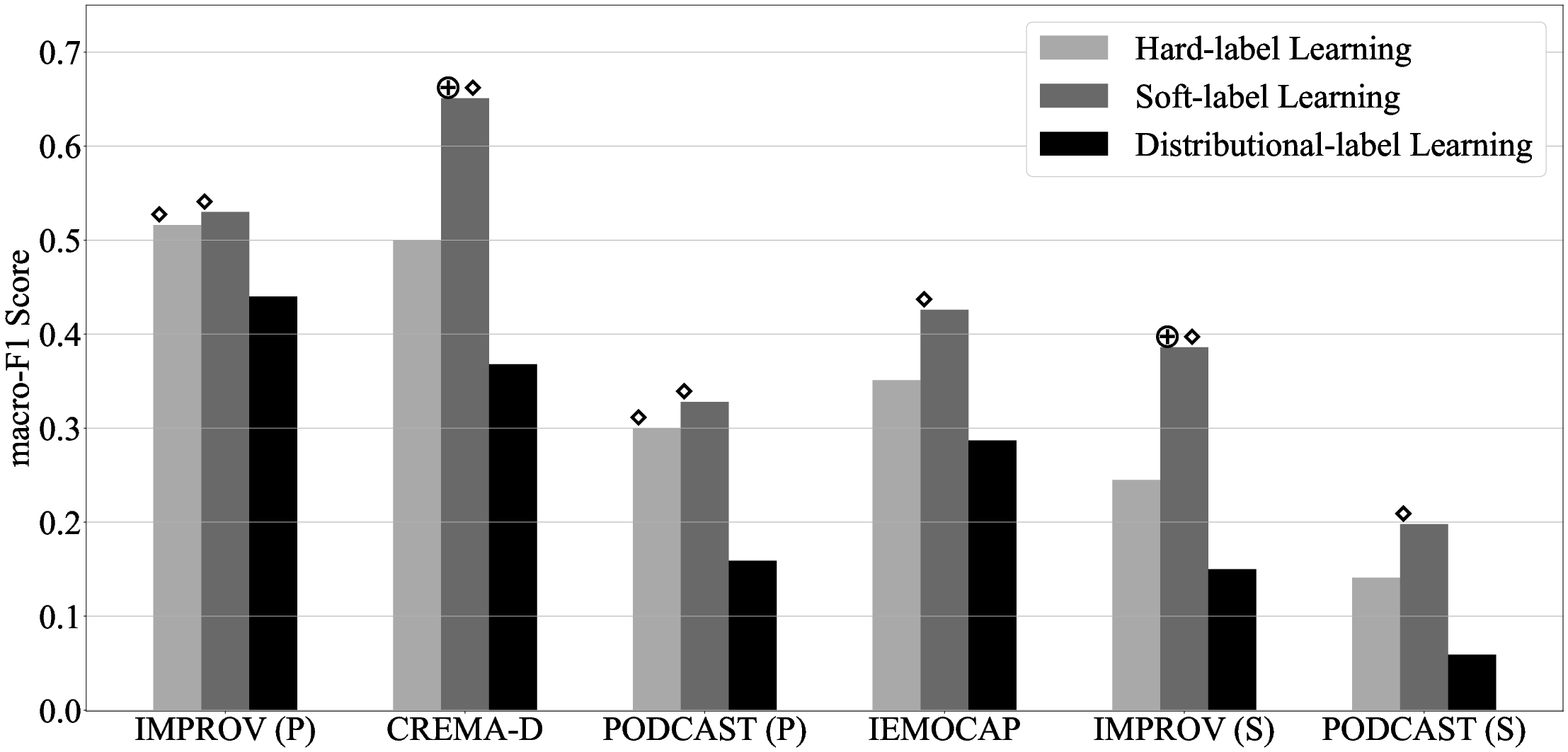}
         \caption{Training with MR.}
         \label{fig:mr_train_AR_PR_test_set}
     \end{subfigure}
     \vfill
     \begin{subfigure}[b]{0.5\textwidth}
         \centering
         \includegraphics[width=0.95\columnwidth]{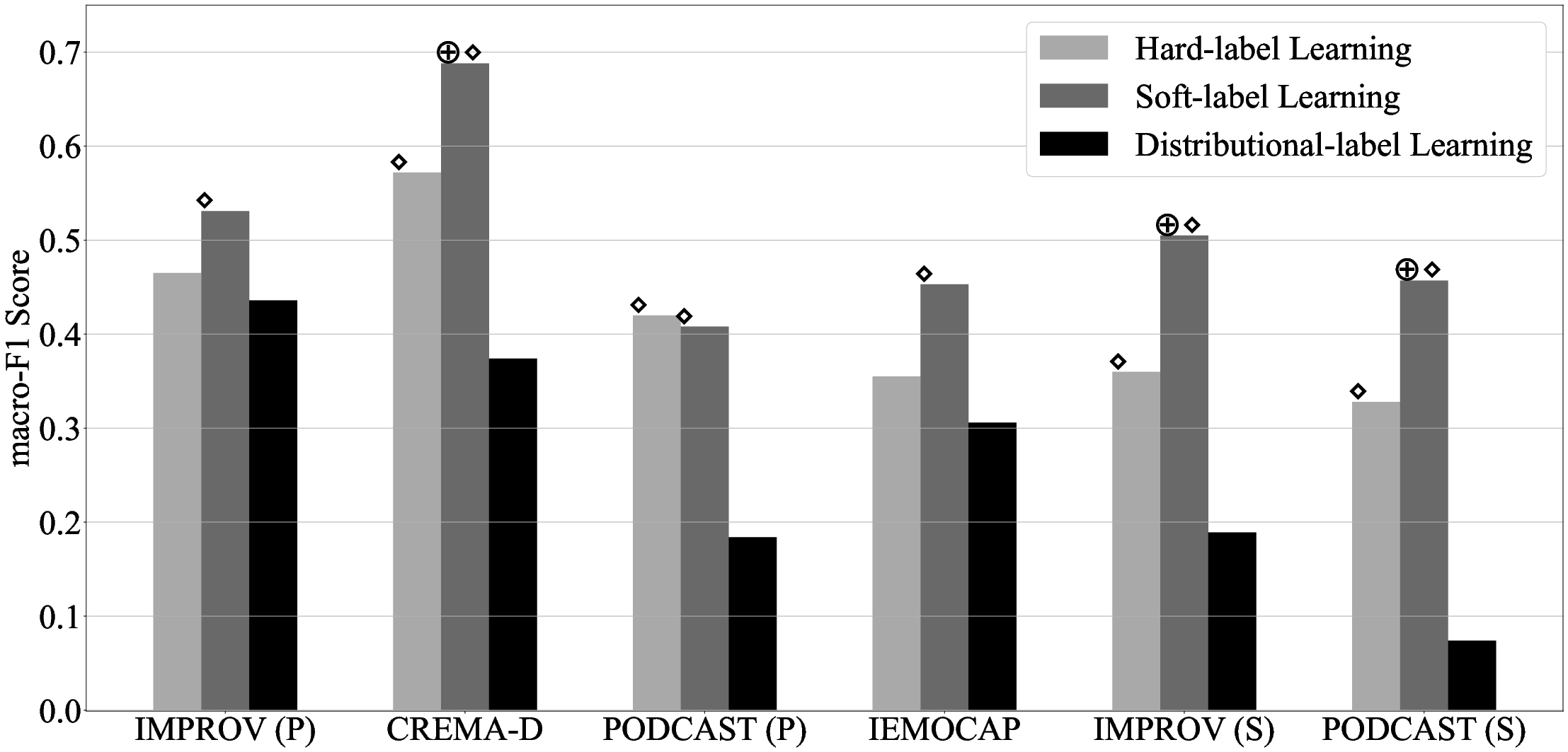}
         \caption{Training with PR.}
         \label{fig:pr_train_AR_PR_test_set}
     \end{subfigure}
     \vfill
     \begin{subfigure}[b]{0.5\textwidth}
         \centering
         \includegraphics[width=0.95\columnwidth]{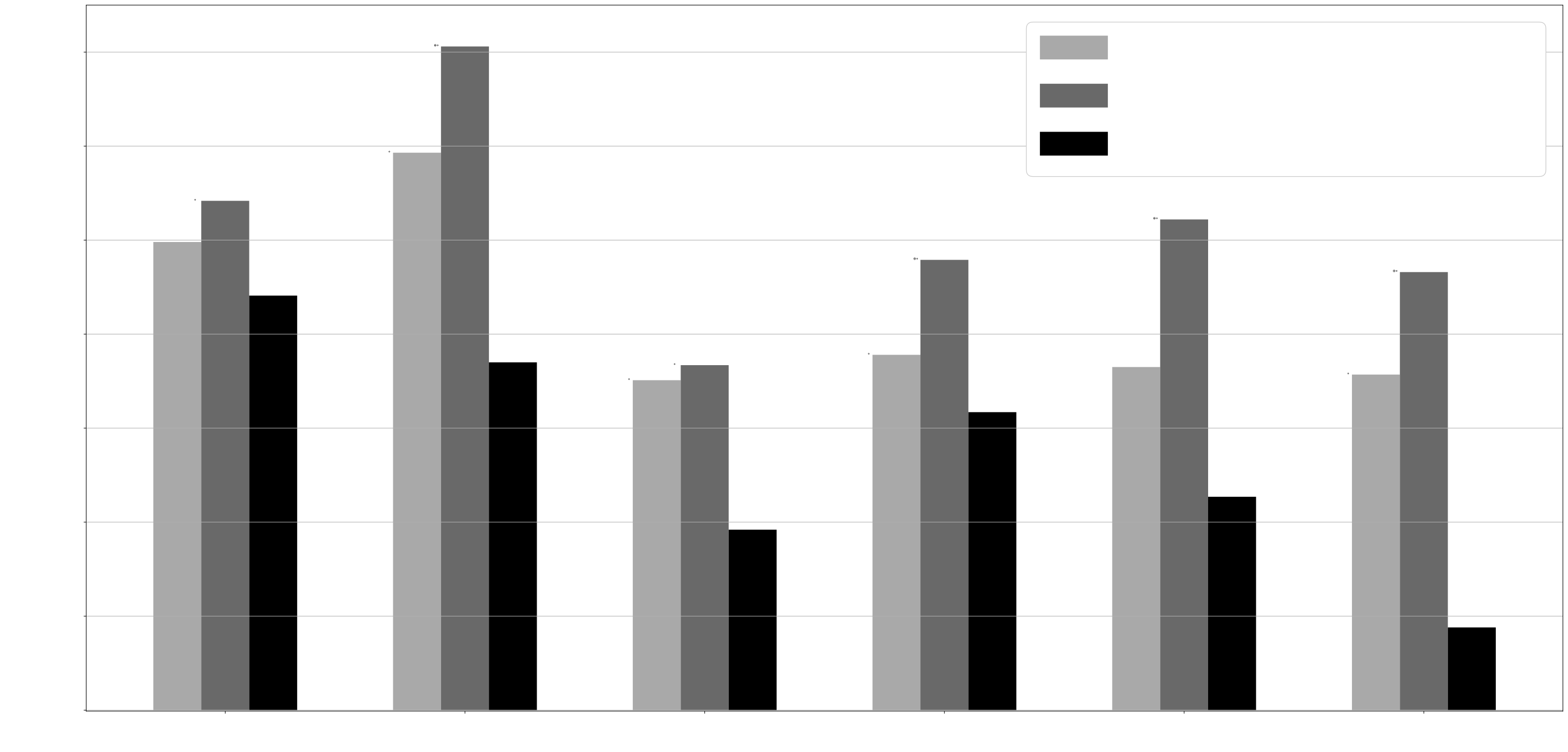}
         \caption{Training with AR.}
         \label{fig:ar_train_AR_PR_test_set}
     \end{subfigure}
        \caption{The macro-F1 scores for each database are provided for distributional-label learning, hard-label learning, and soft-label learning strategies. All models have been evaluated on the \textbf{AR-PR} test set, which includes samples that do not achieve MR or PR consensus. Symbols $\oplus$, $\ddagger$, and $\diamond$ denote instances where a model significantly outperforms those trained using distributional, hard, and soft-label learning strategies, respectively.}
        \label{fig:label_learning_bar_plots_AR_PR_test_set}
\end{figure}

Additionally, Figure~\ref{fig:label_learning_bar_plots_AR_PR_test_set} provides an overview of the macro-F1 scores for each database, comparing three different label learning methods. We focus solely on the $AR-PR$ test set for more precise interpretation, as these samples are the most emotionally ambiguous. Figures \ref{fig:mr_train_AR_PR_test_set} (training with MR), \ref{fig:pr_train_AR_PR_test_set} (training with PL), and \ref{fig:ar_train_AR_PR_test_set} (training with AR) demonstrate that the soft-label learning strategy is the most appropriate method among the existing learning approaches for training SER systems to recognize mixed emotions from the ambiguous samples within the $AR-PR$ set.

\section{Summary}

This paper examined speaker-independent categorical SER systems' performance using an all-inclusive test set, where no data was excluded, and our aggregation rule was applied. Compared to traditional label aggregation methods like the majority rule and plurality rule, the all-inclusive rule allows for the retention of all annotated data and emotional ratings, thereby making it possible to train and evaluate SER systems that are tailored to real-world scenarios.} The initial examination revealed that adhering to the majority or plurality rule excludes a notable portion of the annotated test samples, resulting in a poor representation of expected SER performance in realistic scenarios. In real-world applications, the classifier must recognize emotions in all sentences, regardless of consensus. Experiments with the comprehensive test set demonstrated that employing the all-inclusive aggregation rule to define the ground truth yields more reliable SER performance by incorporating more speech samples with low-agreement annotations. Our findings also indicated a decline in SER model performance as more ambiguous samples were included in the test set, highlighting the significance of using the complete test set. Additionally, we found that training exclusively with high-agreement data is insufficient for predicting ambiguous emotions. Moreover, the results showed that soft-label learning leads to the best performance among label-learning strategies when applied to the entire test set. The average SER performance of models trained with data selected through the all-inclusive aggregation rule consistently surpassed those trained with data specified by the majority or plurality rule on incomplete and comprehensive test sets.

Building upon the findings of this dissertation, future research could broaden the application of the all-inclusive rule to encompass other subjective tasks. Its effectiveness could mainly be assessed in areas like \emph{text-to-speech} (TTS) and textless \emph{speech-to-speech translation} (S2ST) systems. For instance, Zhou et al. \cite{Zhou_2022_2} introduced a system that synthesizes human voices with mixed emotions; however, the limited range of emotions indicates the potential for enhancing emotion embedding. By employing the all-inclusive rule, which utilizes the entire dataset, it is anticipated that more authentic and varied emotional expressions will be achieved in TTS systems, surpassing current approaches. Moreover, despite its critical role in natural human interaction, present S2ST systems still need to incorporate emotional information \cite{Lee_2022, Li_2023_3}. Acknowledging the significance of emotions in effective communication, integrating the all-inclusive rule into S2ST systems is believed to significantly enhance their realism in speech conversion.

\chapter{Training Loss by Using Co-occurrence Frequency of Emotions}
\label{ch:training_loss}

Previous research in SER focusing on categorical emotions has traditionally concentrated on identifying a single emotion per utterance based on the assumption that each emotion is distinct. However, it has emerged that emotions may overlap, particularly in ambiguous emotional expressions that combine elements, such as happiness mixed with surprise. As a result, emotion recognition systems have been updated to predict multiple emotional categories. This approach, though, typically neglects the relationships among different emotions during the training process, treating them independently. This study investigates the interconnected nature of emotional categories and how these relationships impact the training of SER models. Specifically, the co-occurrence frequencies of emotions are assessed based on perceptual evaluations within the training data. This information creates a matrix that assigns penalties according to class dependencies, enforcing harsher penalties for errors between more dissimilar emotions. This matrix combines three established label learning techniques using the modified loss function. Subsequently, SER models are developed using both the newly integrated penalization matrix and the traditional cost functions typically utilized. The newly introduced penalization matrix significantly improves the macro F1-score on the PODCAST dataset, showing increases of 17.12\%, 12.79\%, and 25.8\% for hard-label, multi-label, and distribution-label learning methods, respectively.

\section{Motivation}
SER is crucial for enhancing human-centered computer interactions. Emotional labels for training SER systems are generally obtained from perceptual evaluations. Nonetheless, subjectivity in emotion perception leads to interpreting the same speech, and often the same speech differently \cite{Cowen_2017,Sethu_2019}. Conventionally, SER studies handle the variability in emotional annotations as noise, using label aggregation methods to produce a consensus for training SER systems \cite{Jin_2015,Parthasarathy_2018_3, Aldeneh_2017, Abdelwahab_2018, Yoon_2019}. This methodology tends to disregard the possibility of co-occurring emotions frequently present in emotional expressions (e.g., a sentence might convey both happiness and surprise). Although multi-label learning \cite{Xin_2023,Zhang_2021,Ju_2020} permits ground truths with several valid classes, it still overlooks the relationships among emotions by assuming all emotion classes are disconnected. The dissertation posits that categorical emotions are interrelated and suggests that considering the commonness of emotion co-occurrence in emotional ratings can create better SER systems.

Examining simultaneous emotions that appear in everyday life \cite{Vansteelandt_2005}, Xu et al. \cite{Xu_2020} capitalized on the prevalence of concurrent emotions in perceptual assessments to forge initial linkages within their presented graph-based DNN. Meanwhile, the work \cite{Steidl_2005} leveraged frequent emotional misclassifications by evaluators to evaluate an SER system’s performance. Furthermore, some research works have utilized the soft-label method to capture secondary emotions present in voice \cite{Fayek_2016,Sridhar_2021,Lotfian_2017}. This body of work highlights the importance of accounting for feedback from all evaluators in perceptual evaluations, even when there is variation from agreed-upon labels.

This research explores how incorporating emotional co-occurrence data can improve the training process of a SER system. It involves creating a co-occurrence frequency matrix that captures the relationships between different emotions as determined by perceptual evaluations. This matrix is subsequently converted into a penalization matrix, which adjusts the loss functions by assigning greater penalties for predicting combinations of emotions that seldom occur together. Our approach folds the penalization matrix into existing cost functions as a "penalty loss," thereby increasing the loss value if the model forecasts emotions with low co-occurrence frequencies. For instance, since anger and contempt are seldom co-selected in perceptual evaluations, predicting these two emotions jointly will incur a higher penalty than predicting commonly co-occurring emotions like anger and sadness. This strategy acknowledges the interdependence of emotions, leveraging their relationships to enhance model performance.

\section{Background and Related Works}
\label{sec:background_co_emotion}

\subsection{Contrastive Learning in Emotion Recognition}
To enhance SER systems, Wang et al. \cite{Wang_2023_2} designed an objective function aimed at maximizing inter-class distance while minimizing intra-class distances. Their findings demonstrate the advantages of this approach. Additionally, Zhao \cite{Zhao_2024} introduced a nearest neighbor contrastive learning technique to emphasize distinctions between various emotions by leveraging local neighborhood information, which subsequently boosted SER system performance under cross-corpus testing conditions. Different from the previously mentioned studies, the frequency of co-occurrence of emotions in the training set is used by us to model the relationship between emotion classes.

\subsection{Label Learning in Emotion Recognition}
This work explores the application of co-occurrence frequencies of emotional classes derived from perceptual evaluations to enhance the training process of an SER system. It examines related methods such as distribution-, hard-, multi-label learning, and emotion classification. To illustrate, we use a 4-class emotion classification task that includes angry (A), neutral (N), happy (H), and sad (S) emotions. In this scenario, five annotators independently assess a sentence, each assigning one label. For a single sample, an example set of labels might look like this: "S, N, S, N, S."
\subsubsection{Emotion Recognition using Hard-label Learning}
\label{sssec:hard}
In SER studies, emotion databases annotated via perceptual evaluations often employ consensus labels derived from collating individual annotations through techniques like majority voting or the plurality rule. Such methods produce a definitive hard-label vector indicating a single emotional class, thereby excluding secondary emotions noted by raters who did not align with the majority group. For instance, the constructed one-hot vector could be (0, 0, 1, 0). However, few works have introduced soft labels \cite{Fayek_2016, Sridhar_2021, Lotfian_2017}, enabling each sample to be linked with multiple emotions. The work \cite{Chou_2019} built each annotator individually to capture the subjective nature of perceptual evaluations. These approaches permit the inclusion of sentences in the training dataset even when annotators have differing emotional perceptions. Contrary to systems trained with hard-label learning—which predict a single emotion per utterance and assume that emotions are independent—these advanced techniques consider emotional co-occurrence. My findings reveal that adopting the "penalty loss" method improves performance.

\subsubsection{Emotion Recognition using Multi-label Learning}
\label{sssec:multi-label}

Multi-label learning empowers emotion classifiers to detect several emotions in a single data instance, generating multiple hard vectors that may highlight more than one emotion. For example, the emotions identified could be represented as (1, 0, 1, 0). Prominent examples of this approach in emotion recognition are found in studies like \cite{Ju_2020,Zhang_2020_2, Zhang_2021}. Although multi-label learning facilitates recording various emotions, it assumes these emotions manifest independently. This traditional methodology also fails to signify the relative importance of emotions, such as discerning primary emotions from secondary ones. In the research, we utilize the approach described by the work \cite{Xin_2023}, applying multi-label learning to display enhancements in SER systems through the "penalty loss" concept.

\subsubsection{Emotion Recognition using Distribution-label Learning}
\label{sssec:Distribution}

Distribution-label learning \cite{Geng_2016} aims to produce a distributional output under the assumption that labels are spread across categories so that their probabilities add up to 1. For example, the distributional label might be (0.4, 0.0, 0.6, 0.0). During the training of the SER systems, the objective function can let the model minimize the distance between the actual and predicted distributions. A couple of notable applications of distribution-label learning to SER include the studies by \cite{Chou_2022} and \cite{Wu_2023}. Chou et al. \cite{Chou_2022} tailored distribution-label learning by translating emotional ratings into distribution labels, and we adopted a similar method to train the model. The KLD is typically used as the loss function in traditional distribution-label learning. Our research examines the impact of introducing a "penalty loss" on the accuracy of classification results.

\begin{figure*}[!t]
  \centering
  \resizebox{\textwidth}{!}{%
  \includegraphics[width=\textwidth]{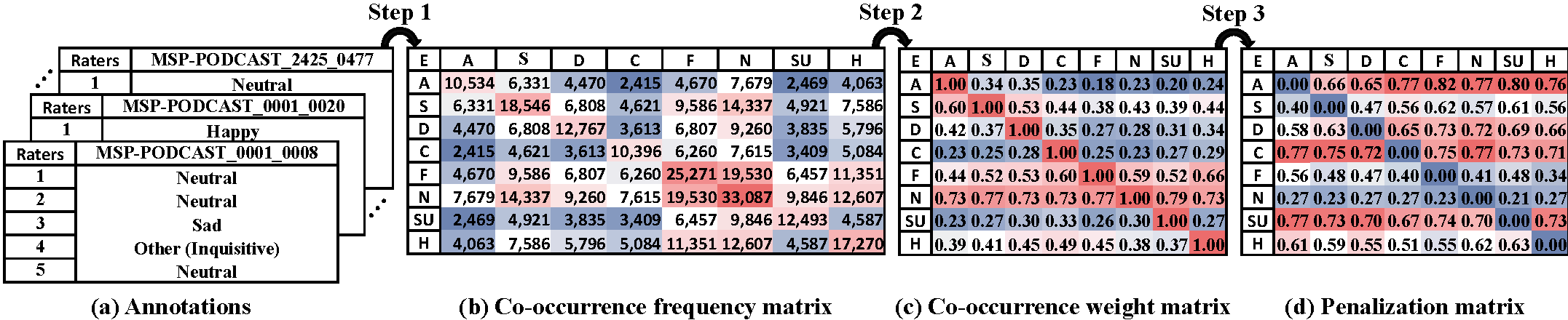}}
  \caption{Illustration of a process for generating the presented penalization matrix. The 8-class emotions involved include contempt (C), neutral (N), sad (S), happy (H), fear (F), disgusted (D), angry (A), and surprised (SU). The procedure in detail can be found in Section \ref{ss:pweights}.} 
  \label{fig:flowchart}
\end{figure*} 

\section{Proposed Method}

\subsection{Penalization Weights based on the Counts of Co-Existing Emotions}
\label{ss:pweights}

Figure \ref{fig:flowchart} details the three-phase process of forming the penalization matrix. In the initial phase, we adopt the method outlined by \cite{Xu_2020} to generate the co-occurring matrix from emotional labels in the training dataset. The proposed matrix indicates how frequently different pairs of emotions appear together, resulting in an \textbf{eight times eight symmetric matrix}. Take Figure \ref{fig:flowchart} (b) as an example. The entry ("S", "S") has the value 18,546, indicating that \emph{angry} was chosen across 18,546 utterances. Among these occurrences, \emph{neutral} was chosen 14,337 times. Consequently, the entries at positions ("S", "N") and ("N", "S") are populated with 14,337, respectively.

In phase 2, we create a matrix indicating the probability of co-existing emotions by dividing each element by the total count of instances marked with the corresponding emotion in each column. Take the second column labeled "S" from Fig.~\ref{fig:flowchart} (b) as an example—the co-occurrence frequencies of \emph{anger} with other emotions are 6,331, 18,456, 12,767, 3,613, 6,807, 9,260, 3,835, and 5,796. By dividing each of these frequencies by the total occurrences of \emph{sad} (6,331), {\color{blue}we} obtain co-occurrence probabilities: 0.34, 1.00, 0.37, 0.25, 0.52, 0.77, 0.27, and 0.41. As an example, the co-existing possibility of disgust and sad (0.37) is greater than that of sad and contempt (0.25). This normalized matrix is termed the co-existing weight matrix (Figure~\ref{fig:flowchart} (c)), which is asymmetric owing to the normalization performed column-wise.

In phase 3, the co-occurrence weight matrix converts into a "penalization matrix." This conversion is essential as it penalizes the SER models for forecasting rare emotion combinations. The conversion process is simple: Each element in the co-existing emotion weight matrix is subtracted from one, resulting in the penalization matrix (Figure~\ref{fig:flowchart} (d)). Employing the method causes an increase in the training loss if the model predicts infrequent combinations of emotions.

\subsection{Label Processing to Train SER Systems}
\label{ssec:LearninLabelProcessing}

We incorporate the penalization matrix into the three labeling approaches outlined in Section \ref{sec:background_co_emotion}: hard-, multi-, and distribution-label learning. Each technique uses a specific cost function. Hard-label learning relies on cross-entropy (CE) loss, multi-one on binary cross-entropy (BCE) loss, and distribution-label learning on the KLD loss. Furthermore, we apply the label smoothing strategy suggested by \cite{Szegedy_2016} to both hard-label and distribution-label learning, setting the parameter at 0.05. A slight value ($10^{-6}$) is also added to entries initially quantified as zero for the multi-label vector.

\subsection{Loss Functions Integrated by the Proposed Penalization Matrix}
\label{ss:pl}

The technique of integrating a penalization matrix into loss functions is proposed. This method requires the definition of $N \times K$ matrices for both the model's predictions ($Y^{P}$) and the actual labels ($Y^{T}$). Here, $C$ signifies the number of emotional categories, while $N$ denotes the number of samples under consideration. The variation in each row of $Y^{T}$ depends on the label learning approach in use, whether it is distribution-, multi-, or hard-label learning. Afterwards, the loss value matrix ($L \in R^{N \times C}$) is determined using the following approach:

\begin{equation}
\begin{array}{l}
\label{eq:loss}
\emph{$L$} = f_{loss}(Y^{T},Y^{P}) 
\end{array}
\end{equation}

\noindent

The loss function, represented by ($f_{loss}$), could be the cross-entropy (CE). The elements of ( L ) are calculated as ( $f_{loss}(Y^{T}{ij}, Y^{P}{ij})$ ), where ( $j \in { 1, \ldots, K }$ ) and ( $i \in { 1, \ldots, N }$ ), to estimate the categorical emotion loss for each utterance. Subsequently, the proposed matrix has been incorporated into Equation \ref{eq:loss}. The matrix introduced, denoted as ( $P$ ), belongs to ( $R^{K \times K}$ ) (refer to Fig. \ref{fig:flowchart}). The integration of the loss function is achieved through the penalization matrix, represented by ( $L_{P+loss}$ ), as follows:

\begin{equation}
\label{eq:pl}
\begin{array}{l}
L_{P+loss} = \sum_{i=1}^{N} ( \emph{$L$}_{i} \cdot \emph{$P$} )\\= \sum_{i=1}^{N}( \sum_{j=1}^{C}  \sum_{z=1}^{C} P_{jz} \cdot f_{loss}(Y^{T}_{ij},Y^{P}_{ij})).
\end{array}
\end{equation}

When we replace $f_{loss}$ with the objective functions, the equations will be:

\begin{align} 
L_{P+CE} &= - \sum_{i=1}^{N}(\sum_{j=1}^{C}\sum_{z=1}^{C} P_{jz} \cdot Y^{T}_{ij} \cdot \log(Y^{P}_{ij})),\label{eq:ce}\\
L_{P+BCE} &= - \sum_{i=1}^{N}(\sum_{j=1}^{C}\sum_{z=1}^{C} (P_{jz} \cdot Y^{T}_{ij} \cdot \log(Y^{P}_{ij}) \label{eq:bce}\\ 
&+ P_{jz} \cdot (1-Y^{T}_{ij}) \cdot \log(1-Y^{P}_{ij}))),  \nonumber
\end{align}

\begin{align} 
L_{P+KLD} &= \sum_{i=1}^{N}(\sum_{j=1}^{C}\sum_{z=1}^{C} P_{jz} \cdot Y^{T}_{ij} \cdot \log(\frac{Y^{T}_{ij}}{Y^{P}_{ij}})).\label{eq:KLD1}
\end{align}

In training, the initial loss is modified by the proposed loss, resulting in the total loss as described by Equation \ref{eq:totalloss}. In this equation, $\alpha \in R$ and $\beta$ are assigned a value of either 1 or 0.

\begin{equation}
\mathscr{L}_{P}^{loss} =   \beta L_{loss} + \alpha \cdot L_{P+loss}.
\label{eq:totalloss}
\end{equation}

\section{Experimental Setup}

\subsection{Resource}

We evaluate the proposed method using the Podcast corpus \cite{Lotfian_2019}, using version 1.9 in this ongoing collection effort. The training set consists of 55,283 speech utterances, the development set includes 9,546, and the test set comprises 16,570 utterances—primary emotions in the perceptual evaluation cover 8 classes mentioned in Section~\ref{s:msp_podcast}. For hard-label learning, we only utilize speaking turns with agreed-upon consensus labels and discard segments lacking agreement. Further details can be found in Section~\ref{s:msp_podcast}.

\subsection{Acoustic Features}
In the research conducted by \cite{Keesing_2021}, the effectiveness of various attributes in SER was assessed across multiple publicly accessible emotional datasets. This analysis identified the wav2vec feature set, introduced by the study \cite{schneider_2019}, as one of the most robust techniques for feature extraction. Consequently, the models incorporate the 512-dimensional wav2vec feature as input. To prepare the data for training, we standardized all features using z-normalization, which was calculated according to the mean and standard deviation from the training dataset.

\subsection{SER Models and Other Details}
We adopted the chunk-level SER models proposed by \cite{Lin_2023} as the primary model. This approach systematically processes sentences of varied lengths by converting them into a specified number of uniformly sized chunks through overlapping adjustments. Based on the paper's suggestions, we utilized LSTM as the feature encoder at the chunk level, along with the RNN-AttenVec model for chunk-level attention \cite{Lin_2023}. This combination allows us to capture emotion-related information at different levels. For further details on the network architecture, see Lin and Busso \cite{Lin_2023}. Model parameters adhered to those in Chou et al. \cite{Chou_2022_2}. Based on previous studies, for the output layer, we used softmax activation for CE \cite{Jin_2015, Aldeneh_2017, Yoon_2019} and KLD \cite{Geng_2016, Chou_2022_2}, with sigmoid activation for BCE \cite{Xin_2023, Fei_2020}. We used Adam optimizer and set the learning rate to $0.0001$; we set the batch size to 128. Then, we train all models with the epochs of 25. Finally, we saved the best-performing model according to their minimum loss on the development set. To analyze the influence of the suggested objective function on the accuracy of SER systems, the parameter $\alpha$ in Equ.~\ref{eq:totalloss} was set to either 0.5 or 1.0. We performed experiments using only $L_{P+loss}$ and using only $L_{loss}$.

\subsection{Evaluation Metrics}
To gauge the accuracy of prediction labels against ground truth, we resort to a suite of tailored metrics based on the type of label learning method in use. For hard-label learning, I used macro-, micro-, and weighted-F1 scores, unweighted average recall, and unweighted average precision. In the case of multi-label and distribution-label learning, I employ measures similar to those referenced in the study by \cite{Fei_2020}: ranking and hamming loss, coverage error, alongside macro F1. Moreover, micro- and weighted-F1 scores are integrated to estimate multi-label tasks specifically. For a performance assessment of multi-label and distribution-label methods, binarization of predictions is necessary. For multi-label learning, predictions' probabilities are changed into "multi-hot" binary vectors using a threshold of 0.5, following \cite{Fei_2020,Xin_2023}'s methodology. For distribution-label learning, a threshold is fixed at 1/convertingion of probability predictions into binary vectors, in line with Chou et al. \cite{Chou_2022_2}.

\subsection{Statistical Significance}

The methodology from Lin and Busso \cite{Lin_2023} is employed, where the original test set is divided at random into 30 smaller subsets of roughly equal size. The average results for all metrics are then reported. A two-tailed t-test is utilized to conduct the statistical significance test, evaluating the difference between the proposed approach and the baseline models. A result is considered statistically significant if the $p$-value is less than 0.05.

\section{Experimental Results and Analyses}
Table \ref{t:results} summarizes the average outcomes across various configurations. Evaluation metrics are marked with $\uparrow$ to show that higher values indicate better performance, while $\downarrow$ signifies that lower values are better. The $*$ symbol is used to flag results that show statistically significant improvements over baseline SER systems where the proposed loss is not applied (i.e., $\alpha=0$ in Eq. \ref{eq:totalloss}). The table is divided into three sections corresponding to different objective functions: CE, BCE, and KLD for hard-, multi-, and distribution-label learning, respectively. The column labeled “$\beta$” denotes whether the loss $L$ (with $\beta=1$) was accounted for in the experiments as opposed to excluding it ($\beta=0$). The $\alpha$ column states the value assigned to the PL loss weight. Specifically, the top-performing values for each loss function are emphasized in bold.

\begin{table}[!t]
\fontsize{7}{9}\selectfont
\centering
\caption{The table overviews the results on distributional-label, multi-hard-label, and single-label tasks for the primary emotion classification task. The mark $*$ denotes that the outcomes for SER systems utilizing the presented matrix have statistical significance compared to the baseline ($\alpha=0$; $\beta=1$).}
\begin{tabular}{@{}l|c@{\hspace{0.05cm}}c@{\hspace{0.05cm}}c@{\hspace{0.1cm}}c@{\hspace{0.1cm}}c@{\hspace{0.05cm}}c@{\hspace{0.05cm}}c@{\hspace{0.05cm}}c@{}}
\toprule
\multicolumn{9}{c}{\textbf{Single-label Task}}                                                                                                                       \\ \midrule
\textbf{$f_{Loss}$}                 & \textbf{$\beta$} & \textbf{$\alpha$}  & \textbf{}      & \textbf{Unweighted Average Recall} $\uparrow$   & \textbf{Unweighted Average Precision} $\uparrow$  & \textbf{Macro F1} $\uparrow$ & \textbf{Micro F1} $\uparrow$ & \textbf{Weighted F1} $\uparrow$  \\ \hline
\multirow{5}{*}{\textbf{CE}}         & \textbf{1} & \textbf{0}   &                & 0.144          & 0.133          & 0.111          & 0.424 & 0.318          \\
\cline{2-9}
                                             & \textbf{1} & \textbf{0.5} &                & \textbf{0.156}* & \textbf{0.137}*          & 0.129*          & \textbf{0.425}          & \textbf{0.347}          \\
                                             & \textbf{1} & \textbf{1}   &                & 0.155*          & 0.136          & \textbf{0.130}* & 0.408          & 0.346 \\
                                             & \textbf{0} & \textbf{1}   &                & 0.154*          & 0.136          & 0.128*          & 0.396          & 0.341          \\ \toprule
\multicolumn{9}{c}{\textbf{Muli-label Task}}                                                                                                                         \\  \midrule 
\textbf{$f_{Loss}$}                 & \textbf{$\beta$} & \textbf{$\alpha$}  & \textbf{Hamming Loss} $\downarrow$   & \textbf{Ranking Loss} $\downarrow$   & \textbf{Coverage Error}$\downarrow$  & \textbf{Macro F1} $\uparrow$   & \textbf{Micro F1} $\uparrow$   & \textbf{Weighted F1} $\uparrow$   \\ \hline
\multirow{5}{*}{\textbf{BCE}}        & \textbf{1} & \textbf{0}   & 0.304          & 0.603          & 6.899          & 0.219          & 0.466          & 0.352          \\ 
\cline{2-9}
                                             & \textbf{1} & \textbf{0.5} & 0.303          & 0.608          & 6.928          & 0.215          & 0.462          & 0.348          \\
                                             & \textbf{1} & \textbf{1}   & \textbf{0.303}          & \textbf{0.587}          & \textbf{6.837}          & 0.235          & \textbf{0.482}          & 0.370          \\
                                             & \textbf{0} & \textbf{1}   & 0.305          & 0.597          & 6.871          & \textbf{0.247}* & 0.477          & \textbf{0.378}          \\ \toprule
\multicolumn{9}{c}{\textbf{Distribution-label Task}}                                                                                                                         \\  \midrule 											 
\textbf{$f_{Loss}$}                 & \textbf{$\beta$} & \textbf{$\alpha$}  & \textbf{Hamming Loss} $\downarrow$   & \textbf{Ranking Loss} $\downarrow$   & \textbf{Coverage Error}$\downarrow$  & \textbf{Macro F1} $\uparrow$   & \textbf{Micro F1} $\uparrow$   & \textbf{Weighted F1} $\uparrow$   \\ \hline
\multirow{5}{*}{\textbf{KLD}} & \textbf{1} & \textbf{0}   & \textbf{0.294} & 0.511          & 6.279          & 0.283          & 0.522          & 0.431          \\
\cline{2-9}
                                             & \textbf{1} & \textbf{0.5} & 0.308          & \textbf{0.507} & 6.220          & 0.322*          & \textbf{0.533} & 0.471*          \\
                                             & \textbf{1} & \textbf{1}   & 0.315          & 0.509          & \textbf{6.214} & 0.330*         & 0.532          & 0.475*          \\
                                             & \textbf{0} & \textbf{1}   & 0.337          & 0.530          & 6.284          & \textbf{0.356}* & 0.526          & \textbf{0.496}* \\  \bottomrule
\end{tabular}
\label{t:results}
\end{table}

\subsection{Does incorporating the penalty loss ($L_{P+loss}$) benefit SER Systems?}
Table \ref{t:results} reveals that models utilizing the proposed penalty loss ($L_{P+loss}$) typically achieve the best outcomes across different label learning strategies. For example, when the results are evaluated on the single-emotion utterances, the model with $L_{P+loss}$ where $\alpha = 0.5$ topped four out of five evaluation metrics and exhibited a 16.22\% relative gain in macro F1-score (maF1) compared to the baseline. In the multi-label classification setting, the model using $L_{P+loss}$ with $\alpha = 1.0$ led in four out of six metrics, showing a 7.31\% relative gain in maF1 over the baseline performance. In terms of the distribution-label task, the majority of the best performances resonate from models integrated with $L_{P+loss}$. Explicitly focusing on maF1, the leading model with KLD reported a remarkable 25.8\% relative enhancement over the baseline and surpasses SOTA results documented by the work \cite{Chou_2022_2} (31.6\% maF1). This suggests that the presented penalty loss ($L_{P+loss}$) significantly improves model accuracy in primary emotion classification tasks. Furthermore, the results exhibit that aggregating the primary loss $L_{loss}$ with the presented loss $L_{P+loss}$ tends to boost overall effectiveness. 

\subsection{Effect of Co-occurrence Matrix}

The aim is to determine if the proposed methodology effectively allows systems to capture the intended co-existing emotions matrix accurately. This is evaluated by measuring the distance between the co-existing emotions matrices derived from the learning targets in the training set and those obtained from model predictions, utilizing the Frobenius norm as the distance metric. The focus is on models that use either $L_{P+loss}$ ($\alpha=1; \beta=0$) or $L_{loss}$ ($\alpha=0; \beta=1$). When $L_{P+loss}$ is implemented, a decrease in the distance was observed from 4.27 to 4.00 using the multi-label approach and from 4.13 to 3.39 using the distribution-label approach. This reduction signifies that the co-existing emotion matrix, as the proposed penalization method predicted, is more closely aligned with the target co-occurrence matrix.

\section{Summary}

This dissertation utilized co-existing emotion counts to create a penalization weight for training a speech emotion recognition (SER) model. The proposed penalty loss considers the relationships between emotions, applying more significant penalties for inaccuracies between more distinct emotions. The effectiveness of this penalty loss was assessed through three distinct label learning approaches: distribution-based, hard-label, and multi-label learning. The results suggest that the proposed loss generally improves recognition performance across most evaluation metrics in eight classes' primary emotion classification tasks.

\chapter{Conclusion}
This dissertation introduces three proposed methods to optimize the pipeline for constructing speech emotion recognition (SER) systems. These methods have been documented in our studies, including rater-modeling \cite{Chou_2019}, all-inclusive \cite{Chou_2024}, and the penalization objective function that accounts for co-currencies of emotions \cite{Chou_2022}. The findings from this dissertation and the respective studies \cite{Chou_2019, Chou_2022, Chou_2024} have facilitated answering the primary research questions.

\textbf{(1) Should we remove the minority of emotional ratings?} The minority of emotional ratings should not be discarded, as retaining them can enhance the recognition of ambiguous utterances with complex emotional perceptions. Adding these minority ratings also boosts the performance of SER systems on comprehensive test sets that mirror real-world conditions. Further analyses reveal that standard approaches, which neglect these minority ratings, show reduced capabilities in clustering samples with dual emotions. Therefore, including the minority of emotional ratings is essential to account for the subjectivity inherent in emotion perception.

\textbf{(2) Should we only let the SER systems learn the emotional perceptions of a few people?} The findings of this dissertation indicate that allowing SER systems to learn emotion perception from a larger group can enhance recognition capabilities through the proposed all-inclusive rule or individual-rater modeling method. This not only improves the performance of SER systems but also offers a unique opportunity for personal growth. By incorporating emotional ratings from a broader range of people, each person's unique sensitivity and emotional background can contribute valuable insights. This should inspire and motivate researchers, developers, and practitioners in the field of SER to continue their work and strive for excellence.

\textbf{(3) Should SER systems only predict one emotion per speech?} To address real-world conditions that involve the simultaneous occurrence of various emotion classes, SER systems need training to predict multiple emotions simultaneously. This dissertation convincingly illustrates that multi-label SER systems can achieve superior performance on test sets featuring a single consensus label determined by the plurality rule and the all-inclusive rule. This underlines the effectiveness of multi-label emotion prediction, which should be incorporated into SER systems rather than focusing on a single emotion.

\section{Discussion and Limitation}
While the dissertation proposes three methods to address challenges in SER, several other factors need consideration. For instance, the effect of various inputs, such as hand-crafted features or raw audio signals, on prediction outcomes should have been analyzed. Additionally, the relationships between layer embeddings and label spaces should be explored. Visualizing these relationships could provide fascinating insights. Furthermore, all the suggested methods are designed for traditional SER emotion databases, while the latest dataset \cite{Zhang_2024} includes intensity scores for each categorical emotion. Using this type of emotion database might mean our methods must be optimized for training and evaluating SER systems. It would be intriguing to see if our methods could be adapted for samples with intensity scores.

Furthermore, the proposed comprehensive rule recommends incorporating all data into the emotion dataset. However, an examination of how the dataset size impacts SER system performance was not conducted. This omission stems from the fact that the largest existing emotion dataset comprises merely around 200,000 utterances, significantly less than databases used in other speech tasks such as Automatic Speech Recognition (ASR). Additionally, extensive experiments under cross-domain conditions were not undertaken, and only a single experiment was performed. It is essential to note that cross-domain experiments are crucial for real-world application of SER systems.

In conclusion, this dissertation recommends utilizing the original emotion classifications within the database. A pertinent question that arises is how many emotion classes are sufficient to accurately capture emotional perception in real-world scenarios. According to \cite{Cowen_2021}, humans are capable of detecting 24 distinct emotions through vocal expressions. Therefore, it would be intriguing to combine multiple emotion databases to encompass these 24 emotions.

\section{Future Works}
All data and emotional ratings in the emotion databases can be utilized with the proposed methods during inference. In future studies, the first benchmark for SER will be proposed based on the partitions defined in this dissertation, allowing for a standardized evaluation of models and easy comparison of performances across different SER models. The intention is to examine performance disparities caused by natural biases in emotional perceptions, including gender, race, and ethnicity. Additionally, the impact of missing modality on SER system performance will be explored, considering that real-world conditions might lead to signal loss. Noisy label learning, such as facial expression recognition \cite{Zhang_2021_2}, is worth investigating to check whether it can be useful in improving SER systems. Multilingual SER systems will be developed in at least ten languages due to slight variations in emotion perception across languages, which might offer complementary information to enhance overall system performance. Finally, personalized SER systems will be constructed based on speaker or user profiles to improve user experience in real-world applications, recognizing that each individual could have unique emotional responses.

\clearpage
\bibliographystyle{IEEEtran}
\bibliography{references}

\end{document}